\documentclass{aa}  
\usepackage{aalongtable,lscape}
\usepackage{graphicx}
\usepackage{txfonts}
\begin{document}
\title{Young stars and brown dwarfs surrounding\\ 
Alnilam ($\epsilon$~Ori) and Mintaka ($\delta$~Ori)}
\titlerunning{Young stars and brown dwarfs surrounding Alnilam and Mintaka}
%
%
\author{J. A. Caballero\inst{1,2}
\and
E. Solano\inst{3}}
\authorrunning{Caballero \& Solano} 
\offprints{Jos\'e Antonio Caballero, \email{caballero@astrax.fis.ucm.es}} 
\institute{
Dpto. de Astrof\'{\i}sica y Ciencias de la Atm\'osfera, Facultad de F\'{\i}sica,
Universidad Complutense de Madrid, E-28040 Madrid, Spain
\and
Max-Planck-Institut f\"ur Astronomie, K\"onigstuhl 17, D-69117 Heidelberg,
Germany 
\and
Laboratorio de Astrof\'{\i}sica Espacial y F\'{\i}sica Fundamental, INSA,
Apdo. 78, E-28691 Villanueva de la Ca\~nada, Madrid, Spain}
\date{Received 17 February 2008; accepted 9 April 2008}

\abstract
{}  
{We look for new regions for the search of substellar objects.}   
{Two circular areas, 45\,arcmin-radius each, centred on the young massive star
systems Alnilam and Mintaka in the Orion Belt, have been explored.
The regions are very young (less than 10\,Ma), have low extinction, and are
neighbours to $\sigma$~Orionis ($\sim$3\,Ma), a young open cluster very rich in
brown dwarfs and planetary-mass objects.
We have used Virtual Observatory tools, the astro-photometric Tycho-2, DENIS and
2MASS catalogues, 10 control fields at similar galactic latitudes, and X-ray,
mid-infrared and spectroscopic data from the literature.} 
{We have compiled exhaustive lists of known young stars and new candidate
members in the Ori~OB1b association, and of fore- and background sources.
A total of {136} stars display features of extreme youth, like early
spectral types, lithium in absorption, or mid-infrared flux excess.
Other two young brown dwarf and {289} star candidates have been identified from
an optical/near-infrared colour-magnitude diagram.
We list additional {74} known objects that might belong to the association.
This catalogue can serve as an input for characterisation of the stellar and
high-mass substellar populations in the Orion Belt.
Finally, we have investigated the surface densities and radial distributions of
young objects surrounding Alnilam and Mintaka, and compared them with those in
the $\sigma$~Orionis cluster.
We report a new open cluster centred on Mintaka.} 
{Both regions can be analogs to the $\sigma$~Orionis cluster, but more massive,
more extended, slightly older, and less radially concentrated.} 
\maketitle
%

\section{Introduction}
\label{introduction}

The knowledge of the frequency and characteristics of brown dwarfs (substellar
objects with masses below the hydrogen burning limit) is essential
for the most advanced scenarios of fragmentation of molecular clouds and very
low-mass star formation (Reipurth \& Clarke 2001; Bate, Bonnell \& Bromm 2003;
Whitworth et~al. 2007).
In particular, they provide valuable information on the bottom of the Initial
Mass Function (e.g. Luhman et~al. 2000). 
Brown dwarfs are found in the field, both as companions to stars (Nakajima
et~al. 1995; Rebolo et~al. 1998; Goldman et~al. 1999) and free-floating
(Delfosse et~al. 1997; Ruiz, Leggett \& Allard 1997; Kirkpatrick et~al. 1999).  
Brown dwarfs are much brighter when younger (Chabrier \& Baraffe 2000); 
they are, thus, very common in young open clusters and star-forming regions,
such as the \object{Pleiades} (Rebolo, Zapatero Osorio \& Mart\'{\i}n 1995),
\object{$\rho$~Ophiuchi} (Luhman, Liebert \& Rieke 1997),
\object{Chamaeleon}~I+II (Neuh\"auser \& Comer\'on 1998), \object{Taurus-Auriga}
(Brice\~no et~al. 1998) or the \object{Orion Nebula Cluster} (Hillenbrand \&
Carpenter 2000). 
There are, however, limitations in the search for brown dwarfs in these regions
and in others: 
in the youngest ones (e.g. Chamaeleon, Ophiuchus), there is variable extinction
that hinders the characterisation of the recently-born brown dwarfs, while in
the others (e.g. Pleiades, Hyades), the relatively old brown dwarfs have dimmed
down to faint magnitudes that force to use very large, expensive, astronomical
facilities. 

There exists, nevertheless, a cornerstone for the search of brown dwarfs and
objects below the deuterium burning-mass limit: the \object{$\sigma$~Orionis}
cluster in the \object{Ori~OB1b} association (Garrison 1967; Lyng\aa~1981;
Walter et~al. 1997).
The cluster is very young ($\sim$3\,Ma), practically free of extinction ($A_V
\lesssim$ 0.3\,mag) and relatively nearby ($d \sim$ 385\,pc).
See Caballero (2007a, 2008c) for extensive bibliographic and data compilations
on the cluster.
It does not only harbour a rich population of OB-type, Herbig Ae/Be and
T~Tauri stars, but also Herbig-Haro objects, X-ray emitters and substellar
objects (Wolk 1996; Reipurth et~al. 1998; Oliveira \& van Loon 2004;
Franciosini, Pallavicini \& Sanz-Forcada 2006). 
Indeed, the $\sigma$~Orionis cluster possesses the best spectroscopically
investigated and most numerous population of brown dwarfs and 
planetary-mass objects down to a few Jupiter masses (Zapatero Osorio et~al.
2000; B\'ejar et~al. 2001; Caballero et~al. 2006). 
Besides, an important fraction of the $\sigma$~Orionis area has been already
covered or is being currently investigated by very deep, wide photometric
surveys, screening the whole brown dwarf and part of the planetary mass regimes
(Gonz\'alez-Garc\'{\i}a et~al. 2006; Caballero et~al. 2007; Bihain et~al.,
in~prep.). 

To compare substellar mass functions, spatial distributions or disc frequencies
of different clusters, and to look for new brown dwarfs and planetary-mass
objects, it is necessary, therefore, to search for new locations. 
Youth, closeness and low extinction, just like in $\sigma$~Orionis, are strongly
required.
Since the new hunting grounds for the search of substellar objects must resemble
$\sigma$~Orionis, it is natural to look for them not far away, just in the
Ori~OB1b~association.

\subsection{Alnilam and Mintaka}
\label{alnilamandmintaka}

A ``clustering of early-type stars elongated roughly parallel to the Galactic
plane'' (Guetter 1979) was firstly noticed by Pannekoek (1929).
Later, in his classical review of nearby O-type associations, Blaauw (1964)
described an \object{Ori~OB1} complex splitted into four divisions, being
Ori~OB1b (the ``Orion Belt'') one of them.
Wide survey observations with Schmidt telescopes have shown a patent overdensity
of H$\alpha$ emission stars in the area (Haro \& Moreno 1953; Wiramihardja
et~al. 1989). 
Canonical age and heliocentric distance are in the intervals 1--7\,Ma and
350--500\,pc (e.g. Anthony-Twarog 1982; Lyng\aa~1987; Blaauw 1991; Brown et~al.
1994; de Zeeuw et al. 1999; Harvin et~al. 2002). 
The most representative stars in the Ori~OB1b association are the bright
O-type supergiants \object{Alnitak}, \object{Alnilam} and \object{Mintaka}
($\zeta$~Ori, $\epsilon$~Ori and $\delta$~Ori, respectively), that constitute
the celebrated asterism of the Orion Belt.
At least one star of the trio (Alnilam) was depicted in the Farnese Atlas and,
therefore, tabulated in the original Hipparchus catalogue (Schaefer 2005).
The easternmost supergiant, Alnitak, is nearly embedded in the core of the
\object{L1630}/Orion~B molecular cloud complex, which contains, among other
nebulosities and H~{\sc ii} regions, the \object{Flame Nebula} (NGC~2024) and
the \object{Horsehead Nebula}.
The high variable extinction and emission in the area (Jaffe et~al. 1994; Lacy
et~al. 1994; Kramer, Stutzki \& Winnewisser 1996) prevent from suitably studying
its stellar and substellar populations {\em \`a~la} $\sigma$~Orionis.

The stellar populations in the Ori~OB1b association have been characterised
after Blaauw's (1964) seminal work by many authors (Hardie, Hesser \& Tolbert
1964; Warren \& Hesser 1978; Guetter 1981; Brown et~al. 1994; Hern\'andez et~al.
2005). 
In contrast to the hypothesis of Sharpless (1962), who claimed a lack of
noticeable excess of members in Ori~OB1b later than A5, at least two nearby
clusters are known within the association: 
$\sigma$~Orionis, centred on the eponym {$\sigma$~Ori} Trapezium-like
star system (see above), and \object{Collinder~70}, centred on Alnilam
(Collinder 1931).
There is an additional open cluster in the background, to the north of Mintaka:
\object{Berkeley~20}.  
It is a unusual, low-metallicity, old open cluster at $d \sim$ 8.4\,kpc, and
about 2.5\,kpc below the Galactic plane (Lyng\aa~1987; MacMinn et~al.~1994).

While the size of $\sigma$~Orionis is well determined at 20--30\,arcmin (B\'ejar
et~al. 2004; Sherry, Walter \& Wolk 2004; Caballero 2008a), the actual size of 
Collinder~70 is not ascertained.
Some authors (e.g. Gieseking 1983; Dias, L\'epine \& Alessi 2001) have
identified the Collinder~70 cluster, sometimes called the ``$\epsilon$~Orionis
cluster'', as the whole Ori~OB1b association.
Markarjan (1951) (and, therefore, Lyng\aa~1987) tabulated an angular diameter of
149\,arcmin, which would make the cluster to comprise the stellar populations
surrounding Mintaka and $\sigma$~Ori.
Subramaniam et~al. (1995) proposed that both Collinder~70 and \object{NGC~1981}
(to the north of the Orion Nebula Cluster) form a ``probable binary open star
cluster in the Galaxy''.
Some classical works also catalogued a bright diffuse galactic nebulae
centred on Alnilam, and with an apparent size of 50\,arcmin 
(Dreyer 1888 -- \object{NGC~1990}; Cederblad 1946 -- \object{Ced~55h}).
The existence of (numerous) small cometary globules, remnant molecular clouds
and giant outflows close to Alnilam and Mintaka, even in larger amount than
close to $\sigma$~Ori (Cernicharo et~al. 1992; Yun et~al. 1997; Ogura \&
Sugitani 1998; Mader et~al. 1999), favours the hypothesis of a wide region with
a rather homogeneous age of no more than $\sim$7\,Ma. 
See also Wolk (1996) and Scholz \& Eisl\"offel (2005) for other age
determinations. 
The recent determinations of heliocentric distances to \object{VV~Ori~AB}, a 
double-lined eclipsing binary in a detached configuration close to Alnilam ($d$
= 388$\pm$30\,pc; Terrell, Munari \& Siviero 2007), and to $\sigma$~Ori~AB, one
of the most massive binaries known ($d \sim$ 385\,pc assuming the hierarchical
triple scenario -- Caballero 2008b; D.~M. Peterson et al., in~prep.), seem
to support the homogeneity of the region. 
There are hints, nonetheless, of subestructure and overlapping populations
within the association, as firstly noticed by Hardie et~al. (1964) and Warren \&
Hesser (1977), and later by Jeffries et~al. (2006), Caballero (2007a) and
Caballero \& Dinis~(2008). 

Searches for low-mass young stars in the central and western regions of
the Ori~OB1b association (the $\sigma$~Orionis cluster is to the east) have
been already carried out by Sherry, Walter \& Wolk (2000), Sherry (2003) and
Brice\~no et~al. (2005).
Several very low-mass star and brown dwarf candidates have also been
identified surrounding Alnilam by B\'ejar (2001), P\'erez-Garrido,
D\'{\i}az-S\'anchez \& Villo (2005) and Scholz \& Eisl\"offel (2005). 
These works are, however, biased towards very low-mass objects or incomplete. 
On the one hand, P\'erez-Garrido et~al. (2005) selected brown dwarf candidates
in the 2MASS catalogue (Skrutskie et~al. 2006) without optical counterpart in
the USNO-A2.0 catalogue (Monet et~al. 1998); 
the majority, if not all, of their 23 objects are, nonetheless, identified in
the red optical passbands of the most recent, deeper USNO-B1.0 and DENIS
catalogues (Epchtein et~al 1999; Monet et~al. 2003).
On the other hand, B\'ejar (2001) and Scholz \& Eisl\"offel (2005)
selected their cluster member candidates from deep optical surveys ($IZ$, $RI$)
in $\sim$1000\,arcmin$^2$-wide areas to the southeast and the northwest of
Alnilam, respectively.
Each of them might have surveyed less than one quarter of the minimum size of
Collinder~70, and only for objects fainter than $I$ = 14--16\,mag.
B\'ejar et~al. (2003a) and B\'ejar, Caballero \& Rebolo (2003b) carried out the
spectroscopic follow-up of nine ``$\epsilon$~Orionis cluster'' photometric
member candidates presented in B\'ejar (2001). 
Derived spectral types ranged between M4.5 and M8, at the expected star-brown
dwarf boundary.
Some of the spectra also showed features of extreme youth (Li~{\sc i}
$\lambda$6707.8\,\AA~in absorption, H$\alpha$ $\lambda$6652.8\,\AA~in emission,
and/or faint alkali lines -- a signpost of low surface gravity, low~$g$).
No searches for brown dwarfs have been carried out in the Mintaka region~yet.

In the present work, we compile lists of confirmed very young stars, association
member candidates and fore- and background sources in two wide survey areas
centred on Alnilam and Mintaka, using tools of the Virtual Observatory.
The final catalogue, ranging from the two massive OB-type supergiants to
intermediate M-type substellar objects, is useful for next studies on
characterisation of stars and brown dwarfs, initial mass function, frequency and
properties of discs and multiplicity in the Ori~OB1b association, and
complements past and future works (e.g. B\'ejar et~al., in~prep.;
P\'erez-Garrido et~al., in~prep.). 
Finally, we present a preliminar study of the spatial distribution of
association members and~candidates.

\section{Analysis and results}

\subsection{Data and survey areas}

\begin{figure}
\centering
\includegraphics[width=0.49\textwidth]{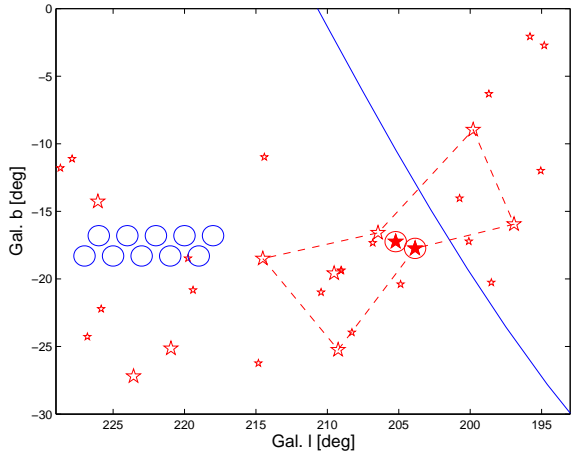}
\caption{Pictogramme showing the Orion-Lepus region and the survey area.
{\em Stars}: some of the brightest stars of Orion, Lepus and adjacent
constellations (sizes are roughly proportional to brightness). 
The filled stars are Alnilam and Mintaka.
{\em Circles}: comparison fields and the survey area around Alnilam and Mintaka.
{\em Solid line}: parallel of constant declination $\delta$ = +2\,deg.
There are no DENIS data northern (i.e. to the right) of the parallel.
Colour versions of all our figures are available in the electronic publication.}
\label{felb}
\end{figure}

\begin{figure*}
\centering
\includegraphics[height=0.32\textwidth]{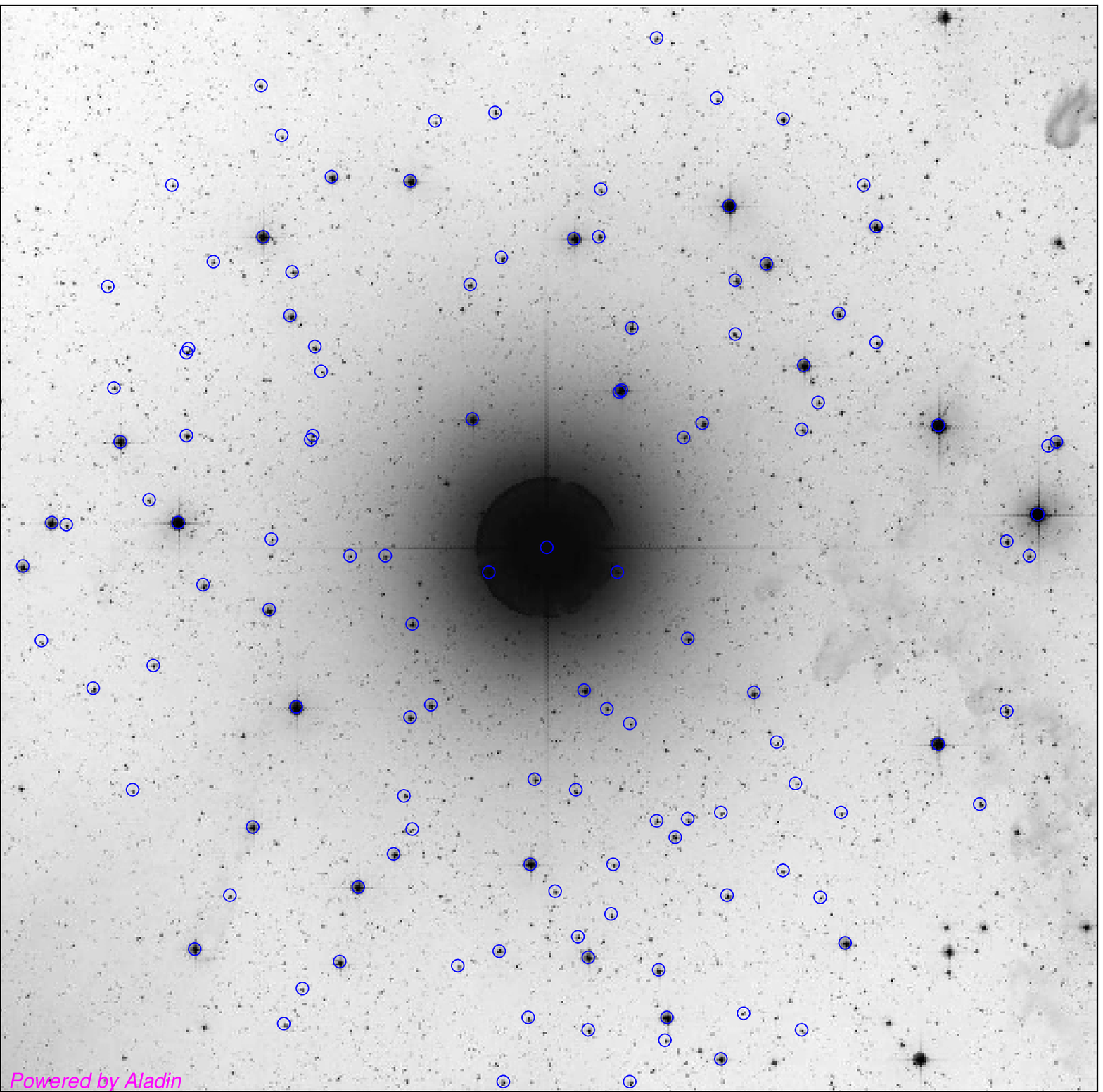}
\includegraphics[height=0.32\textwidth]{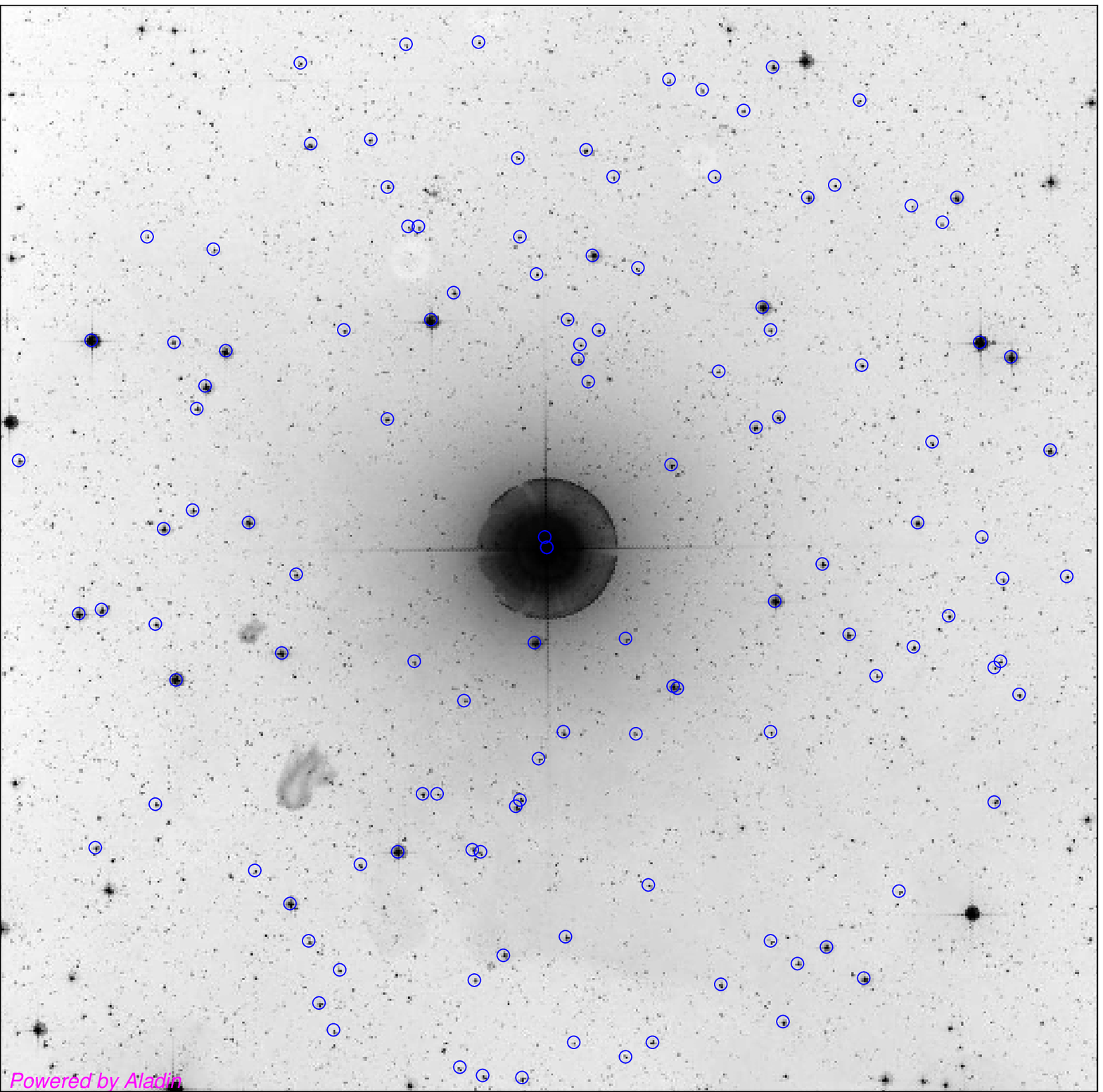}
\includegraphics[height=0.32\textwidth]{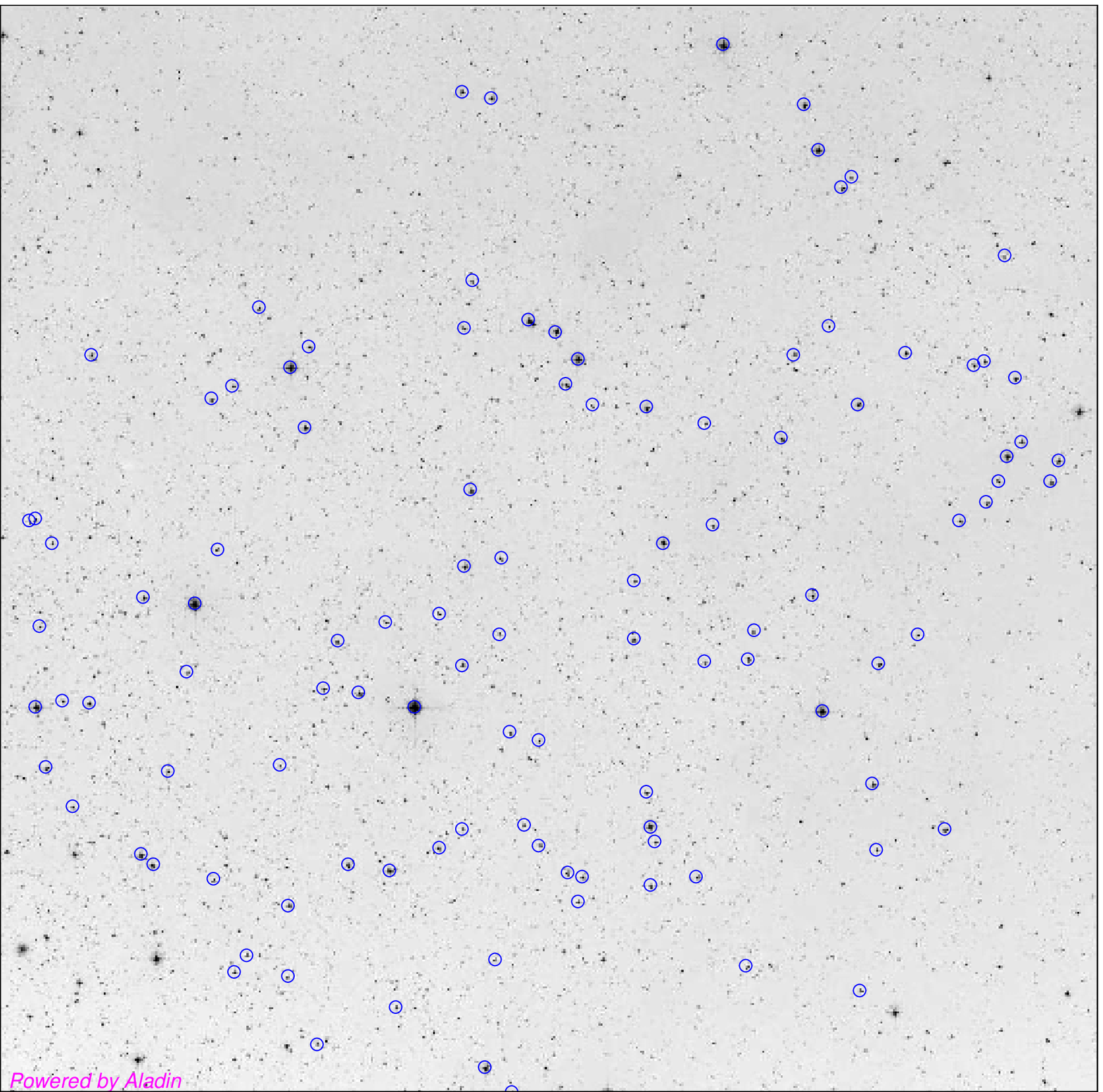}
\caption{Inverted-colour, SERC $B_J$ (photographic blue) DSS1 images of the
Alnilam (left), Mintaka (centre) and $c_1$ (right) fields. 
Circles (in blue) indicate Tycho-2/2MASS stars.
Approximate sizes are 1.5\,$\times$\,1.5\,deg$^2$.
North is up, east is left.
A tonge-shaped nebula appears in both Alnilam and Mintaka images
(\object{IC~434}; see Fig.~\ref{therollingstones}).}
\label{alnilamandmintakaandc1}
\end{figure*}

The search for stellar and substellar members of the Ori~OB1b association
around Alnilam and Mintaka were conducted using the Tycho-2 (H{\o}g et al.
2000), DENIS and 2MASS catalogues and photometric, spectroscopic and astrometric
information from the literature.
Two main survey fields were defined as 45\,arcmin-radius circles centred on the
stars.  
Since the two circular fields are almost tangent, the size choice simultaneously
allows exploring a large manageable area of about 3.5\,deg$^2$, and prevents
overlapping between them
(there is, however, a tiny overlapping of a few arcmin$^2$ between both circular
fields). 
The radius of investigation is roughly twice the radius of the core of the
$\sigma$~Orionis cluster, that is expected to be a model.
To discard background and foreground objects in the two main fields, we selected
ten nearby, same-size comparison fields, $c_i$ ($i$ = 1, 2... 10), fulfilling
the following requirements:
\begin{itemize}
\item data availability (i.e. $\delta <$ +2\,deg; there is no DENIS data for
larger declinations);
\item separation from the Orion region to avoid gas clouds and dispersed young
stellar populations;
\item same galactic latitude to the main fields ($b \sim$ --17.5\,deg);
\item similar galactic longitude to the main fields, but keeping away from the
Galactic bulge to minimise the number of background giant stars ($l \sim$
215--230\,deg). 
\end{itemize}
\noindent
Central coordinates (equatorial and Galactic) of the twelve fields are given in
Table~\ref{fields}. 
Both main and comparison fields were explored down to limiting magnitude
$V_{T}$ = 11.5\,mag (Tycho-2), $i$ = 18.0\,mag (DENIS), $J,~H,~K_{\rm s}$ =
17.1, 16.4, 14.3\,mag (2MASS) 
[in Tables~\ref{TYC2Malnilam} and~\ref{TYC2Mmintaka}, there are Tycho-2 stars
fainter than the indicated $V_{T}$-band limiting magnitude]. 
A pictogramme of the survey areas is given in Fig,~\ref{felb}, while
Fig.~\ref{alnilamandmintakaandc1} shows the images of the fields surrounding
Alnilam and Mintaka and of one comparison~field. 

   \begin{table}
      \caption[]{Observed fields (45\,arcmin-radius each).} 
         \label{fields}
     $$ 
         \begin{tabular}{lccccc}
            \hline
            \hline
            \noalign{\smallskip}
Name	& $\alpha$ 		& $\delta$ 		& $l$ 	& $b$		\\  
	& (J2000)		& (J2000)		& [deg]	& [deg]		\\  
            \noalign{\smallskip}
            \hline
            \noalign{\smallskip}
Mintaka	& 05 32 00.40		& --00 17 56.7		& 203.9	& --17.7	\\  
Alnilam	& 05 36 12.81		& --01 12 06.9		& 205.2	& --17.2	\\  
$c_1$	& 05 59 46.51		& --11 57 00.3  	& 218.0 & --16.8	\\  
$c_2$	& 05 55 49.69		& --13 26 39.1   	& 219.0 & --18.3	\\    
$c_3$	& 06 03 04.36		& --13 41 16.2   	& 220.0 & --16.8	\\ 
$c_4$	& 05 59 03.14		& --15 10 29.9  	& 221.0 & --18.3 	\\
$c_5$	& 06 06 20.78		& --15 25 51.3   	& 222.0 & --16.8 	\\
$c_6$	& 06 02 15.03		& --16 54 41.1  	& 223.0 & --18.3 	\\
$c_7$	& 06 09 36.11 		& --17 10 44.3	  	& 224.0 & --16.8 	\\
$c_8$	& 06 05 25.67 		& --18 39 11.3	  	& 225.0 & --18.3 	\\
$c_9$	& 06 12 50.69 		& --18 55 53.7	  	& 226.0 & --16.8 	\\
$c_{10}$& 06 08 35.37 		& --20 23 59.1	  	& 227.0 & --18.3 	\\
            \noalign{\smallskip}
            \hline
         \end{tabular}
     $$ 
   \end{table}

In what follows, information retrieval, data manipulation, filtering and
selection has been done taking advantage of Virtual
Observatory\footnote{http://www.ivoa.net} standards and tools, in particular
Aladin\footnote{http://aladin.u-strasbg.fr/aladin.gml} 
(Bonnarel et~al. 2000) and
TOPCAT\footnote{http://www.star.bris.ac.uk/~mbt/topcat/}.   
The Virtual Observatory is an international, community-based initiative to
provide seamless access to the data available from astronomical archive and
services, as well as to develop state-of-the-art tools for the efficient
analysis of this huge amount of information.  
Padovani et~al. (2004), Tsalmantza et~al. (2006), and Caballero \& Solano (2007)
are examples of the efficiency of such tools in helping astronomers to produce
scientific~results. 

\subsection{Bright early-type stars}
\label{earlytype}

\begin{figure*}
\centering
\includegraphics[width=0.49\textwidth]{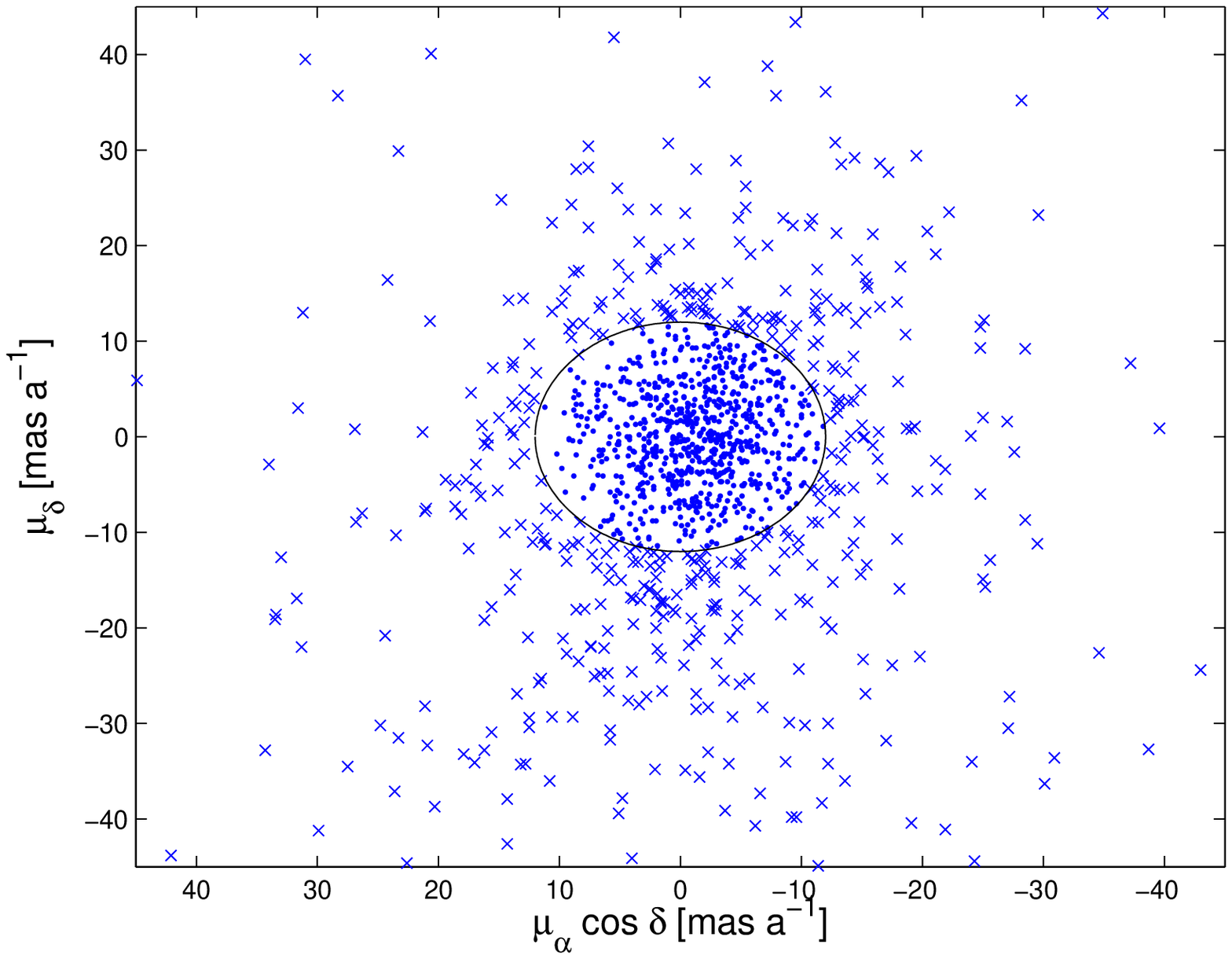}
\includegraphics[width=0.49\textwidth]{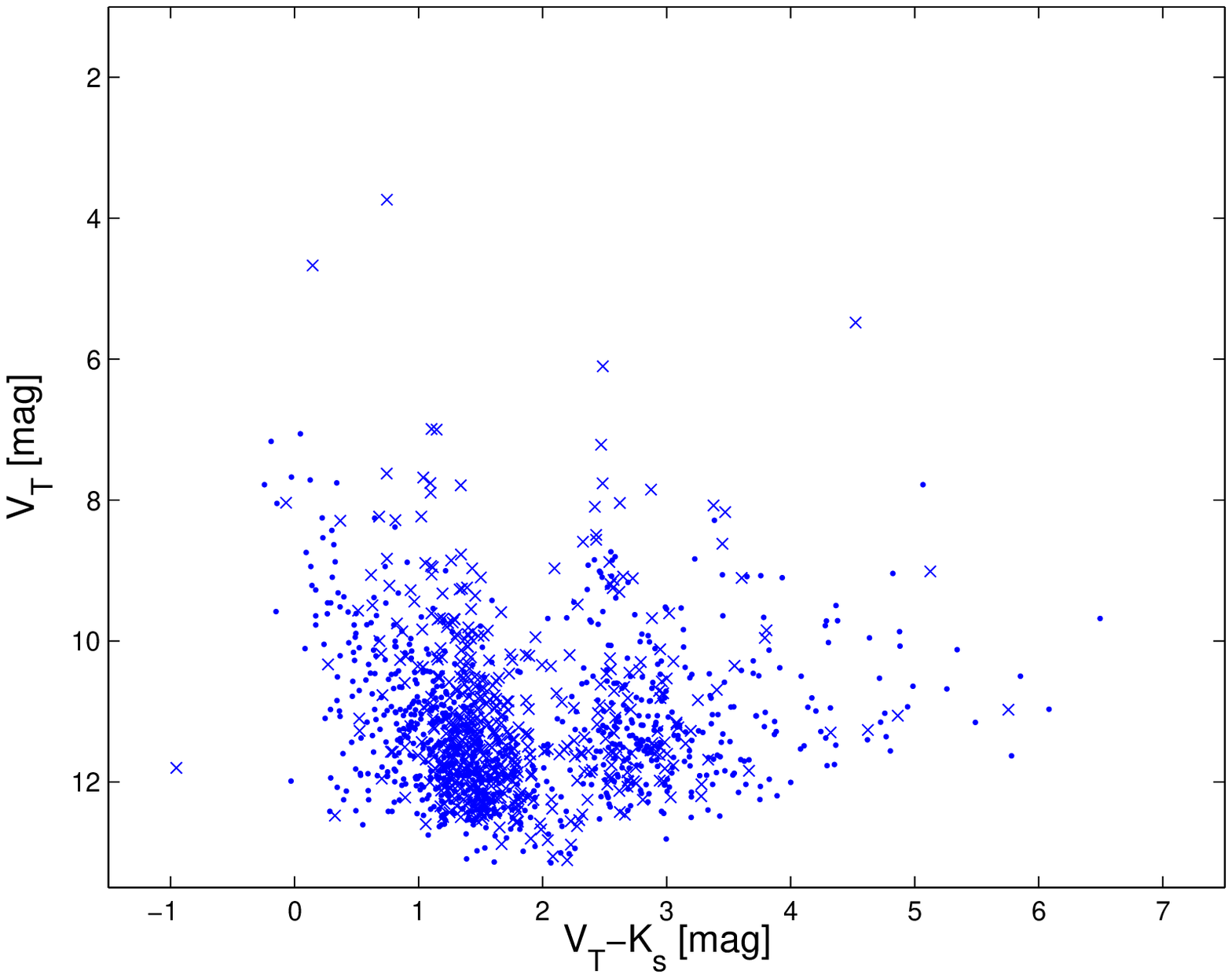}
\includegraphics[width=0.49\textwidth]{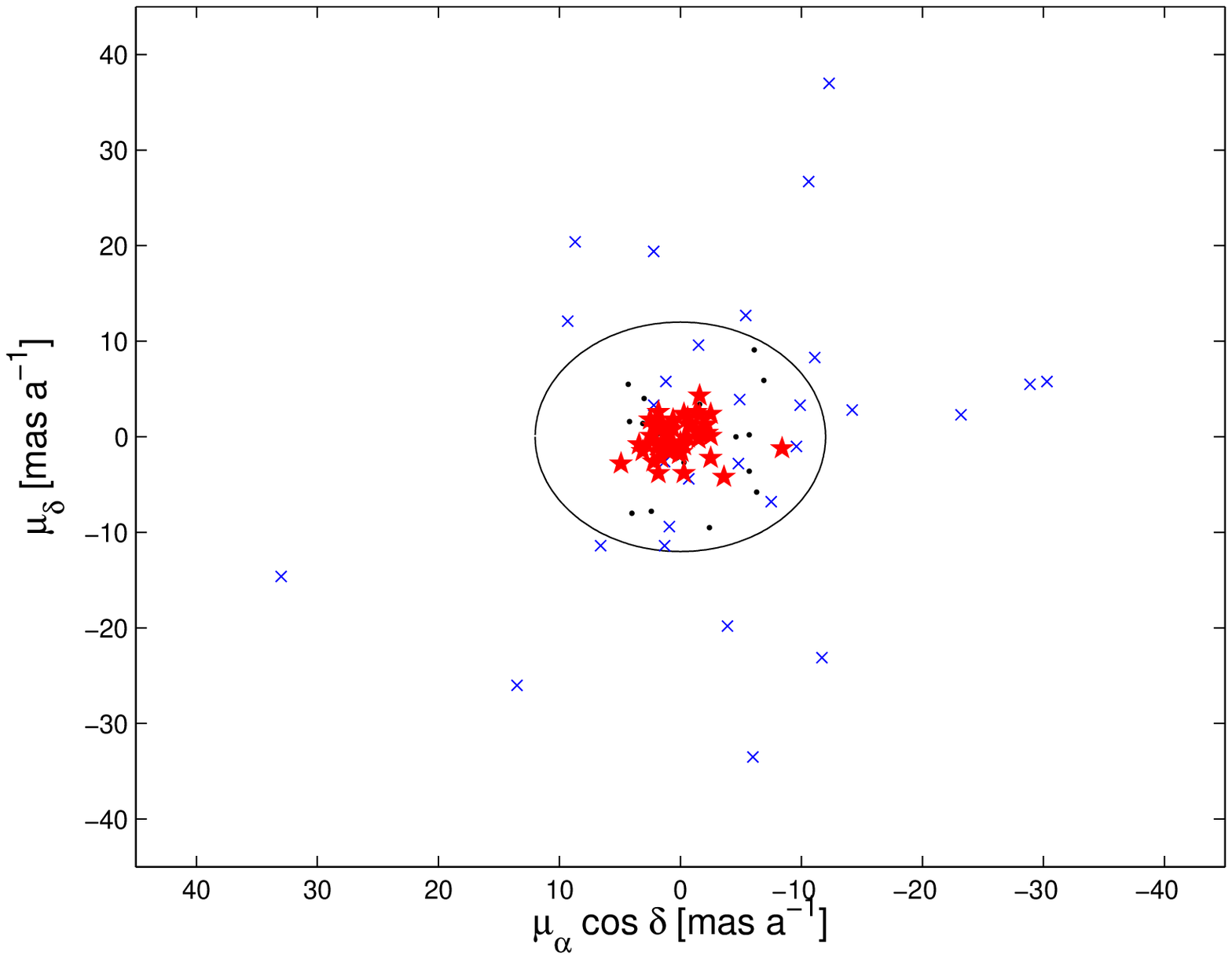}
\includegraphics[width=0.49\textwidth]{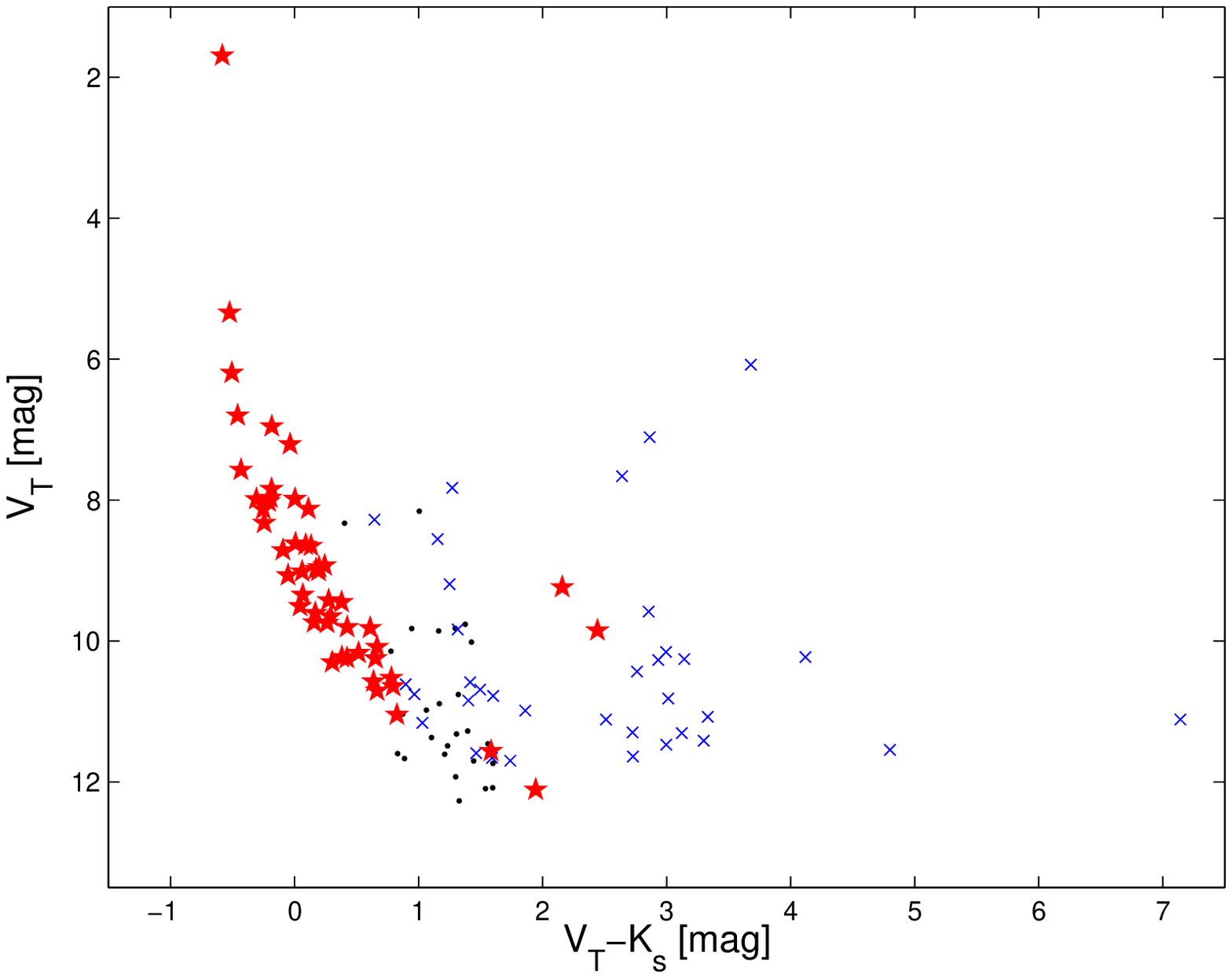}
\includegraphics[width=0.49\textwidth]{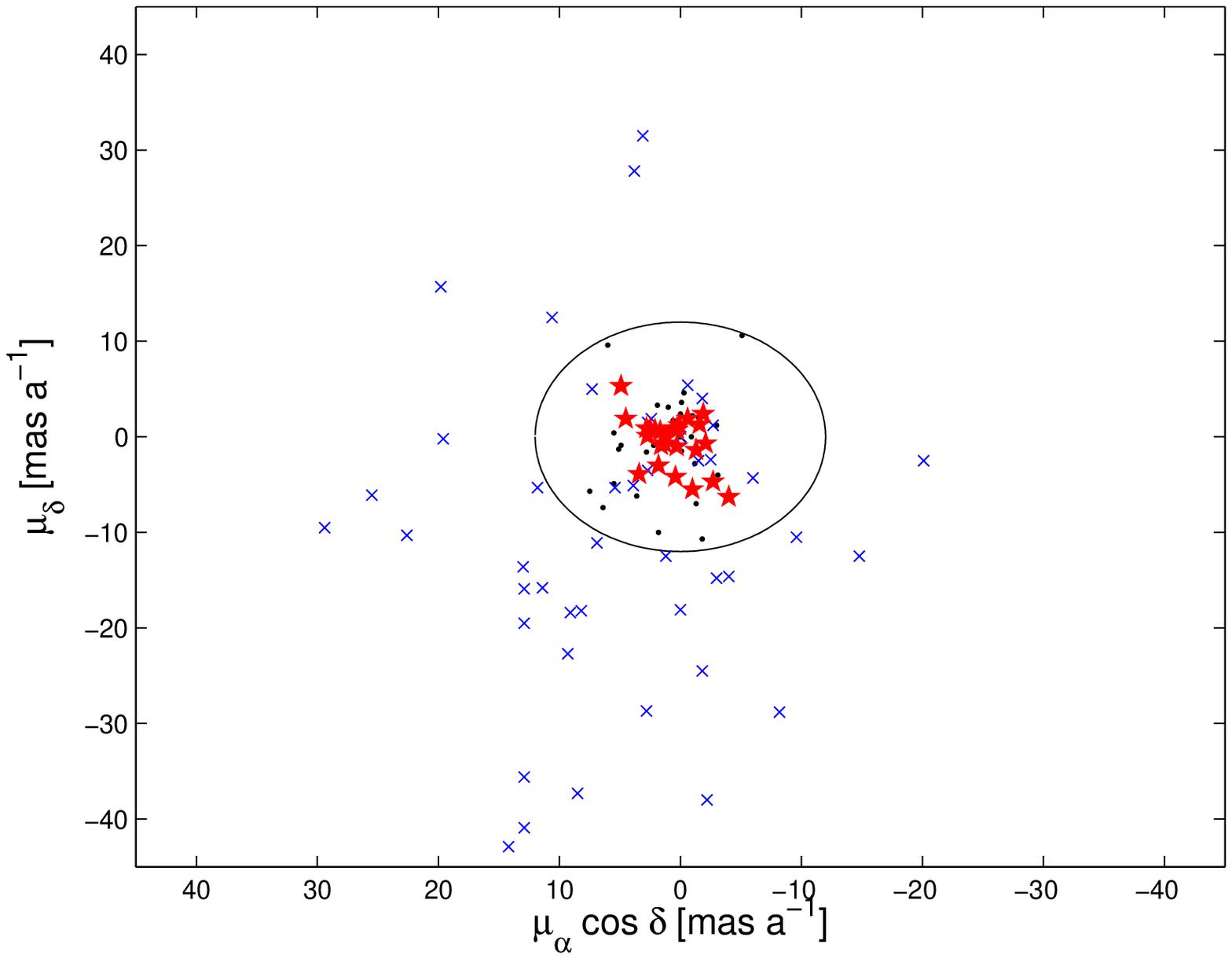}
\includegraphics[width=0.49\textwidth]{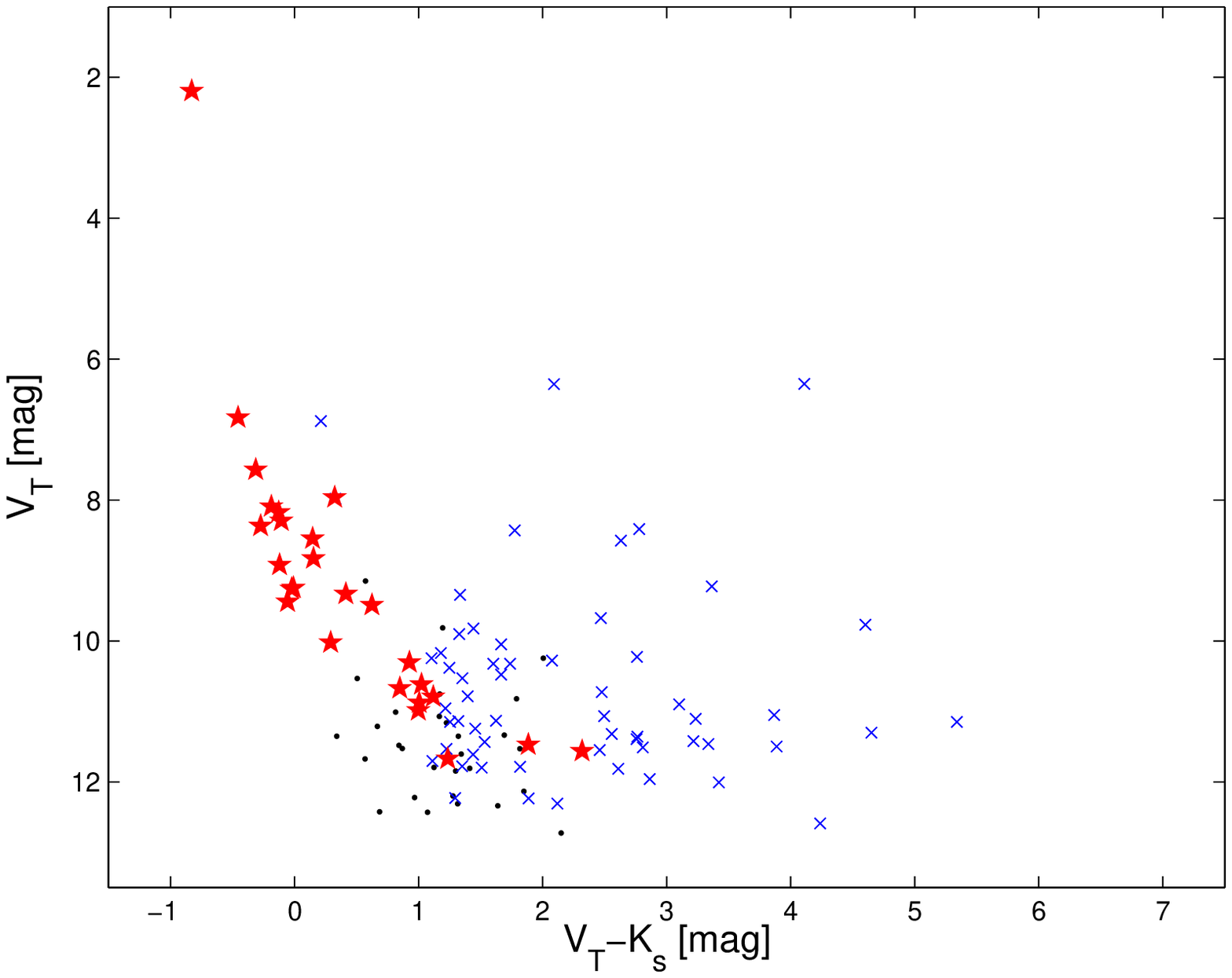}
\caption{The Tycho-2/2MASS cross-match in the 10 comparison fields (top), and
Alnilam (middle) and Mintaka (bottom) fields.
Stars classified as association Ori~OB1b non-members are marked with (blue)
crosses. 
Remaining stars are marked with (blue) dots, except for stars with signposts of
youth, that are marked with (red) filled stars.
{\em Left panel}: proper motion diagrams.
Big circles separate stars with proper motions larger than 12\,mas\,a$^{-1}$.
{\em Right panel}: $V_T$ vs. $V_T-K_{\rm s}$ colour-magnitude diagrams.}
\label{TYC2M}
\end{figure*}
%
%

We have followed a procedure very similar to that carried out in Caballero
(2007a) to identify bright very young stars using Tycho-2 and 2MASS astrometric
and photometric data.  
First, we loaded the data within the twelve 45\,arcmin-radius fields.
Secondly, a cross-correlation between Tycho-2 and 2MASS was done using the
Catalog Cross Match tool implemented in Aladin.
For each Tycho-2 source, the 2MASS counterpart was defined as the nearest source
found in a circle centred on the Tycho-2 source and radius of 4\,arcsec (the
default threshold). 
A total of 1276 Tycho-2 sources with 2MASS counterpart were identified within
the ten comparison fields, which yielded $\overline{N}$ = 130 objects per field
[$\sigma(N)$ = 30]. 
The actual number of Tycho-2/2MASS sources per $c_i$ field varies between
83 ($c_2$) and 169 ($c_7$), while there are $N$ = {111} and {113}
stars in the Alnilam and Mintaka fields, respectively.
The $B_T$, $V_T$, $J$, $H$ and $K_{\rm s}$ magnitudes of the {224} stars are
provided in Tables~\ref{allTYC2Malnilam} and~\ref{allTYC2Mmintaka}, one for
each main field. 

Although the number $N$ of Tycho-2/2MASS stars surrounding Alnilam and Mintaka
does not {\em a priori} support the hypothesis of a stellar overdensity, there
is an evident overdensity of bright {\em blue} stars surrounding the two
supergiants (as expected from an OB association).   
In the ten comparison fields, there are only six stars bluer than $V_T-K_{\rm
s}$ = 0.0\,mag and brighter than $V_T$ = 10.0\,mag. 
Two of them have accurate determinations of the parallax ($\pi / \delta \pi
> 3$) and are significantly closer to the Sun than the Ori~OB1b association:
\object{HD~40071} ($d$ = 250$\pm$70\,pc) and \object{HD~40355~AB} ($d$ = 
180$\pm$30\,pc). 
The other four stars (HD~41367~AB, HD~41488, HD~41737 and HD~42263) are B8--9
dwarfs, subdwarfs and subgiants located at tens of degrees from the young Orion
associations that seem to populate the interstellar field.  
Therefore, $\sim$0.6 (between 0 and 1) interloper stars bluer than $V_T-K_{\rm
s}$ = 0.0\,mag and brighter than $V_T$ = 10.0\,mag are expected to be in each of
the Alnilam and Mintaka fields.
However, there are actually 17 and 11 such stars, respectively. 
Hence, there are between 20 and 30 more bright blue stars surrounding the
two supergiants than in other regions at the same Galactic latitude and far from
the Orion star-forming region. 
Given the Tycho-2 limiting magnitudes and the expected spectral types (i.e.
colours) of young stars at $d \sim$ 385\,pc, we could only detect an overdensity
of blue (i.e. early-type) stars.
This calculated overdensity is similar to that in the $\sigma$~Orionis
cluster (Caballero 2007a) and justifies next~steps.

We have added other {21} Tycho-2/2MASS stars with blanks in the Tycho-2 proper
motion that were not considered by the Aladin Catalog Cross Match tool to the
list of {224} correlated Tycho-2/2MASS stars.  
In those cases, we have taken the proper motions from the USNO-B1 and Tycho-1
catalogues (H{\o}g et~al. 1998).
This addition makes the sample of bright stars to increase up to~245.

With the help of optical-near infrared colour-magnitude diagrams, the Tycho-2
proper motions, {\em IRAS} fluxes, spectroscopic data from the literature
and data from Vizier catalogues obtained using UCD\footnote{Unified Content
Descriptor.}-based searches (Ochsenbein, Bauer \& Marcout 2000), we have
classified the stars in the Alnilam and Mintaka fields into three groups,
separated by decreasing probability of membership in the Ori~OB1b association. 
On the one hand, the classification is illustrated with the six panels in
Fig.~\ref{TYC2M}. 
On the other hand, Tables~\ref{TYC2Malnilam} to \ref{TYC2Nmintaka} show the
results of the classification. 
The three groups are:

\begin{itemize}
\item stars with signposts of youth (Tables~\ref{TYC2Malnilam} and
\ref{TYC2Mmintaka}).
By features of youth we understand early spectral types (O and~B),
Li~{\sc i} in absorption, {\em strong} X-ray or H$\alpha$ emission (possibly
associated to accretion processes), and infrared excesses by circumstellar
material.  
We have also included low proper motion early A-type stars that follow the
sequence defined by the remaining young stars in the colour-magnitude diagrams.
There are only two stars with previously determined spectral type later than A: 
RX~J0535.6--0152~AB and SS~28.
Both of them display, however, features common of the T~Tauri phase (see notes
in Tables~\ref{TYC2Malnilam} and~\ref{TYC2Mmintaka}).
We have also considered additional youth features, like the star being in the
Herbig~Ae/Be phase (HAeBe) or having a Vega-like~disc.
\item stars with unknown association membership status
(Tables~\ref{TYC2Oalnilam} and \ref{TYC2Omintaka}).
They are stars with proper motions $\mu <$ 12\,mas\,a$^{-1}$ that do not deviate
very much from the young-star sequence in the colour-magnitude diagrams 
but have no known, clear, signposts of youth.
Some A-type stars with $\mu \sim$ 5--12\,mas\,a$^{-1}$ that do not follow the
sequence of the confirmed young stars are in this class.
The threshold for the maximum $\mu$ of 12\,mas\,a$^{-1}$ is larger than in
Caballero (2007a), who used $\mu >$ 10\,mas\,a$^{-1}$ to identify stars with
high tangential velocities among a list of photometric member candidates of the
Ori~OB1b association (the difference in the mean proper motion of the
association, almost null, is not significant).  
The new conservative threshold allows us to recognize some bona-fide
young stars with relatively large proper motions;
\item stars that do not belong to the association (Tables~\ref{TYC2Nalnilam} and
\ref{TYC2Nmintaka}).
This class comprises: 
($i$) foreground stars with proper motions $\mu >$ 12\,mas\,a$^{-1}$; 
($ii$) foreground G-, K- and M-type stars with spectral type determination;
($iii$) {\em Hipparcos} stars with $\pi / \delta \pi > 3$ and heliocentric
distances less than 250\,pc;
($iv$) red stars without spectral type determination, and with colours
$V_T-K_{\rm s} >$ 2.5\,mag and spectral energy distributions of K--M-type stars;
($v$) very red objects with colours $V_T-K_{\rm s} >$ 4.5\,mag and without
flux excess due to discs in the mid-infrared (see notes in
Appendix~\ref{catalogue}).  
There are stars that simultaneously satisfy more than one criterion.
For example, HD~36443 in the Mintaka field is a high proper motion ($\mu
\approx$ 488\,mas\,a$^{-1}$) G5V star located at $d \sim$ 38\,pc and with a
radial velocity $V_r \approx$ 9\,km\,s$^{-1}$ discordant with association
membership (Adams et~al. 1929; Roman 1955; Perryman et~al.~1997). 
\end{itemize}

   \begin{table}
      \caption[]{Number of bright stars per group and field.} 
         \label{classification}
     $$ 
         \begin{tabular}{lcccc}
            \hline
            \hline
            \noalign{\smallskip}
Field	& $N$ 		& $N$ 		& $N$ 		& $N$	\\  
	& members	& non-members	& unknown	& all	\\  
            \noalign{\smallskip}
            \hline
            \noalign{\smallskip}
Alnilam	& 49		& 39		& 34		& 122	\\  
Mintaka	& 29		& 63		& 31		& 123	\\  
            \noalign{\smallskip}
            \hline
            \noalign{\smallskip}
Total	& 78		& 102		& 65		& 245	\\  
            \noalign{\smallskip}
            \hline
         \end{tabular}
     $$ 
   \end{table}

Provided coordinates and proper motions in Tables~\ref{TYC2Malnilam}
to~\ref{TYC2Nmintaka} are from Tycho-2 (some exceptions are indicated).
The majority of the spectral types and features of youth listed in
the tables have been borrowed from the literature. 
The references are indicated in the last column.
Abundant notes and remarks on the discussed stars are also provided in
Appendix~\ref{catalogue}.
The most used works have been the spectroscopic studies in the Orion Belt by
Guetter (1976, 1979) and the Henry Draper Extension Charts by Nesterov et~al.
(1995).
The latter authors compiled positions, proper motions, photographic
magnitudes and spectral types from the original works by Cannon and Cannon \&
Pickering (e.g. Cannon \& Pickering 1918--1924).
Table~\ref{classification} summarizes the number of stars in each class and the
total number of correlated Tycho-2/2MASS stars.

From the comparison with the $c_i$ fields, between 0 and 1 late B-type stars
surrounding Alnilam and Mintaka may actually be fore- or background B8--9-type
stars. 
The contamination rate in the group of stars with signposts of youth 
is, therefore, $\sim$3.5--5.5\,\% for this spectral types, and null for late O-
and early B-type stars.
On the other hand, from the number of stars in the comparison fields with
colours 0.0\,mag $< V_T-K_{\rm s} <$ 0.5\,mag (28 of the 1276 stars in the ten
fields), it is expected that only $\sim$3 early A-type stars in the foreground
contaminate the Alnilam and Mintaka fields.
It leads to estimate the contamination by such stars in
Tables~\ref{TYC2Malnilam} to \ref{TYC2Omintaka} at less than~10\,\%.

\subsection{Intermediate- and late-type stars and brown~dwarfs}
\label{midlatetype}

We have performed a correlation between the DENIS and 2MASS catalogues identical
to the Tycho-2/2MASS one.
In this case, we have analysed the optical $i$ (DENIS) and near-infrared
$JHK_{\rm s}$ (2MASS) counterparts of more than 10$^5$ sources distributed
amongst the ten comparison and the two Orion Belt fields.
In particular, in the Alnilam and Mintaka fields, we have compiled the
coordinates and four-band photometry of 10523 and 8288 sources, respectively.
%

A total of {50} DENIS/2MASS stars with spectroscopic features of youth (i.e.
with Li~{\sc i}, H$\alpha$, low~$g$; see Section~\ref{alnilamandmintaka})
found in the literature have been identified in the Orion fields.
Most of them are located surrounding Alnilam and come from wide
prism-objective surveys of H$\alpha$ emitters by, e.g., Haro \& Moreno (1953;
Haro objects) and Wiramihardja et~al. (1989, 1991; Kiso objects), and from the
spectroscopic analyses by B\'ejar et~al. (2003a, 2003b; E\,Ori objects). 
We have not been able to identify \object{V1299~Ori}, an hypothetical B-type
star close to the remnant molecular cloud \object{[OS98]~40B}.
On the contrary, we have identified {8} DENIS/2MASS stars without known
spectroscopic features of youth, but with {\em strong} X-ray emission detected
by the {\em Einstein} and {\em ROSAT} satellites, that can be ascribed to a
young age. 
Only three of them had been previously catalogued (MacDowell 1994; Ueda
et~al. 2001).
Their colours are typical of young stars close to the main sequence.
The results of {\em XMM-Newton} observations centred on Alnilam will
be described in Caballero et~al. (in~prep.).

The names, coordinates, $i$- and $K_{\rm s}$-band magnitudes, features of youth
and references of the {58} young stars are provided in
Tables~\ref{DE2Malnilam} and~\ref{DE2Mmintaka}.
We list other spectro-photometric signposts of youth, such as Ca~{\sc ii}
$\lambda\lambda$3933.7,3968.5\,\AA~(H and K lines), H$\beta$
$\lambda$4861.3\,\AA, and [O~{\sc i}] $\lambda$6300.3\,\AA~in emission
(indicative of strong magnetic activity and/or outflows), and mid-infrared flux
excess (mIR; suggestive of the presence of a circumstellar disc).
Other {17} DENIS/2MASS stars in the literature have unreliable features of
youth, such as faint H$\alpha$ emission with $i-K_{\rm s}$ colour typical of
field stars, conservative upper limits of the strength of the Na~{\sc i} doublet
$\lambda\lambda$8183.3,8184.8\,\AA, or {\em faint} X-ray counterparts with
a large uncertainty ellipse.  
They are tabulated in Table~\ref{DE2O}.
Among them, we have not accounted for \object{Kiso~A--0904~62}, a
star-like source with blue colours and very faint H$\alpha$ emission in only one
epoch out of three in Wiramihardja et~al.~(1989).

\begin{figure}
\centering
\includegraphics[width=0.49\textwidth]{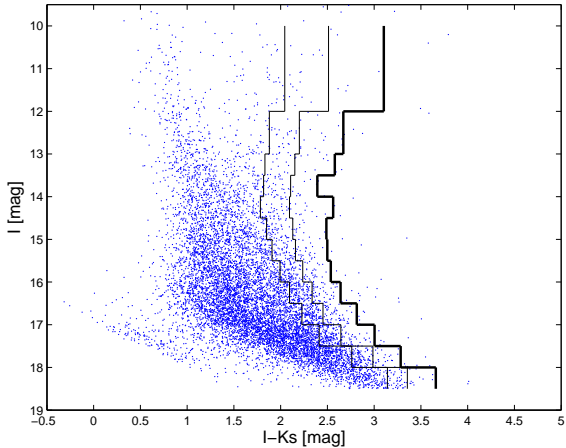}
\caption{$i$ vs. $i-K_{\rm s}$ colour-magnitude diagrams of one comparison
field.  
Solid lines are for $i-K_{\rm s}$ percentiles $x$ = 0.90, 0.97, 0.995, from left
to~right.} 
\label{feijksc}
\end{figure}
%
%

\begin{figure*}
\centering
\includegraphics[width=0.49\textwidth]{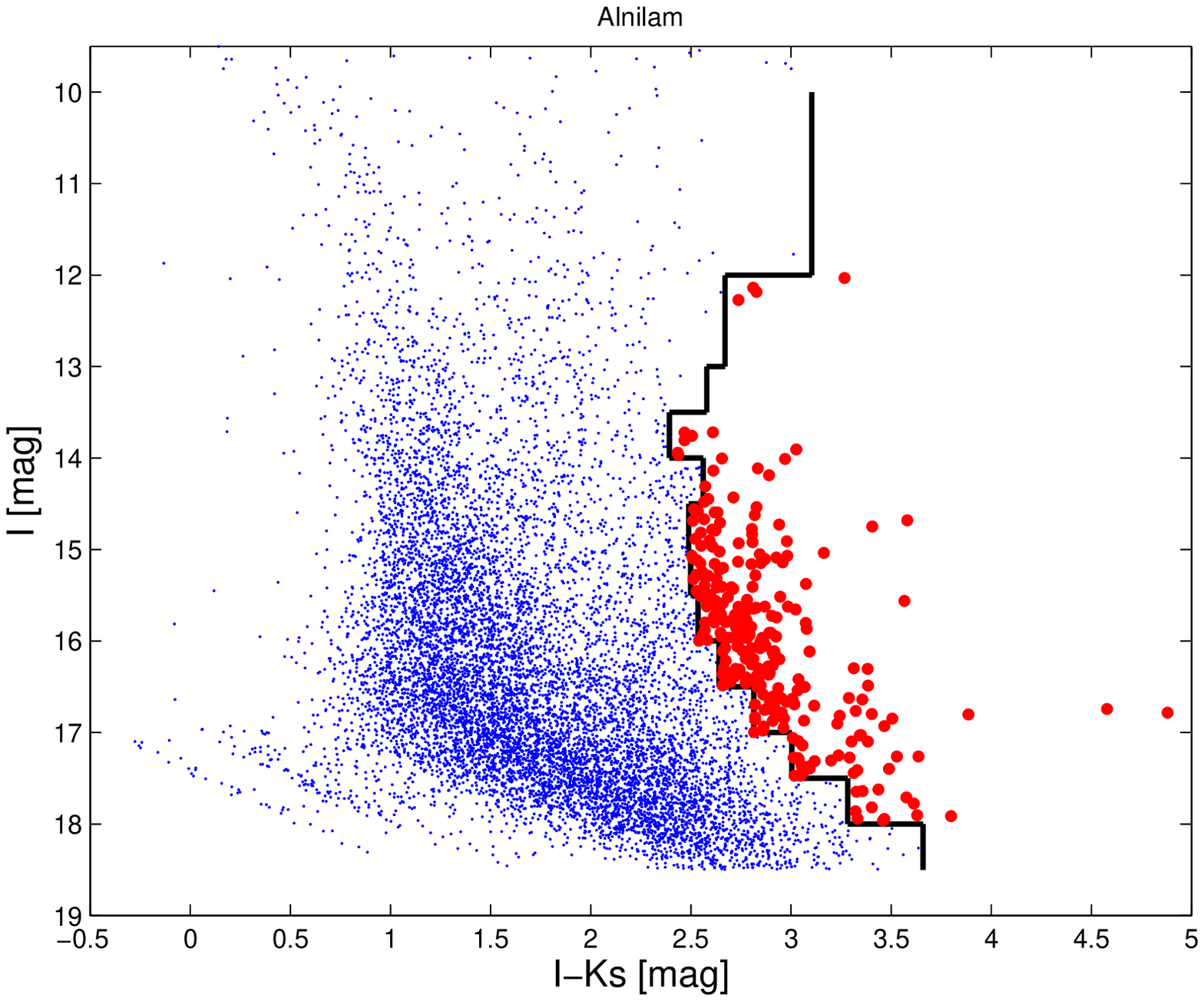}
\includegraphics[width=0.49\textwidth]{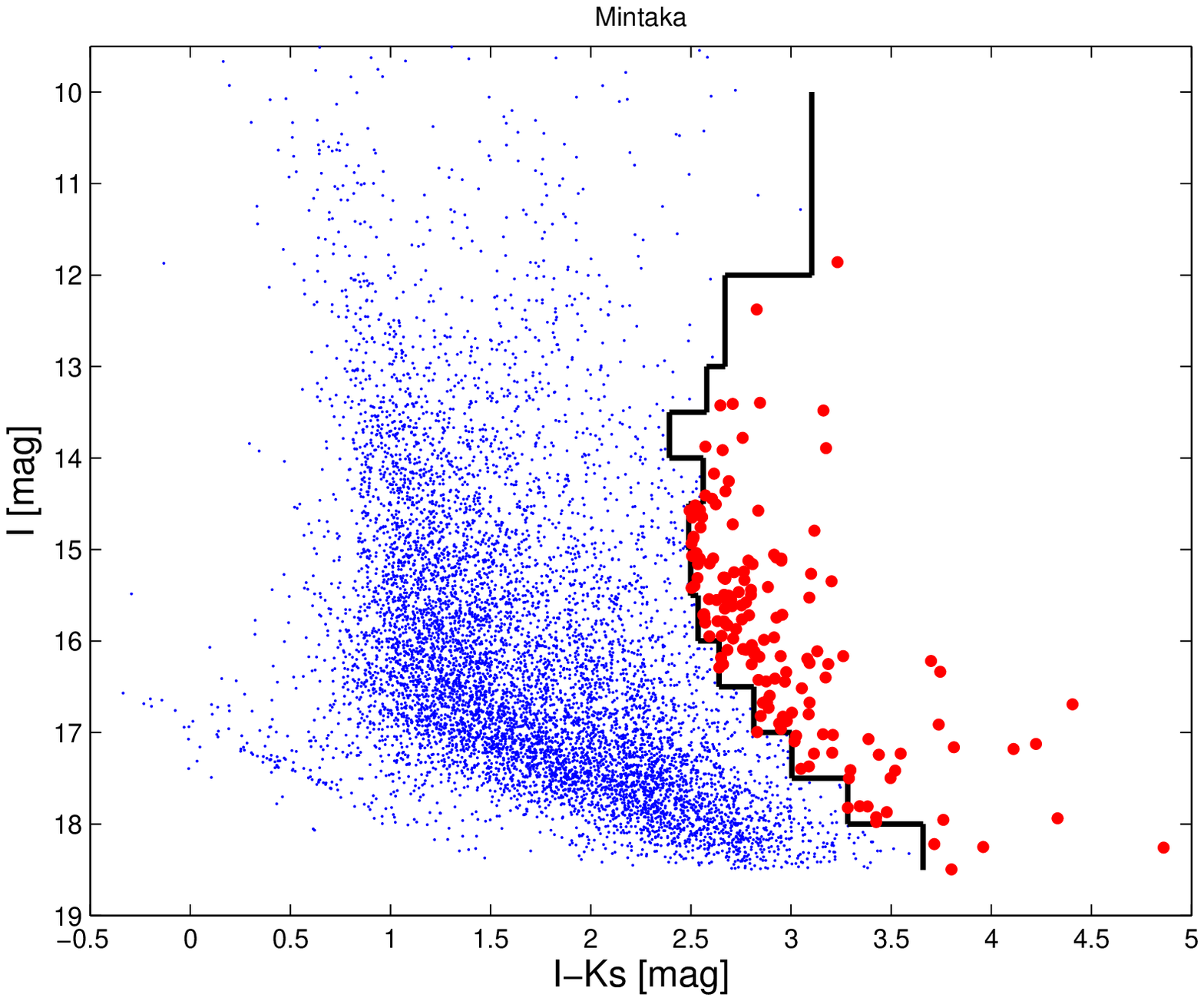}
\caption{Same as Fig.~\ref{feijksc}, but for the Alnilam (left) and Mintaka
(right) fields. 
Solid lines are for $i-K_{\rm s}$ percentile $x$ = 0.995.  
Objects to the red of this line are marked with (red) filled~circles.}
\label{feijks}
\end{figure*}

We have looked for new cluster member candidates without known features of
youth. 
Since the number of confirmed young objects in the analysed regions ({58}) is 
relatively low, we cannot define a lower envelope of association members as
Caballero (2008c) did in the $\sigma$~Orionis cluster (he used 241 young
objects). 
Besides, we also want to avoid the uncertainties at very young ages
associated to theoretical models (Baraffe et~al. 2003), and choose a selection
criterion as conservative, neutral, reproducible, and objective as possible.
For the photometric selection, we have used the data of the $\sim$87000 sources
in the comparison fields to determine the locations in the $i$ vs. $i-K_{\rm s}$
colour-magnitude diagrams of the Alnilam and Mintaka fields where the
probability of contamination by fore- and background sources is minimum.
Figs.~\ref{feijksc} and~\ref{feijks} illustrate the selection procedure.

First, we have divided the $i$ vs. $i-K_{\rm s}$ diagram (Fig.~\ref{feijksc}) of
all the comparison fields in eleven horizontal strips of width $\Delta i$ =
0.5\,mag between $i$ = 13.0 and 18.5\,mag, and two wider strips between $i$ =
10.0 and 13.0\,mag.  
Since the DENIS catalogue fails to provide accurate photometry for the brightest
stars (due to saturation and non-linear effects in their detectors), we will
only investigate sources with $i >$ 10.0\,mag in this section.
Secondly, we have computed for each strip the $i-K_{\rm s}$ colour of the source
that lefts redwards of it the $100 (1-x)$\,\% of the remaining objects, where
$x \lesssim 1$ (e.g. the percentiles $x$ = 0.90, 0.97, 0.995 separate the 10, 3
and 0.5\,\% reddest objects, respectively).
For a colour-magnitude diagram and a fixed value of $x$, there are 13 different
values of $i-K_{\rm s}$, one for each strip.
The collection of the 13 values determines a boundary for the selection of
association member candidates.
Thirdly, we have counted the number of objects redder than the selection
boundary for different values of the percentile $x$.
We plot in Figs.~\ref{feijksc} and~\ref{feijks} the boundaries for $x$ = 0.90,
0.97, 0.995 in the comparison colour-magnitude diagram, and only for $x$ = 0.995
in the Alnilam and Mintaka diagrams. 
This is the value actually used for the selection.
The percentile $x$ = 0.995 maximises the ratio between the number of objects
redder than the boundary in the Orion Belt fields and the number of expected
contaminants.
As a first order approximation, there should be about 44 objects redder than the
$x$ = 0.995 boundary in each of the Alnilam and Mintaka fields ($\sim 87000
\frac{1}{10} 0.005 \approx 44$).
The actual figures of sources redder than the $x$ = 0.995 boundary are {272} and
{157} in the Alnilam and Mintaka fields, respectively.
Accounting for the incomplete coverage of the DENIS survey (see
Section~\ref{spatialdistribution}), we estimate average 
frequencies of contamination at $\sim$25 and $\sim$33\,\% for the two Orion Belt
fields. 
Using larger (smaller) values of $x$ would lead to lower (larger) frequencies of
contamination, but also to smaller (larger) number of photometric association
member candidates.

%
%

The frequencies of contamination actually are lower, since many of the sources
to the red of the selection criterion are at different heliocentric
distances to the Ori~OB1b association.
We have complemented our DENIS/2MASS data with information in the literature,
astro-photometric data from the USNO-B2.0 catalogue, and visual inspection of
digitized phtographic plates.
The 429 sources have been investigated, one by one, to ascertain their
membership in association.
Eventually, we classify {167} of them as DENIS/2MASS fore- and background
sources in the Alnilam and Mintaka areas based on different criteria
(optical/near-infrared colours, proper motions, location in a cometary globule,
extended point spread functions): 
\begin{itemize}

\item Table~\ref{DE2N} shows three intermediate and late F-type stars in the
foreground, one nearby high-proper motion star (G~99--18), one distant Mira~Cet 
variable star (X~Ori) and two previously unknown sources with very red colours
(Ruber~1 and~2).
While the former five stars were already known to contaminate the Orion field,
the latter two stars are identified here for the first time.
One of them (\object{Ruber~1}) is a very late M dwarf in the foreground,
with an appreciable proper motion, while the other star (\object{Ruber~2}) seems
to be a pulsating giant in the distant Berkeley~20 open cluster (see details in
the notes to Table~\ref{DE2N}). 

\item Seven probable reddened sources in the direction of two dense cometary
globulae are listed in Table~\ref{DE2Mred}. 
Six of them fall in the direction of the IC~423 Bok globule.
This molecular cloud harbours the T~Tauri star IRAS~05307--0038 and has been
also classified as a reflection nebula ([RK68]~29; see note on
IRAS~05307--0038 in Table~\ref{DE2Mmintaka}).
The globule and the corresponding reddened sources are shown in
Fig.~\ref{therollingstones}. 
The remaining probable reddened source lies close to the centre of the
Ori~I--2 globule. 
There is not enough information to determine whether the seven sources are
reddened background stars or very young (Class~I/II-like) objects embedded in
the globules. 
A spectroscopic follow-up is necessary to ascertain their actual status.

\item Tables~\ref{gx2Malnilam} and~\ref{gx2Mmintaka} provide the
2MASS/2MASX designations of {152} galaxies.
The vast majority of them appear tabulated in the Two-Micron All Sky Survey
Catalog of Extended Sources, 2MASX (Jarret et~al. 2000), and are, therefore,
catalogued in the NASA/IPAC Extragalactic Database (NED).
Some of them also appeared in the works by Paturel et~al. (1989) and Monnier
Ragaigne et~al. (2003) or in radio catalogues (see notes on PMN~J0534--0044 in
Table~\ref{gx2Mmintaka}).
The red colours of many galaxies can be ascribed to their intrinsic nature
(starsbursts, ellipticals, bulges of spirals, and mergers). 
The extragalactic radio source complex \object{4C--01.06} could not be
identified by~us. 

\item Finally, there is one additional red DENIS/2MASS source with poor
photometry, \object{2MASS~J05380010--0122377}.
It was rejected during a visual inspection: it is a binary object
partially resolved in the Digital Sky Survey images that probably does not
belong to the Ori~OB1b association.
\end{itemize}

Accounting for the known young stars in the association, foreground dwarfs,
background giants, reddened stars and galaxies in Tables~\ref{DE2Malnilam}
to~\ref{gx2Mmintaka} that are redwards of the $i-K_{\rm s}$ percentile $x$ =
0.995, there remain {189} and {102} photometric association member
candidates in the Alnilam and Mintaka fields, respectively (all published stars
of unknown status --Table~\ref{DE2O}-- are bluewards of the selection
criterion).
The {291} sources are tabulated in Tables~\ref{DE2Oalnilam}
and~\ref{DE2Omintaka}\footnote{Since there are known young stars in the
association and foreground dwarfs bluewards of the selection criterion, and
photometric association member candidates in the overlapping region between
survey areas, the count of DENIS/MASS sources does not {\em seem} to coincide: 
$50 + 167 + 291 \neq 429$.}. 

Ten of them were photometric member candidates of the ``$\epsilon$~Orionis
cluster'' in Scholz \& Eisl\"offel (2005).
They classified three of these sources, with identification numbers 44, 120 and
126, as candidates with significant periodic variability (see details in notes
to Tables~\ref{DE2Oalnilam} and~\ref{DE2Omintaka}). 
V993~Ori is also a bright photometric variable (V993~Ori; Luyten 1932).
Photometric variability is a very common feature in young stars in general and
T~Tauri stars in particular (e.g. Bertout 1989). 
There are also variable T~Tauri substellar analogs (Caballero et~al. 2006).
The {280} remaining red DENIS/2MASS sources are firstly shown in our work.
We use the acronyms ``Annizam'' and ``Mantaqah'' plus running numbers for
naming the objects in the Alnilam and Mintaka fields,
respectively\footnote{The name Alnilam derives from the Arabic {\em
an-ni\.z\=am}, related to the word {\em na\.zm}, ``string of pearls''.
The name Mintaka comes from the Arabic {\em man{\c t}aqah}, ``belt''.}.
The running numbers indicate the position of the association members and
candidates with respect to the two supergiants.
The three last digits are for the position angle, while the three or four first
digits are for the angular separation (for example, Annizam~1751268 is located
at $\rho \approx$ 1751\,arcsec [20.8\,arcmin] and $\theta \approx$ 268\,arcsec
with respect to Alnilam).
This designation is similar to the Mayrit nomenclature for $\sigma$~Orionis
cluster members and candidates (Caballero~2008c).

\subsection{Remarkable fore- and background objects}
\label{remarkablesection}

For completeness, in Table~\ref{remarkable} we list four very bright nearby
stars ($K_{\rm s} \le$ 4.5\,mag; \object{19~Lep}, \object{HD~43429},
\object{$\theta$~Lep}, \object{$\eta$~Lep}), a recently-identified very bright
He-B subdwarf (\object{Albus~1} -- Caballero \& Solano 2007; Vennes, Kawka \&
Allyn Smith 2007), two unknown Tycho-2 high proper motion stars with
$\mu >$ 120\,mas\,a$^{-1}$, and six optical counterparts of {\em IRAS} sources
with very red colours ($V_T-K_{\rm s} >$ 5.4\,mag) which fall in the comparison
fields.
Only one of the {\em IRAS} sources had previously been investigated in the
literature, \object{CSS~205}, which was classified as a S-type star by
Stephenson (1984). 
Some of the other five stars are even redder than CSS~205, indicating that they
could be C- or S-type stars as well.
Their very red colours and strong mid-infrared flux excess indicate that they
might be stars in the last stages of the AGB phase, or close to the post-AGB
stage and evolving into the planetary nebula phase (van~der~Veen, Habing \&
Geballe 1989; Trams et~al. 1991; Riera et~al.~1995).

\section{Discussion}

\begin{figure*}
\centering
\includegraphics[height=0.32\textwidth]{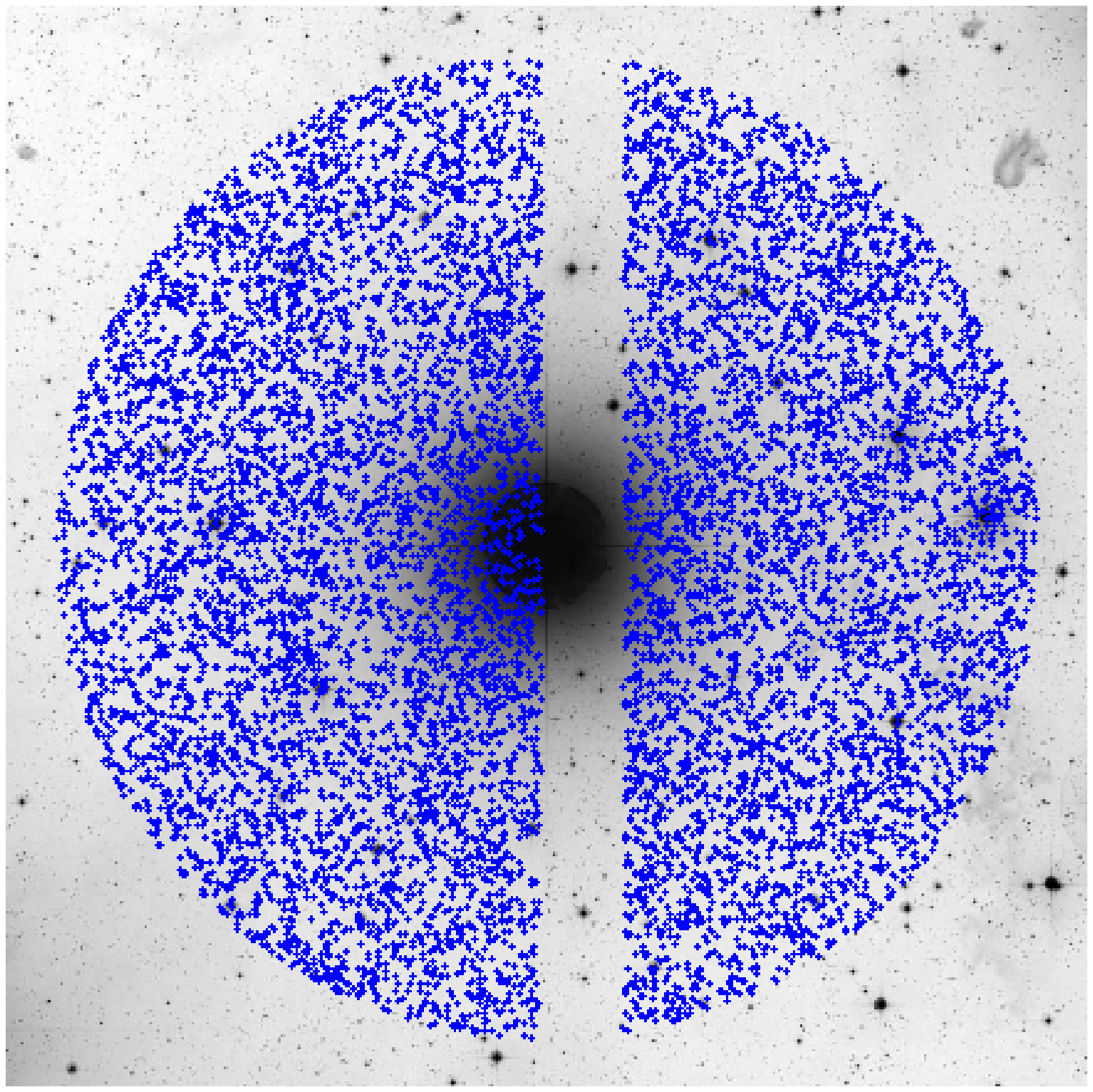}
\includegraphics[height=0.32\textwidth]{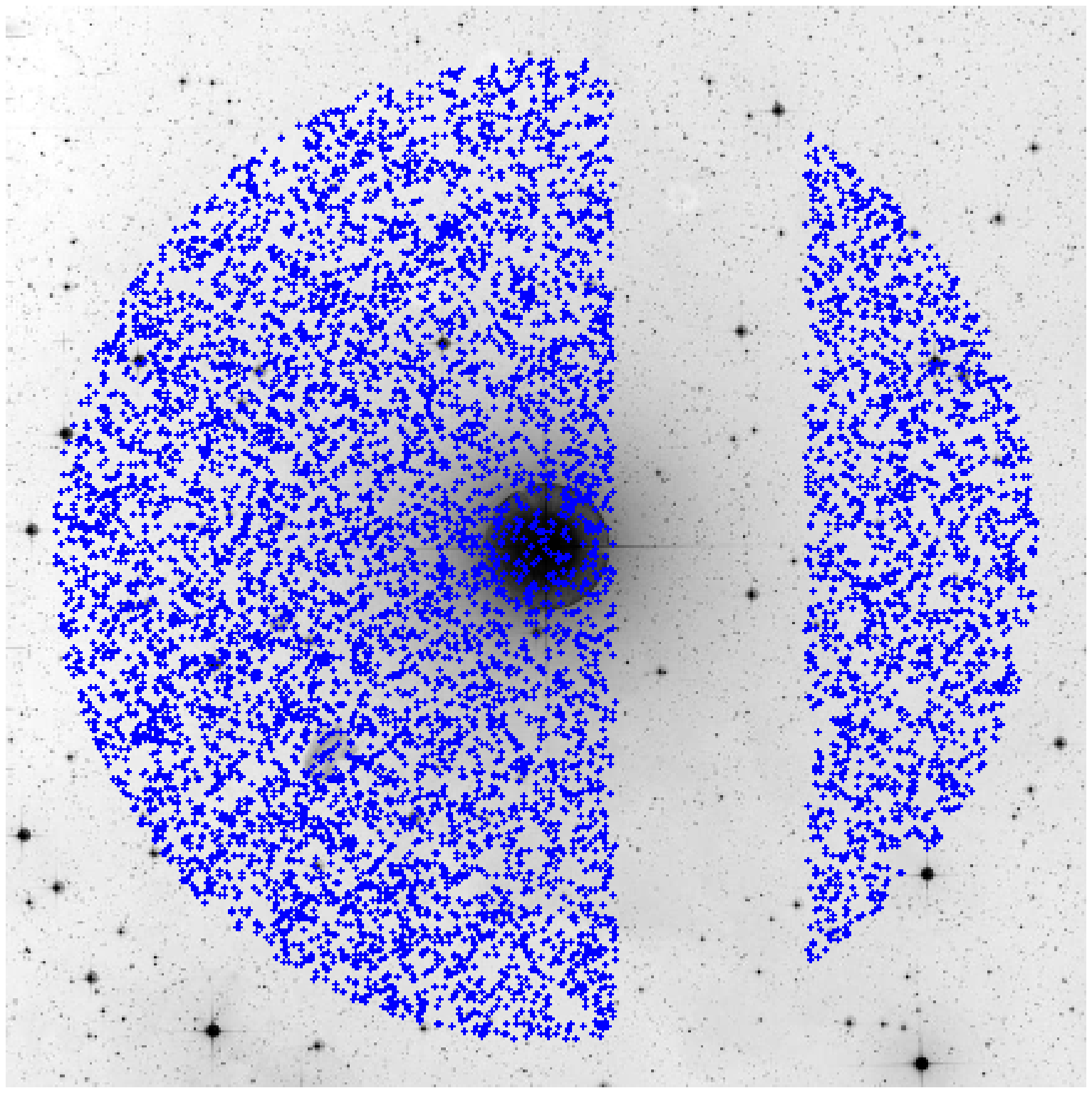}
\includegraphics[height=0.32\textwidth]{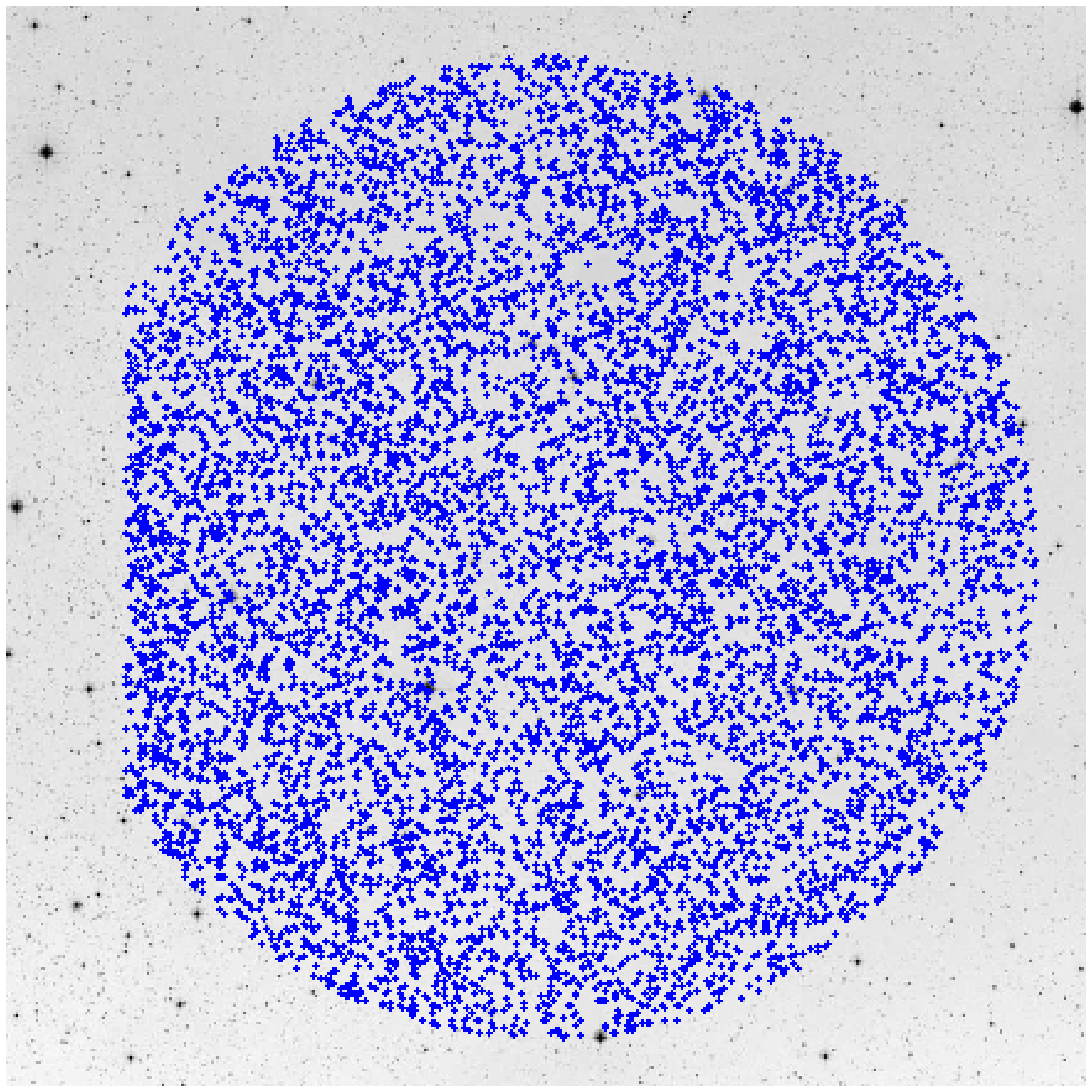}
\caption{Same as Fig.~\ref{alnilamandmintakaandc1}, but for the DENIS/2MASS
sources.
Note the missing DENIS strips.}
\label{alnilamandmintakaandc1DE2M}
\end{figure*}

Accounting for stars in the tiny overlapping region between the Alnilam and
Mintaka fields, we have identified {78} bright early-type
(Tables~\ref{TYC2Malnilam} and~\ref{TYC2Mmintaka}) and {58} intermediate and
late-type stars (Tables~\ref{DE2Malnilam} and~\ref{DE2Mmintaka}) with signatures
of youth.  
Many of the {65} Tycho-2/2MASS (Tables~\ref{TYC2Oalnilam}
and~\ref{TYC2Omintaka}) and {17} DENIS/2MASS (Table~\ref{DE2O}) published
stars of unknown association membership status may also be young. 
Together with the {291} DENIS/2MASS photometric member candidates of the
Ori~OB1b association (Tables~\ref{DE2Oalnilam} and~\ref{DE2Omintaka}), this
makes a catalogue of {509} confirmed and candidate young stars and
brown~dwarfs.  

For the canonical age of 5\,Ma for the Ori~OB1b association, the heliocentric
distance of $d$ = 388$\pm$30\,pc to the spectroscopic eclipsing binary VV~Ori
(Terrell et~al. 2007) and the colour excess of the supergiant Alnilam $E(B-V)$ =
0.09\,mag (Lee 1968), and using the {\sc Dusty00} models of the Lyon group
(Chabrier et~al. 2000), we estimate that the star-brown dwarf boundary (at $M
\approx$ 0.072\,$M_\odot$ for solar metallicity) in the region is at $J \approx$
15.5$\pm$0.2\,mag.
The estimation is identical if the {\sc NextGen98} models are used (Baraffe
et~al. 1998). 
The objects \object{[SE2005]~126} and \object{Mantaqah~2691223}, which are
fainter than this magnitude, are the only candidate young brown dwarfs in our
work. 
The star-brown dwarf boundary in the Alnilam-Mintaka region is $\sim$1\,mag
fainter than in the $\sigma$~Orionis cluster (at about $J \sim$ 14.5\,mag;
Caballero et~al. 2007), which is younger and {\em supposed to be} slightly
closer. 
A different heliocentric distance to Ori~OB1b, $d'$ (e.g. 330\,pc, 440\,pc;
see Sherry [2003]), would simply shift the star-brown dwarf boundary by a factor
$\Delta J = 5 \log{d/d'}$ (about 0.3\,mag in the examples above), which is of
the order of the uncertainty in the magnitude limit.
Since we have {\em not} used the heliocentric distance in any step of the
association member selection, a different $d'$ would only affect the actual
number of substellar objects in our catalogue (the closest distance would lead
to have 11 and 5 brown dwarfs in the Alnilam and Mintaka regions, respectively;
there would be no brown dwarfs for the farthest distance).

Apart from being 20--40\,\% more massive than $\sigma$~Orionis members of the
same apparent magnitude, the young objects in the Alnilam-Mintaka region are
also distributed along a wider area.
Caballero (2008c) identified 75 very low-mass stars, brown dwarfs, and
candidates fainter than $J$ = 14.0\,mag from a DENIS/2MASS correlation, very
similar to that presented here, but in a smaller area (a circle of radius
30\,arcmin) centred on the Trapezium-like $\sigma$~Ori system.
Accounting for the factor 2.25 of the different survey areas ($\pi 45^2 / \pi
30^2$), we expected to have found within the completeness $\sim$170 young
objects with $J >$ 14.0\,mag in each Orion Belt field {\em if} the surface
densities there and in $\sigma$~Orionis were identical.
Actually, a total of {131} and {63} intermediate- and late-type
photometric member candidates of the Ori~OB1b association in
Tables~\ref{DE2Oalnilam} and~\ref{DE2Omintaka} are fainter than $J$ = 14.0\,mag
(i.e. $M \lesssim$ 0.09\,$M_\odot$). 
We have identified, therefore, $\sim$70 and $\sim$40\,\% less of the expected
number of low-mass stars in the Alnilam and Mintaka fields, respectively.
Assuming also a similarity in mass functions, it is deduced, therefore, that the
surface density of brown dwarfs in Alnilam-Mintaka should be $\sim$70--40\,\% of
that in $\sigma$~Orionis.
In this computation, we have not taken into account the different location of
the star-brown dwarf boundaries. 
Far from being pessimistic, the lower (sub)stellar density surrounding the two
supergiants suggests to survey only $\sim$1.3 (Alnilam) and $\sim$2.7 (Mintaka)
times more area to find the same number of brown dwarfs than in
$\sigma$~Orionis. 
A coarse extrapolation of the number of brown dwarf candidates and possible
contaminants in B\'ejar et~al. (2003b) and Scholz \& Eisl\"offel (2005) 
supports our estimations of a relatively high substellar surface density
surrounding Alnilam.

The depth of the DENIS survey ($i_{5\sigma} \sim$ 18.0\,mag) and the expected
red colours of young brown dwarfs in the Ori~OB~1b association ($i-J \ge$
2.5\,mag) has allowed our search to be complete only down to $M \sim$
0.08--0.07\,$M_\odot$ ($M \sim$ 0.05 and 0.10\,$M_\odot$ for $d'$ = 330 and
440\,pc, respectively).  
Besides, the spatial coverage is incomplete (some strips of the
DENIS survey are absent; see Fig.~\ref{alnilamandmintakaandc1DE2M}) and an
important fraction of the actual intermediate- and late-type members of the
association may lie bluewards of the conservative $i-K_{\rm s}$ selection
criterion used in Section~\ref{midlatetype}. 
Even accounting for these incompletenesses and for possible contamination among
the photometric association member candidates, our work is by far the
most comprehensive star compilation in the Alnilam-Minataka region.

Our compilation is not only useful for probing the stellar and substellar
populations in the Alnilam-Mintaka region, but also for investigating the mass
function in the whole stellar domain from $\sim$15 to $\sim$
0.08--0.07\,$M_\odot$, the spatial distribution, or the frequency of discs
(there are a large amount of association members and candidates with {\em IRAS}
flux excess and/or red near-infrared colours, $J-K_{\rm s} >$ 1.15\,mag).
Some of these works will be carried out in the near future (Caballero \& Solano,
in~prep.).
In the next Section, we present a preliminar study of the radial distribution of
young stars and candidates surrounding Alnilam and~Mintaka.

\subsection{Spatial distribution}
\label{spatialdistribution}

\begin{figure*}
\centering
\includegraphics[width=0.49\textwidth]{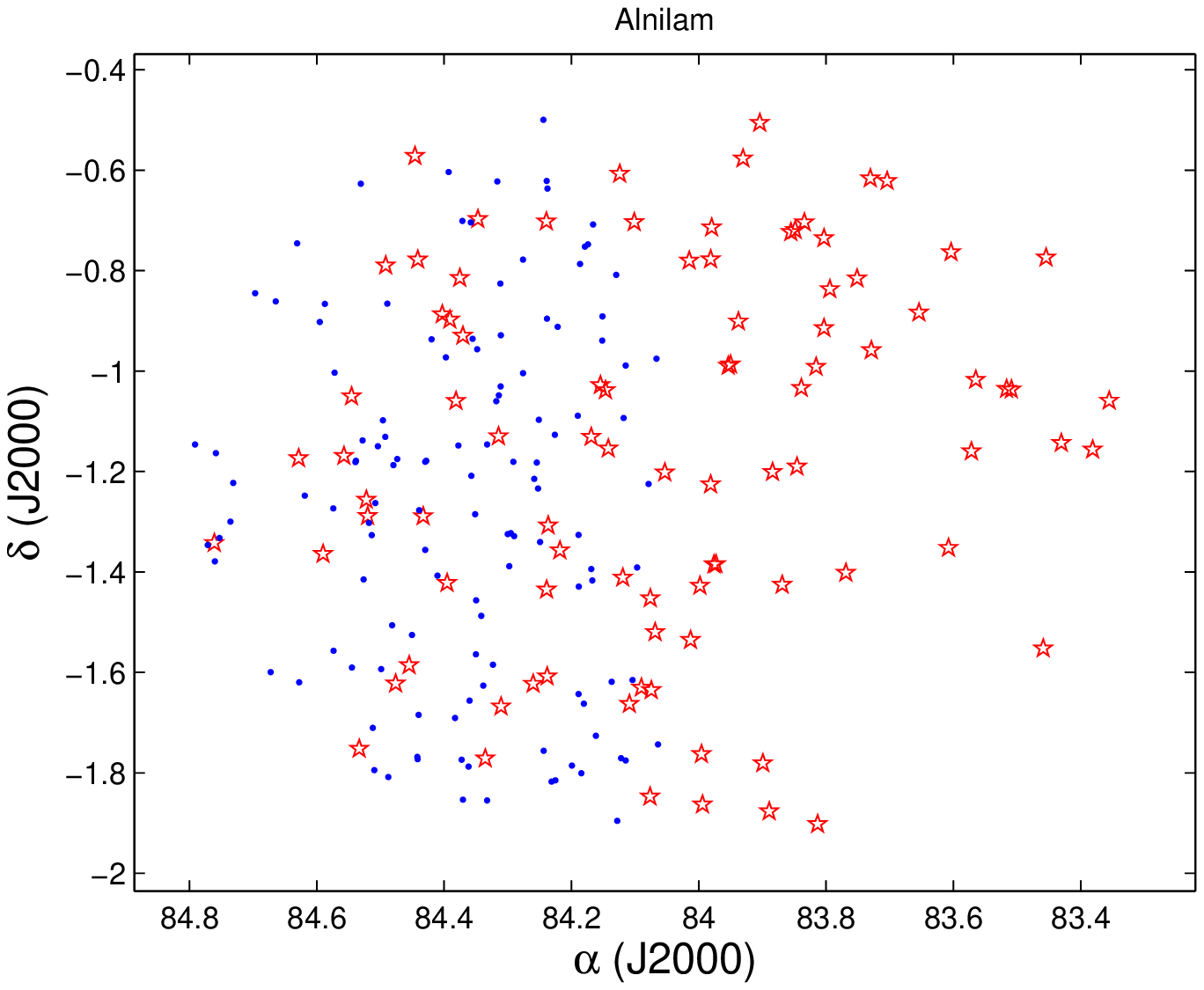}
\includegraphics[width=0.49\textwidth]{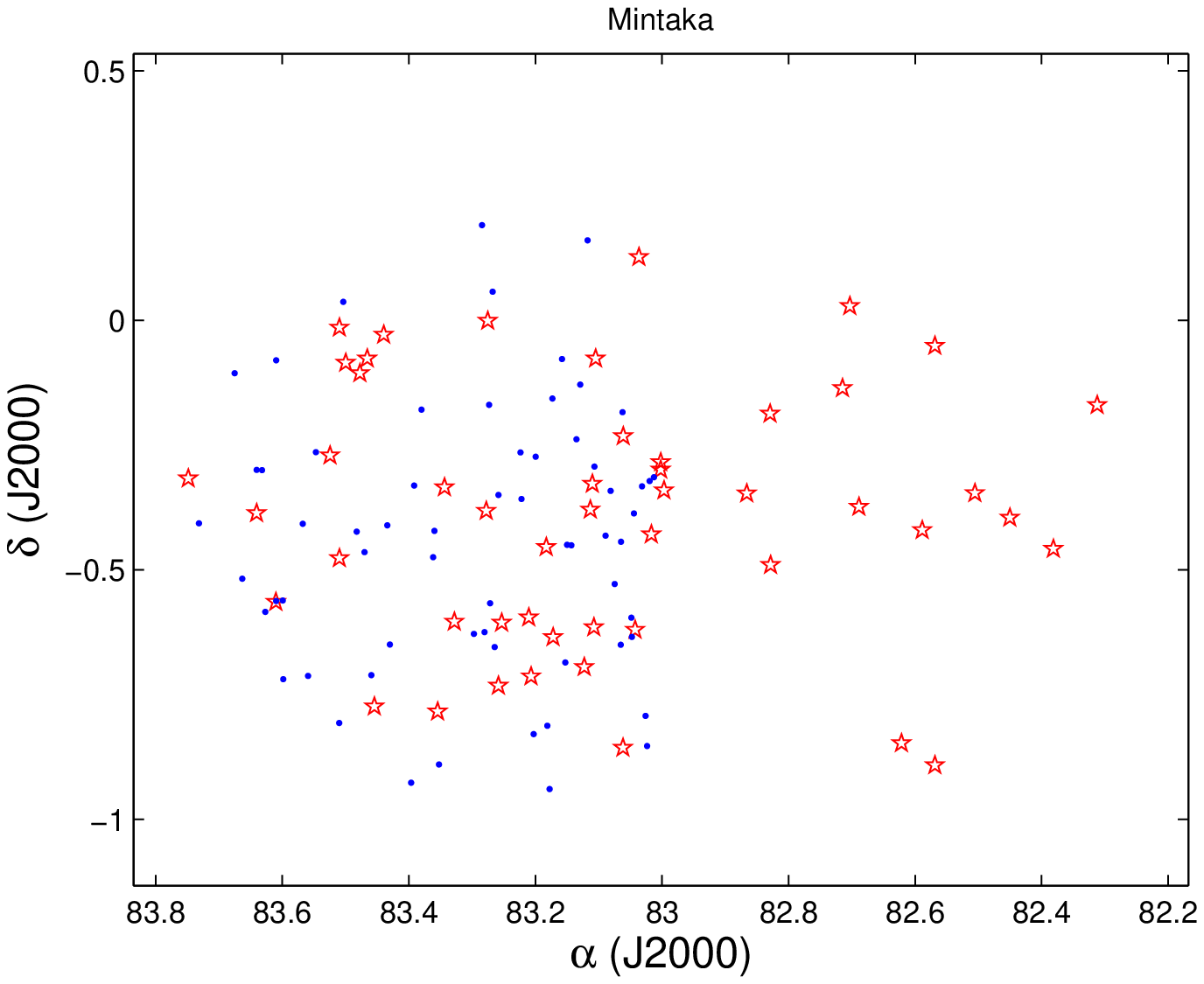}
\includegraphics[width=0.49\textwidth]{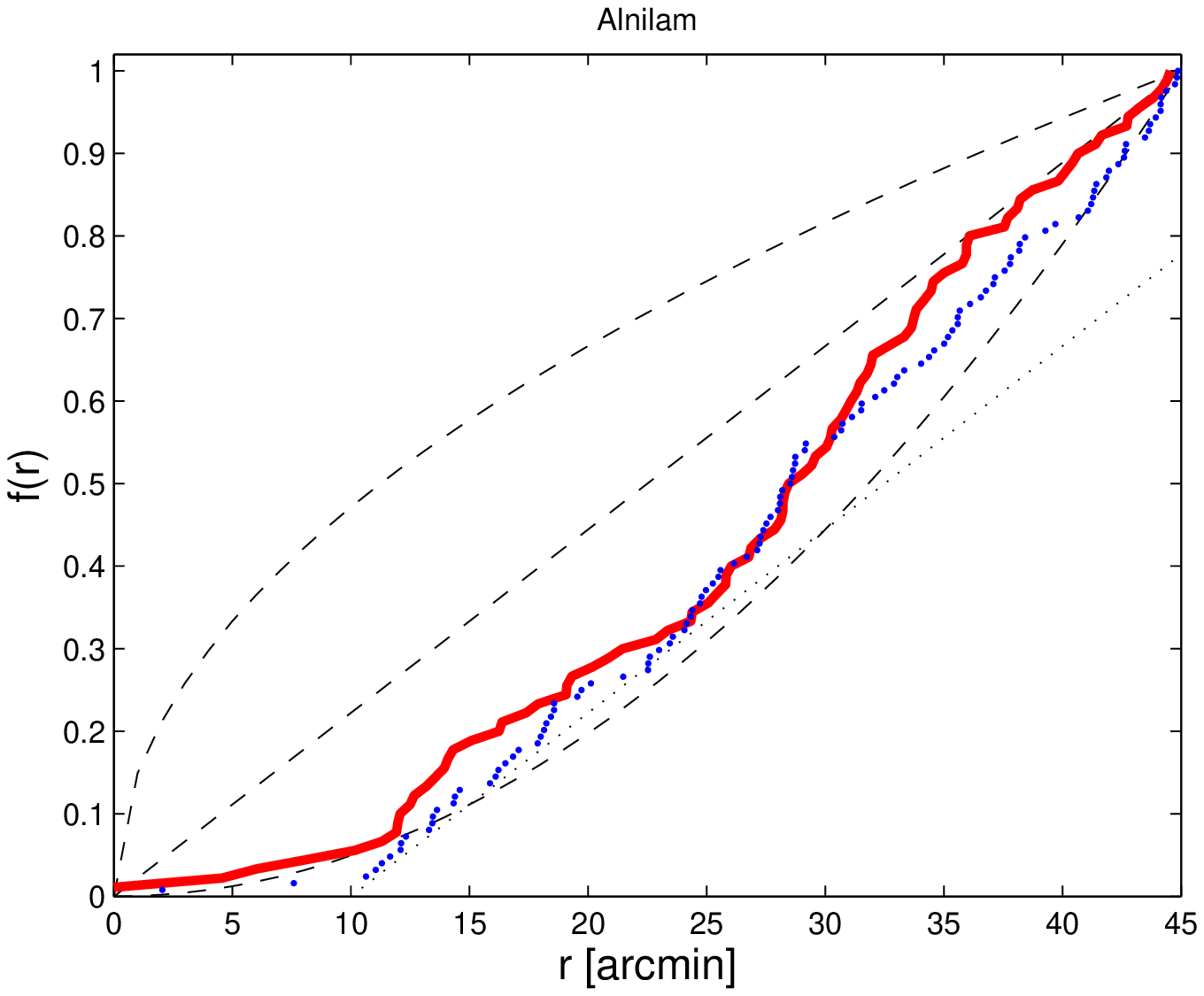}
\includegraphics[width=0.49\textwidth]{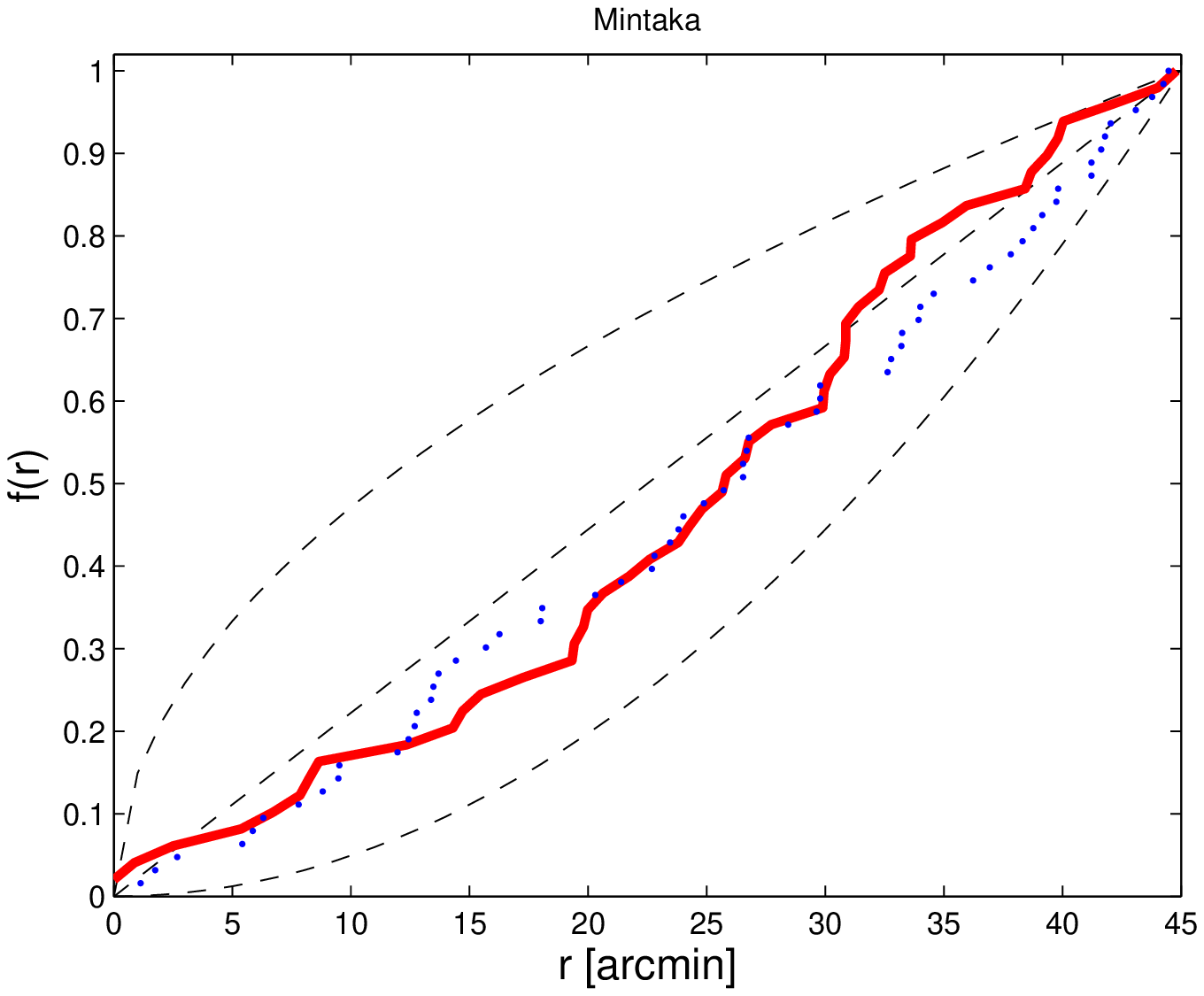}
\caption{Radial distribution of young stars and brown dwarfs surrounding Alnilam
(left) and Mintaka (right).
{\em Top panels:} spatial distribution of confirmed young stars (--red-- open
stars) and photometric candidates to the east of the
supergiants (--blue-- dots). 
{\em Bottom panels:} normalized cumulative number, $f(r)$, of confirmed young
stars (--red-- thick solid line) and photometric candidates (--blue-- dotted
line). 
The --black-- dashed lines indicate the theoretical power-law
distributions for $f(r) \propto r^{1/2}$, $r^{1}$, $r^{2}$ (from top to bottom).
Compare with figs.~3 and~4 in Caballero~(2008a).}
\label{spadis}
\end{figure*}

In total, we catalogue {89} confirmed stars and {189} DENIS/2MASS photometric
association member candidates in the Alnilam field;
in the Mintaka field, we catalogue {47} confirmed stars and {102} DENIS/2MASS
photometric association member candidates.
The spatial distribution of both type of objects, displayed in top panels in
Fig.~\ref{spadis}, shows no {\em clear} radial concentration towards the
supergiants. 
Following the $\sigma$~Orionis cluster parallelism, we expected a radial density
gradient centred on the massive OB-type stars.

We have investigated the normalized cumulative number of confirmed stars and
photometric association member candidates in projection within a distance $r$ to
Alnilam and Mintaka, $f(r)$.
Here, we follow the procedure detailed by Caballero (2008a).
Given the incomplete coverage of DENIS in the survey fields (evident in
Fig.~\ref{alnilamandmintakaandc1DE2M}), and to maintain the {\em radial} 
symmetry, we have only accounted for the association member candidates to the
east of the supergiants.
The radial distributions for both confirmed and candidate young objects, shown
in the bottom panels of Fig.~\ref{spadis}, resemble each other within Poissonian
errorbars (not shown for clarity). 
This resemblance suggests that there is no clear bias in our data compilation
(e.g. there has not been a tendency in the literature to survey close to the
central OB-type stars, like in $\sigma$~Orionis).
In the bottom panels, we also plot three theoretical power-law distributions.
The distribution $f(r) \propto r^{2}$ corresponds to a uniform distribution of
objects in the survey area.
The observed radial distribution in $\sigma$~Orionis is much more radially
concentrated, with $f(r) \propto r^{1}$ in the innermost 20\,arcmin-radius
cluster core.
This distribution corresponds to a volume density proportional to $r^{-2}$,
which is consistent with the collapse of an isothermal spherical molecular cloud
(see again details in Caballero [2008a]). 
Next, we show our results for the Alnilam and Mintaka regions.

\subsubsection{Alnilam}
 
Within the uncertainties and the incompleteness of our catalogue,
the distribution of young stars and candidates surrounding Alnilam clearly
departs from a radially concentrated distribution, as found in $\sigma$~Orionis.
The distribution at less than 25\,arcmin to Alnilam is fairly fit by a uniform
spacing, while there is a hint of an overdensity of young stars at larger
separations.
This result is in accord with the classical view of Collinder~70 (see
Section~\ref{alnilamandmintaka}) being a sparse, very wide clustering that
might extend to, and overlap with, neighbouring regions (e.g. Mintaka or
the ``halo'' of the $\sigma$~Orionis cluster -- Caballero 2008a). 
Clues of a $\sigma$~Orionis-like cluster around Alnilam were not found
either by Sherry (2003); 
his survey had, however, a less extensive spatial coverage, which would increase
the difficulty in identifying a ``weak~cluster''.

It is important to notice the great difference between the ``cores'' of
Collinder~70 and $\sigma$~Orionis.
While there are $\sim$130 known stars and brown dwarfs in the innermost
10\,arcmin of the latter cluster (most of them with features of youth), we
estimate that there are no more than 50 stars in the same area centred on
Alnilam. 
This is not an observational bias, because this abrupt deficiency is not
detected in the Mintaka field (Alnilam and Mintaka have roughly the same
magnitudes, and so do the sizes of their optical glares -- an intense background
by a nearby bright star may prevent the detection of sources within the
completeness in a photometric survey).
However, the frequency of young stars and brown dwarfs at intermediate
separations from Alnilam (e.g. 25--45\,arcmin) can be larger than in the same
corona centred on $\sigma$~Ori. 
In any case, a wider study of the radial distribution of young stars, covering
the whole Orion Belt, is needed to ascertain the real nature of the
Collinder~70~cluster.

\subsubsection{Mintaka}
 
The radial distribution of young stars surrounding Mintaka follows, in
contrast to Collinder~70, a power-law with an odd exponent
intermediate between 1 and~2.
This radial concentration may suggest that there is actually a clustering
of young stars surrounding the supergiant.
This is the first time to propose the existence of a cluster in the area,
to which we call ``Mintaka cluster''. 
A radial distribution with a power-law $f(r) \propto r^{1.5}$ would correspond
to a volume density proportional to $r^{-1.5}$.
The lower central concentration than in the scenario of collapse of an
isothermal spherical molecular cloud may suggest that $(i)$ the Mintaka cluster
formed from a non-isothermal molecular cloud, $(ii)$ it formed from an
isothermal molecular cloud but next suffered from dynamical evolution (and
the Mintaka cluster would be, in this scenario, a dynamically-evolved analog to
$\sigma$~Orionis), or $(iii)$ there is significant overlapping between the
stellar population of Ori~OB1b and Ori~OB1a. 
The Ori~OB1a sub-association is expected to have a spatial distribution
that  completely overlaps with the region near Mintaka.  
Ori~OB1a is older than Ori~OB1b, but also up to $\sim$100\,pc nearer (Sherry
2003).  
As a result, the isochrones for Ori~OB1a and~b roughly fall on the same location
in the colour-magnitude diagram. 
Stars from Ori~OB1a would have indicators of youth as well, but would not follow
the clustering around Mintaka.
They would also be difficult to disentangle from proper motions due to the
unfavourable direction of Orion and the Sun's relative motions. 
The combination of a clustered population (the ``Mintaka cluster'' in Ori~OB1b)
with a nearly uniformly distributed population (Ori~OB1a) would take an $f(r)$
spatial distribution intermediate between $r^1$ and $r^2$. 
Obviously, further work is required to derive any conclusion.

\section{Summary}

In search for new hunting grounds for substellar objects, we have
investigated the stellar populations surrounding two bright supergiants in the
Ori~OB~1b association (the Orion Belt). 
The two very young supergiants are Alnilam ($\epsilon$~Ori) and Mintaka
($\delta$~Ori).
Accounting for all the Tycho-2, DENIS and 2MASS sources in the twelve
45\,arcmin-radius main and comparison fields, we have examined 107\,434 sources
in total. 
After a comprehensive, inclusive, massive Virtual Observatory analysis and
bibliographic data compilation, we~list:

\begin{itemize}
\item 78 bright (Tycho-2/2MASS) stars with features of extreme youth (i.e. with
very early spectral types or HAeBe signatures), 
\item 58 intermediate- and late-type (DENIS/2MASS) stars with features of
extreme youth (i.e. with Li~{\sc i} in absorption, H$\alpha$ in emission,
low~$g$ spectroscopic signatures, X-ray emission),
\item 289 intermediate- and late-type (DENIS/2MASS) young star candidates with
very red $i-K_{\rm s}$ colours,
\item 2 brown dwarf candidates with very red $i-K_{\rm s}$ colours:
[SE2005]~126 (Mantaqah 1582164) and Mantaqah 2691223 (their actual
substellar nature depends on the heliocentric distances and ages),
\item 82 (Tycho-2/2MASS and DENIS/2MASS) stars without clear signposts of youth
that might also belong to the Ori~OB~1b association.
\item 117 (Tycho-2/2MASS and DENIS/2MASS) stars in the fore- or the background
based on their proper motions, spectral types, parallactic heliocentric
distances, radial velocities and/or colours,
\item 152 extended galaxies, and
\item 13 remarkable fore- and backgound stars in the comparison~fields.
\end{itemize}

We report for the first time X-ray emission, {\em IRAS} flux excess and possible
resolved multiplicity for dozens young stars and candidates in the
Alnilam-Mintaka region.
This abundance represents an excellent compilation of candidates for
further follow-up dedicated studies. 
The vast majority of the listed association member candidates are new.
A wealth of detailed information is provided in Appendix~\ref{catalogue} for
about one hundred investigated sources.

Finally, we investigate the spatial distribution of stars surrounding Alnilam
and Mintaka, and discuss on the possibilities for searching for brown dwarfs.
Collinder~70, the cluster that surrounds Alnilam, {\em if} it exists, must be
larger than our search radius of 45\,arcmin.
Its (sub)stellar population may, therefore, spatially overlap with neighbouring
star-forming regions, like the $\sigma$~Orionis cluster, which is one of the
richest regions in substellar objects.
The evidence for a real cluster surrounding Mintaka is, however, more apparent
but not conclusive from our data analysis.
The ``Mintaka cluster'', that is presented here for the first time, is less
concentrated than the $\sigma$~Orionis cluster and might represent a next
evolutionary stage of it.
Accounting for the fainter star-brown dwarf boundary and the lower spatial
density of stars very close to the supergiants with respect to $\sigma$~Orionis,
the clusters surrounding Alnilam and Mintaka can be considered to be ``elder
brothers'' (in contraposition to ``fraternal twins'') of $\sigma$~Orionis.

\begin{acknowledgements}

We appreciate the skillfull referee report by W.~H.~Sherry.
J.A.C. formerly was an Alexander von Humboldt Fellow at the MPIA and currently
is an Investigador Juan de la Cierva at the UCM.
We thank V.~J.~S.~B\'ejar, F.~Fontanot, S.~More, R.~Mundt, A.~Sicilia-Aguilar
and S.~Wolf for helpful comments. 
Partial financial support was provided by the Universidad Complutense de Madrid,
the Spanish Virtual Observatory and the Spanish Ministerio Educaci\'on y Ciencia
under grants AyA2005--02750, AyA2005--04286 and AyA2005--24102--E of the
Programa Nacional de Astronom\'{\i}a y Astrof\'{\i}sica and by the Comunidad
Aut\'onoma de Madrid under grant CCG07--UCM/ESP--2679 and PRICIT project
S--0505/ESP--0237 (AstroCAM).  
This research has made use of: 
the SIMBAD, operated at Centre de Donn\'ees astronomiques de Strasbourg, France;
TOPCAT (Tool for OPerations on Tables And Catalogues), provided by the AstroGrid
Virtual Observatory project; 
the NASA's Astrophysics Data System as bibliographic service; and
the NASA/IPAC Extragalactic Database (NED) which is operated by the Jet
Propulsion Laboratory, California Institute of Technology, under contract with
NASA.
\end{acknowledgements}

\appendix

\section{The Annizam/Mantaqah catalogue}
\label{catalogue}

{
\paragraph{Notes to Table~\ref{TYC2Malnilam}:}
\begin{list}{}{}
\item[*] HD~36980~AB is a close binary with $\rho \approx$ 0.7\,arcsec, 
$\theta \approx$ 61\,deg (catalogue of Components of Double and Multiple stars; 
Dommanget \& Nys~1994 -- there are no fundamental differences between this
edition and the second one [Dommanget \& Nys~2002], except for the number of
considered sources).
\item[*] RX J0535.6--0152~AB is a T~Tauri star with a red $V_T-K_{\rm s}$
colour.
It is a G6V-type spectroscopic binary with lithium in absorption (pEW(Li~{\sc
i}) = +0.32\,\AA), partially filled H$\alpha$ line (pEW(H$\alpha$) = +2.40\,\AA)
and X-ray in emission (Alcal\'a et~al.~1996, 2000). 
RX~J0535.6--0152~AB was the third strongest X-ray source in the investigation of
40 weak-line T~Tauri stars in Orion by Marilli et~al. (2007), with $\log{L_X} =
30.7^{+0.2}_{-0.7}$ (erg\,cm$^{-2}$\,s$^{-1}$).
These authors found it to be a photometric variable with a period of 1.74\,d.
\item[*] HD~37285~AB is a visual binary star with $\rho \approx$ 0.4\,arcsec, 
$\theta \approx$ 263\,deg (Dommanget \& Nys 1994).
\item[*] HD~37389 is embedded in the        {Ori~I--2} Cometary Globule 
(Ho, Martin \& Barrett 1978; Cernicharo et~al. 1992; Mader et~al. 1999).
Oudmaijer et~al. (1992) and Coulson, Walther \& Dent (1998) have reported 
infrared and submillimetre flux excesses due to a Vega-like disc.
The star has appreciable polarization in the optical (Bhatt \& Manoj~2000).
Some catalogues tabulate a hypothetical companion,        {BD--01~985B}, at 
$\rho \sim$ 5.0\,arcsec, $\theta \sim$ 350\,deg.
\item[*] HD~37149 is a helium-weak star (Bernacca \& Ciatti 1972; Renson 1988) 
with H$\alpha$ in medium emission (Bidelman 1965). 
It is likely the UV-emission source        {[SC93b]~328} (Schmidt \& 
Carruthers~1993).
\item[*] HD~290770 was discovered as an emission-line star by Bidelman
(1965) and has been classified as a B8--9Ve Herbig Ae/Be star by many other
authors (Guetter 1976; Gieseking 1983; Dong \& Hu 1991; Nesterov et~al. 1995;
The et~al. 1994). 
Vieria et~al. (2003)  found [O~{\sc i}]+[S~{\sc ii}] in emission and measured
the H$\alpha$ line in double-peak emission, with the secondary peak having more
than half the strength of the primary. 
Yudin \& Evans (1998) found negligible polarization in the optical.
Previously unnoticed, HD~290770 has one of the most apparent flux excesses at 
the {\em IRAS} passbands in the Ori~OB~1~b association (Caballero \& Solano, 
in prep.).
Here we report a close visual companion to the star at $\rho \sim$ 6.8\,arcsec,
$\theta \sim$ 340\,deg, and $\Delta K_{\rm s}$ = 4.46$\pm$0.03\,mag fainter.
From its $I-J$ and $J-K_{\rm s}$ colours from DENIS and 2MASS, it seems to be a
F--G:-type star in the fore-/background. 
\item[*] HD~37344 is embedded in the bright-rimmed cloud complex 
{Ori~I--2N}, close to cloud {[OS98]~40C} (Ogura \& Sugitani~1998).
\item[*] HD~290602 is also {BD--01~947}.
\item[*] HD~290674 is also {BD--01~977}.
\item[*] HD~37321~AB is a well-known helium-weak star with a high rotational 
velocity ($v \sin{i} \approx$ 100\,km\,s$^{-1}$; Mermilliod 1983) and
spectrum variability (Molnar 1972; Garrison 1994 -- but see Pedersen \& Thomsen 
1977).
Blaauw \& van Albada (1963) proposed the star to be a long-period spectroscopic
binary;
Morrell \& Levato (1991) measured, however, a constant radial velocity of 
24$\pm$6\,km\,s$^{-1}$ during their monitoring.
It is a close binary with $\rho$ = 0.756$\pm$0.002\,arcsec, $\theta$ = 
14$\pm$1\,deg ($\Delta H_p$ = 1.62$\pm$0.01\,mag; Perryman et~al. 1987).
It was only resolved by Tycho-2.
We accounted for the $B_T V_T$ magnitude of the A component and the near
infrared $K_{\rm s}$ magnitude of both A and B components as a single object.
The system could have a faint, red, third component ($H$ = 10.98$\pm$0.02\,mag), 
at $\rho \approx$ 17.7\,arcsec, $\theta \approx$ 32\,deg.
HD~37321~AB may also be the far-ultraviolet emission source        {[SC93b]~341}
(Schmidt \& Carruthers~1993).
\item[*] HD~36955 is a peculiar magnetic star with abnormal abundances of Si, 
Cr and Eu (Gray \& Corbally 1993; Kudryavtsev et~al. 2006).
\item[*] V1247~Ori is a Herbig He/Be star (Garc\'{\i}a-Lario et~al. 1997; Fujii
et~al. 2002) whose non-banded H$\alpha$ emission was found by McConell (1982). 
Spectral types from A5III, through A7, to F0V have been provided (Schild \&
Cowley 1971; Nesterov et~al. 1995; Vieira et~al. 2003).
Vieira et~al. (2003) found no forbidden lines in its optical spectrum, but
identified an H$\alpha$ symmetric profile without, or with only very shallow,
absorption features.
No H$_2$O, NH$_3$ or CO radio lines were found by Wouterloot et~al.
(1986,~1988,~1989).  
V1247~Ori is, besides, a well-studied $\delta$~Scuti star, with $P$ =
0.096967\,d and peak-to-peak amplitude in the $V$ band of 0.050\,mag (Lampens \&
Rufener 1990; Garc\'{\i}a et~al. 1995; Handler 1999).
Its SED shows clear excesses from the $J$ band to 60--100\,$\mu$ (Caballero \& 
Solano, in prep.), and is composed of two components, one warm 
(1.2--2.2\,$\mu$m) and other cool (12--100\,$\mu$m).
\item[*] Alnilam ($\epsilon$~Ori, 46~Ori, HD~37128; $V$ = 1.70\,mag) is 
one of the brightest supergiants in the sky and, therefore, one of the best 
known stars.
The first spectroscopic study was carried out more than a century ago by 
Campbell (1894) and Keeler (1894).
It is a hot, massive, single star in the hydrogen shell burning phase (Lamers
1974; Jarad, Hilditch \& Skillen 1989) with photometric and spectroscopic
variability (Stebbins 1915; Ebbets 1982; Prinja et~al.
2004), H$\alpha$, X-ray and radio emission (Cherrington 1937; Abbott et~al.
1980; Bergh\"ofer et~al. 1996; Blomme et~al. 2002), and strong stellar wind and
mass loss (Groenewegen \& Lamers 1989; Prinja et~al. 2001; Crowther, Lennon \& 
Walborn 2006). 
Alnilam is one of the few early-type stars with determination of the angular
diameter using optical interferometry (Hanbury Brown, Davis \& Allen 1974).
It illuminates the {NGC~1990} reflection nebula.
Last, it has been used as a bright spectrophotometric standard (B0Ia in  the MK
classification by Johnson \& Morgan 1953), to investigate the interstellar
extinction (e.g. Whitford 1958; Bohlin, Savage \& Drake 1978) and for comparison
with other early-type supergiants (Humphrey 1978). 
A review of ``classic'' works on Alnilam can be found in Lamers (1972).
The status of ``Alnilam~B'' (BD--01~969B; see~Table~\ref{DE2O}), at
3\,arcmin to the northeast, is~unknown.
\item[*] HD~37397 is a low-amplitude variable star ($P$ = 0.572885\,d, 
$A(V)$ = 0.00089\,mag; Koen \& Eyer~2002) with a constant radial velocity of 
22--23\,km\,s$^{-1}$ (Morrell \& Levato 1991; Duflot et~al. 1995; Grenier 
et~al.~1999).
\item[*] VV~Ori~AB is a double-lined eclipsing binary in a detached
configuration (Miller Barr 1904; Adams 1912; Daniel 1916; Struve \& Luyten
1949). 
The two early-type stars, B1.0V+B4.5V, are separated by 
13.49$\pm$0.05\,$R_\odot$ ($P \approx$ 1.48\,d; Terrell et~al. 2007). 
The mid-infrared source {IRAS~05309--0111} is located at $\rho \sim$
16\,arcsec, $\theta \sim$ 180\,deg, to VV~Ori.
Friedemann, G\"urtler \& L\"owe (1996) had been the only investigators before us
to notice the {\em IRAS} thermal emission of VV~Ori, and attributed it to
circumstellar dust.
This is very important, because: 
($i$)~stars with very early spectral types, just as the primary in VV~Ori, are
not expected to have circumstellar discs at the age of the Ori~OB~1~b
association, and
($ii$)~the disc would surround the binary system (i.e. the inner part of the
disc would be at several --tens-- astronomical units, while the binary
components are separated by $\sim$0.64\,AU).
VV~Ori may also be associated to the X-ray sources {[NYS99]~A--01} and
{1AXG~J053331--0110} (Nakano et~al. 1999; Ueda et~al.~2001).
\item[*] HD~36684~AB is a close binary with $\rho \approx$ 0.2\,arcsec, 
$\theta \approx$ 200\,deg (Dommanget \& Nys~1994).
It also has a high rotational velocity ($v \sin{i} \approx$ 160\,km\,s$^{-1}$; 
Sharpless 1974).
\item[*] HD~290750 is a low-amplitude suspected variable (Rufener \& 
Bartholdi~1982).
\item[*] HD~36779~AC forms together with the post-T~Tauri star 
HD~36779~B a likely Lindroos system (Lindroos 1985).
HD~36779~AC is, in its turn, a spectroscopic binary (Morrell \& Levato~1991).
\item[*] HD~37187 might form a new (very wide) Lindroos system together with
V583~Ori ($\rho \approx$ 29.0\,arcsec, $\theta \approx$ 212\,deg).
\item[*] HD~37076 and HD~290671 form the STF~751 double system, with 
$\rho \approx$ 15.6\,arcsec, $\theta \approx$ 124\,deg (Dommanget \& Nys~1994).
They share Tycho-2 proper motion within the uncertainties.
The X-ray emission found by {\em ROSAT} (with HRI and PSPC) is associated to the
faintest component (HD~290671, B9.5V; Caballero et~al. in prep.).
\item[*] HD~290665 is a SiCrEuSr chemically peculiar star 
(Bartaya 1974; Schild \& Cowley 1971; Guetter 1976; Joncas \& Borra 1981; 
Gieseking 1983).
It is also a strong magnetic star ($\langle B_z \rangle \approx$ --0.17\,T; 
Bagnulo et~al.~2006).
Last, HD~290665 has a radial velocity discordant with association membership
(Gieseking 1983).
It could be, however, a spectroscopic binary.
\item[*] V1379~Ori is a slowly pulsating B star (Waelkens et~al.~1998).
\item[*] HD~290662, a peculiar Vega-like star, was proposed to be a
spectroscopic binary by Gieseking (1983) based on low quality data.
\item[*] HD~36954~AB is a spectroscopic binary (SB1) with a period $P \sim$ 
4.6\,d (Neubauer 1936; Morrell \& Levato~1991). 
\item[*] HD~37235 is a spectroscopic variable B7--A0:V according to 
Bernacca \& Ciatti~(1972).
Renson (1992) tabulates it as a He-weak star.
The origin of the mid-infrared source {IRAS~05344--0044}, located at a
separation $\rho \sim$ 35\,arcsec to the south of the star, probably lies on the
extended source 2MASX~J05365804--0042413, whose spectral energy distribution
resembles those of starburst galaxies with large dust content (see, e.g., Chary
\& Elbaz~2001) 
\item[*] HD~290648, HD~290660 and HD~290650 are also {BD--00~1004}, 
{BD--00~1020} and {BD--00~1012}, respectively.
\end{list}

\paragraph{Notes to Table~\ref{TYC2Mmintaka}:}
\begin{list}{}{}
\item[*] HD~290515 is at a projected angular separation of 34\,arcsec to the 
background RR~Lyr star {[VZA2004]~28} (Vivas et~al. 2004).
\item[*] HD~290492~AB is a close binary with $\rho$ = 
0.739$\pm$0.005\,arcsec, $\theta$ = 63.9$\pm$1.9\,deg ($\Delta b$ = 
0.6$\pm$0.2\,mag -- Rossiter 1955; Marchetti, Faraggiana \& Bonifacio~2001).
It is a non-variable, mild $\lambda$~Boo star candidate (Paunzen \& Gray 1997; 
Paunzen et~al. 2002 -- but see Gerbaldi, Faraggiana \& Lai [2003] and Faraggiana
et~al. [2004]).
Paunzen (2001) derived a photometric distance of $d$ = 279$\pm$20\,pc, but he 
erroneously assumed no contamination by the secondary in the spectrum of the 
primary. 
HD~290492~AB and the G8III star {GSC~04766--02124}, which is $\sim$3\,mag
fainter in the $V$ band and at $\sim$24\,arcsec to the west, do not share a
common proper motion, as has been proposed in the literature.
\item[*] SS~28 is a T~Tauri star.
It has been only investigated by Stephenson \& Sanduleak (1977), Bopp (1988), 
Wiramihardja et~al. (1989) and Kogure et~al. (1989).
It has a double-peaked H$\alpha$ emission line with intensity about three times
that of the continuum, H$\beta$ also in emission and an apparently ``filled in''
Ca~{\sc ii} H and K. 
According to Bopp (1988), SS~28 resembles some unusual interacting F+B binary
systems.
It is, besides, an {\em Einstein} Observatory soft X-ray source.
The {\em ROSAT} Satellite measured afterwards 18 events associated to SS~28
(ROSAT 2000; White et~al.~2000).
SIMBAD tabulates a quadruple identification (it is also {Kiso~A--0904~6},
{Kiso~A--0903~234} and {2E~1299}). 
\item[*] BD--00~984 is a chemically peculiar star based on their abnormal
abundances of Hg, Mn (Woolf \& Lambert 1999 -- they classified it as one of the
youngest HgMn stars) and Si (Brown \& Shore~1986).
It forms a curious perfect alignement with two nearby radio sources:
{[LPZ94]~147} ($\rho \approx$ 31.5\,arcsec, $\theta \approx$ 177.0\,deg 
-- $S_\nu$(3.9\,GHz) = 48$\pm$24\,mJy, Larionov et~al. 1994) and 
{TXS~0529--004} ($\rho \approx$ 1.8\,arcmin, $\theta \approx$ 178\,deg -- 
$S_\nu$(1.4\,GHz) = 82.2$\pm$2.5\,mJy, Condon et~al. 1998; 
$S_\nu$(0.365\,GHz) = 214$\pm$24\,mJy, Douglas et~al. 1996).
The spectral index of TXS~0529--004 is --0.9, consistent within the 
uncertainties with thermal Bremsstrahlung emission.
\item[*] HD~290500 was classified as a Herbig Ae/Be star by Vieira et~al.
(2003). 
They derived A2 spectral type and detected H$\alpha$ in double-peaked emission,
with the secondary peak having less than half the strength of the primary.
No forbidden lines were identified.
Codella et~al. (1995) gave an upper limit for the 22.2\,GHz H$_2$O maser
emission of the star.
HD~290500 has a mid-infrared flux excess, as measured by {\em IRAS} (Caballero \&
Solano, in prep.). 
\item[*] HD~36841 has been widely used for determining ultraviolet 
interstellar extinction curves (e.g. Savage et~al. 1985; Barbaro et~al.~2001).
It was formerly considered a late-O-type star (Mannino \& Humblet 1955; Goy 
1973; Cruz-Gonz\'alez et~al.~1974).
\item[*] V1093~Ori~AB (HD~36313) is a variable of $\alpha^2$~CVn type (North 
1984; Catalano \& Renson 1998), a helium-weak, silicon, magnetic peculiar 
(Guetter 1976; Borra 1981; Bychkov, Bychkova \& Madej 2005) and a close binary 
star ($\rho \approx$ 0.22\,arcsec, $\theta \approx$ 172.6\,deg -- 
Couteau 1962; van Biesbroeck 1974).
\item[*] Mintaka AE--D ($\delta$~Ori, 34~Ori, HD~36486; $V$ = 2.23\,mag) 
is the most famous star in the Orion belt.
It is a very bright triple within a hierarchical quintuple system.
Mintaka~D ($\delta$~Ori~Ab) is an early-B-type star at $\rho$ =
0.267\,arcsec, $\theta$ = 140\,deg ($\Delta H_P$ = 1.35$\pm$0.02\,mag), from the
tight AE binary (Heintz 1980; McCalister \& Hendry 1982; Perryman et~al. 1997);
it may be a rapid rotator or a spectroscopic binary (Harvin et~al. 2002).
Mintaka~A (O9.5II, $\delta$~Ori~Aa1) and E (B0.5III, $\delta$~Ori~Aa2) form an
eclipsing spectroscopic binary with a peak-to-peak amplitude $A(H_P)$ =
0.01\,mag and a period $P$ = 5.7325\,d
(Hartmann 1904; Jordan 1914; Pismis, Haro \& Struve 1950; Koch \& Hrivnak 1981;
Harvey et~al. 1987; Harvin et~al. 2002; Kholtygin et~al. 2006);
the binary has suffered from an intense mass loss. 
The other two components, Mintaka~B and C, are described below.
\item[*] Mintaka~C ({$\delta$~Ori~C}, HD~36485) is at 
$\rho \approx$ 52.2\,arcsec, $\theta \approx$ 0.1\,deg to Mintaka~AE 
($\Delta V_T$ = 4.417$\pm$0.012\,mag; H{\o}g et~al.~2000).
It is a helium-strong star (Morgan, Abt \& Tapscott 1978; Walborn 1983; 
Bohlender 1989) with nonthermal radio emission (Drake et~al.~1987).
\item[*] HD~290491 is also {TYC~4766~2264~1}.
\item[*] HD~36726~A is a $\lambda$~Boo star (Abt \& Levato 1977; Paunzen 2001)
and has a quite high rotational velocity, $v \sin{i}$ (about 120\,km\,s$^{-1}$; 
Abt~1979). 
It is the brightest component of a triple system tabulated by Aitken \&
Doolittle (1932).
The secondary is BD--00~993B (Table~\ref{DE2O}).
The system could be quadruple, since there is an additional 2MASS source at 
$\rho \approx$ 5.7\,arcsec, $\theta \approx$ 3\,deg to HD~36726~A with $J$ = 
15.109$\pm$0.137\,mag.
\item[*] HD~290572 is a B8V and a K0V (sic) according to Cannon \& Pickering 
(1924) and Nesterov et~al. (1995), respectively.
An intermediate~A spectral type better matches the $B_T-K_{\rm s}$ 
colour.
\item[*] HD~290569 is an A0V according to Nesterov et~al. (1995);
however, the relatively red colour $B_T-K_{\rm s}$ = 1.29$\pm$0.12\,mag better 
matches with a later spectral type (i.e. intermediate~A).
\end{list}

\paragraph{Notes to Table~\ref{TYC2Oalnilam}:}
\begin{list}{}{}
\item[*] HD~37038~AB is a binary whose components are separated by $\rho 
\approx$ 0.6\,arcsec, $\theta \approx$ 265\,deg.
The secondary is 2\,mag fainter than the primary (Dommanget \& Nys~1994).
It could actually be a hierarchical triple, since Nordstr\"om et~al. (1997) 
found the F-type dwarf to be a double-lined spectroscopic binary with evident
radial-velocity variations in scales of a few days 
(the resolved binary cannot be responsible of such~variations).
\item[*] HD~36863 has a radial velocity that deviates more than 25\,km\,s$^{-1}$
with respect to the average radial velocity of the association (Gieseking 1983).
It satisfies, however, the photometric and proper motion criteria of very young 
stars in Orion.
HD~36863 might be a very young single-line spectroscopic binary (SB1).
Guetter (1976) classified it as an A7-type~star. 
\item[*] [NYS99]~A--06 is an X-ray source identified by {\em ASCA}
and~{\em ROSAT} ({1AXG~J053448--0131}, {1RXS~J053450.5--013120}).
We estimate F: spectral type from its $V_T-K_{\rm s}$ colour.
\item[*] HD~290673 is also {BD--01~975}.
\item[*] TYC~4767~1130~1 seems to be a close ($\rho \sim$ 3\,arcsec) binary from
the 2MASS data.
\item[*] HD~37172 is a probable non-member of the association according to
Guetter (1976), although other authors consider it to be a real member (Warren 
\& Hesser 1977, 1978; Hesser, McClintock \& Henry 1977; Gieseking 1983; de~Geus 
\& van~de~Grift 1990).   
It has a peculiar Mn~{\sc i} $\lambda$4030--4035\,\AA~blend (Gray \& Corbally
1993) and falls slightly to the red of the association sequence in the 
$V_T-K_{\rm s}$ vs. $V_T$ diagram (Fig.~\ref{TYC2M}).
It is at $\rho \approx$ 5.0\,arcmin, $\theta \approx$ 115\,deg, to~Alnilam.
\item[*] TYC~4766~2371~1 is the third closest bright star to~Alnilam ($\rho 
\approx$ 6.0\,arcmin, $\theta \approx$ 250\,deg).
\item[*] TYC~4766~2150~1 ({CCDM~J05375--0103B}) has been repeatedly
proposed to form a binary system together with HD~290749 ($\rho \approx$
29.0\,arcsec, $\theta \approx$ 344\,deg -- Burnham 1906; Dommanget \& Nys 1994).
They do not share, however, a common proper~motion.
\item[*] TYC~4766~542~1 might be the X-ray source 
{1RXS~J053447.9--010224} ({1AXG~J053446--0102}; Ueda et~al. 2001).
The nearby binary {2MASS~J05344642--0102340} ($\rho \approx$ 4.0\,arcsec,
$\theta \approx$ 220\,deg, $\Delta H$ = 1.89$\pm$0.05\,mag), at $\sim$35\,arcsec
to the west of TYC~4766~542~1 could also be the X-ray~source.
\item[*] HD 290746 is a G0V according to Nesterov et~al. (1995);
however, the relatively blue colour $B_T-K_{\rm s}$ = 1.38$\pm$0.09\,mag better 
matches with an earlier spectral type (i.e. late~A or early~F).
\end{list}

\paragraph{Notes to Table~\ref{TYC2Omintaka}:}
\begin{list}{}{}
\item[*] HD 290585 has a debris disc according to MIPS/{\em Spitzer} 
observations at 24\,$\mu$m by Hern\'andez et~al. (2006).
It is a double binary resolved by 2MASS ($\rho \approx$ 5.6\,arcsec, 
$\theta \approx$ 132\,deg, $\Delta H$ = 2.05$\pm$0.05\,mag);
the secondary is not in the Tycho-2 catalogue.
\item[*] HD 290513 is a F0V dwarf according to Nesterov et~al. (1995);
a G--K spectral type better matches the observed colours.
\item[*] TYC~4766~790~1 has a visual companion at $\rho \approx$ 8.5\,arcsec, 
$\theta \approx$ 190\,deg ($\Delta H$ = 0.69$\pm$0.04\,mag).
\item[*] HD~290583~A has a visual companion of roughly the same brightness.
{HD~290583~B}, not identified by Tycho-2, is at $\rho \approx$ 
7.44\,arcsec, $\theta \approx$ 3.8\,deg ($\Delta H$ = 0.40$\pm$0.03\,mag).
\item[*] HD~290507 and HD~290504 are A5V dwarfs according to Nesterov et~al. 
(1995);
F--G spectral types better match the observed colours.
\item[*] TYC~4766~528~1 is the brightest component of a visual triple system.
The other two components are located at $\rho \approx$ 3.4\,arcsec, 
$\theta \approx$ 165\,deg ($\Delta H$ = 0.36$\pm$0.07\,mag), and $\rho \approx$ 
7.4\,arcsec, $\theta \approx$ 195\,deg ($\Delta H$ = 2.07$\pm$0.06\,mag).
\item[*] TYC~4766~1168~1 has a visual companion at $\rho \approx$ 9.4\,arcsec, 
$\theta \approx$ 10\,deg ($\Delta H$ = 0.98$\pm$0.04\,mag).
\item[*] TYC~4766~2424~1 has a low a proper motion of less than 5\,mas\,a$^{-1}$
and a blue colour $B_T-K_{\rm s}$ = 0.63$\pm$0.09\,mag, typical of early A-type 
stars in the association.
\item[*] TYC~4753~49~1 is at $\rho \sim$ 46\,arcsec, $\theta \sim$ 288\,deg
to the radio source {[LPZ94] 146} (Larionov et~al.~1994). 
\end{list}

\paragraph{Notes to Table~\ref{TYC2Nalnilam}:}
\begin{list}{}{}
\item[*] IRAS~05354--0142 has the reddest $V_T-K_{\rm s}$ colour among the 
$\sim$1500 Tycho-2/2MASS investigated stars in the survey area ($V_T-K_{\rm s}$ 
= 7.1$\pm$0.3\,mag). 
The closeness of IRAS~05354--0142 to the Ori~I--2 globule may explain part of 
its reddening, but not all. 
It might be a S-type or a C-type giant with a very late spectral type and very
low effective temperature. 
The absence of a mid-infrared excess (Kraemer et~al. 2003) rules out the 
hypothesis of IRAS~05354--0142 being a proto-star in the upper part of the 
Hayashi track of collapse associated to the Bok~globule. 
\item[*] HD~290680 has a $V_T-K_{\rm s}$ colour that better matches with a K
spectral type.
\item[*] HD~290679 is also        {GSC~04766--01466}.
\item[*] HD~36780 is a K5III giant as tabulated in SIMBAD.
It has a (variable) radial velocity inconsistent with association membership 
(V$_r$ = +91\,km\,s$^{-1}$, Griffin 1972; V$_r$ = +73.2\,km\,s$^{-1}$, Bois et 
al. 1988). 
The star has an {\em Hipparcos} distance of 260$\pm$50\,pc.
\item[*] HD~290675 has a discordant radial velocity (Gieseking 1983).
It is also        {BD--01~967}.
\item[*] HD~37491 is likely associated to the mid-infrared source 
       {IRAS~05363--0111}.
\item[*] HD 290749 is a B8V star according to Nesterov et~al. (1995).
With proper motion of 13.8\,mas\,a$^{-1}$ and colour $V_T-K_{\rm s} \sim$ 
1.0\,mag, it is likely a late A- or an early F-type star in the foreground.
See also the note for TYC~4766~2150~1 (Table~\ref{TYC2Oalnilam}).
\item[*] HD~36882 is at $d$\,=\,220$\pm$50\,pc (Perryman et~al. 1997).
Because of a transcription error, SIMBAD tabulates HD~36882 as one of the 
early-type stars associated to the H~{\sc ii} region {Sh~2--264}, close to 
{$\lambda$~Ori}, in Sharpless (1959);
the actual early-type star is {$\phi^{01}$~Ori} (HD~36822, B0III).
\item[*] HD~290667 and StHA~46 were catalogued by Stephenson (1986) as 
H$\alpha$ emission stars.
However, Downes \& Keyes (1988) and Maheswar et~al. (2003) failed to detect the 
(sporadic?)~emission.
StHA~46 is at 18\,arcsec to the southwest of the early-A star AG--00~669.
\item[*] HD~290647 falls in the tiny overlapping region between the Alnilam and
Mintaka fields.
It is also        {BD--00~1001}.
\item[*] TYC~4767~2257~1 has a colour $V_T-K_{\rm s} >$ 4.5\,mag, typical of 
intermediate M stars.
Its proper motion is, however, very low ($\mu \lesssim$ 1\,mas\,s$^{-1}$).
\end{list}

\paragraph{Notes to Table~\ref{TYC2Nmintaka}:}
\begin{list}{}{}
\item[*] HD~290647 falls in the tiny overlapping region between Alnilam and
Mintaka fields.
It is also BD--00~1001.

\item[*] TYC~4766~516~1 has a very red colour of $V_T-K_{\rm s}$ =
5.44$\pm$0.09\,mag and no {\em IRAS} excess.
It could be a mid-M-type giant or subgiant in back-/foreground.

\item[*] HD~290576 is also {BPM~71736}.

\item[*] TYC~4766~2124~1 is also {GSC~04766--02124}.

\item[*] HD~290568 is also {BD--00~987} and {IRAS~05303--0009}.

\item[*] HD~36117 is a nearby ($d$ = 170$\pm$30\,pc; Perryman et~al. 1997), 
peculiar, A-type star (Gray \& Corbally 1993) with X-ray emission 

\item[*] HD~36139 is a nearby ($d$ = 124$\pm$13\,pc; Perryman et~al. 1997), 
high rotation-velocity, radial-velocity variable (Morrell \& Levato~1991), 
A-type star with no known companion.

\item[*] HD~36840 is at an {\em Hipparcos} distance of $d$ = 380$\pm$120\,pc,
which is probably incorrect, given the spectral type of the star (G5V).

\item[*] HD~36558 has a discordant radial velocity of $V_r$ = 
+42.1$\pm$4.8\,km\,s$^{-1}$ (Nordstr\"om et~al.~2004).

\item[*] HD~36443 (LHS~5107, G~99--16; Roman 1995) is a well-known, solar-like, 
high-velocity star at only $d$ = 38.2$\pm$1.9\,pc and with radial velocity 
$V_r \approx$ --9.1\,km\,s$^{-1}$ (Wilson~1953).

\item[*] HD~290486~AB is a visual binary star with $\rho \approx$ 1.7\,arcsec, 
$\theta \approx$ 304\,deg (Dommanget \& Nys 1994).
\end{list}

\paragraph{Notes to Table~\ref{DE2Malnilam}:}
\begin{list}{}{}
\item[*] E\,Ori~2--1328 is a young M4.5-type very low-mass star with lithium in 
absorption (pEW(Li~{\sc i}) = +0.40$\pm$0.05\,\AA) and Balmer lines in faint 
(chromospheric) emission (pEW(H$\alpha$) = --8.2$\pm$0.5\,\AA).
The sodium line in the red optical is weak in comparison with field dwarfs of 
the same spectral type (pEW(Na~{\sc i}) = +3.4$\pm$0.5\,\AA; B\'ejar 
et~al.~2003a). 

\item[*] E\,Ori~2--1868 is a young M6.0-type very low-mass star with faint 
alkali lines (pEW(Na~{\sc i}) $<$ 3\,\AA; B\'ejar et~al.~2003b). 
The $i$-band magnitude has been taken from the SuperCOSMOS Science Archive 
(Hambly et~al.~2001).

\item[*] E\,Ori~1--388 is a young M6.0-type very low-mass star with faint alkali
lines (pEW(Na~{\sc i}) = +3.9$\pm$0.5\,\AA; B\'ejar et~al.~2003a) and Balmer 
lines in faint (chromospheric) emission (pEW(H$\alpha$) = --6.5$\pm$2.0\,\AA).
It is embedded in the [OS98]~40B remnant molecular cloud. 

\item[*] E\,Ori~1--1644 is a young M5.0-type very low-mass star with Balmer 
lines in faint (chromospheric) emission (pEW(H$\alpha$) = --7.6$\pm$1.0\,\AA; 
B\'ejar et~al.~2003a). 

\item[*] E\,Ori~2--878 is a young M5.5-type very low-mass star with faint 
alkali lines (pEW(Na~{\sc i}) $<$ 3\,\AA; B\'ejar et~al.~2003b). 

\item[*] E\,Ori~2--705 is a young M5.0-type very low-mass star with faint 
alkali lines (pEW(Na~{\sc i}) $<$ 4\,\AA; B\'ejar et~al.~2003b). 
The $i$-band magnitude is from SuperCOSMOS.

\item[*] E\,Ori~2--603 is a young M5.5-type very low-mass star with faint 
alkali lines (pEW(Na~{\sc i}) $<$ 3\,\AA; B\'ejar et~al.~2003b). 

\item[*] Kiso~A--0904~41 and Kiso~A--0904~42 form a binary system with 
$\rho \approx$ 11.0\,arcsec, $\theta \approx$ 277\,deg.
The $i$-band magnitudes are from SuperCOSMOS.
They hay been identified in {\em XMM-Newton} observations (Caballero et~al.,
in~prep). 

\item[*] V469~Ori is proably associated to the {[OS98]~29J}, 
{[OS98]~29H} and {[OS98]~29K} remnant molecular clouds.

\item[*] Kiso~A--0904~76 is a K6-type variable ($\Delta V$ = 0.33\,mag) star 
with pEW(H$\alpha$) = --20.2\,\AA~and pEW(Li~{\sc i}) = +0.5\,\AA~(Brice\~no 
et~al.~2005).
It has a visual companion at $\rho \approx$ 5.3\,arcsec, $\theta \approx$ 
321\,deg ($\Delta H$ = 2.91$\pm$0.05\,mag).

\item[*] Haro 5--80 is a variable, emission-line star (Haro \& Moreno 1953; 
Fedorovich 1960; Wiramihardja et~al. 1989).
It has a very nearby ($\rho \sim$ 3\,arcsec, $\theta \sim$ 15\,deg) companion or
small jet (possibly associated to an unknown Herbig-Haro~object)

\item[*] 2E~1398 is an X-ray source tabulated in at least six catalogs from
{\em Einstein}, {\em ROSAT} and {\em XMM-Newton} data (Harris et~al. 1994;
McDowell 1994; Moran et~al. 1996; Voges et~al. 1999; {\em ROSAT} Consortium
2000; {\em XMM-Newton} Survey Science Centre Consortium 2007).
It is located at 4.5\,arcmin to the west of Alnilam.
The $i$-band magnitude is from SuperCOSMOS.

\item[*] Kiso~A--0904~37 has rather blue $I-J$ and $J-K_{\rm s}$ colours.
Besides, it has a faint, red, visual companion at $\rho \approx$ 
5.6\,arcsec, $\theta \approx$ 144\,deg, unresolved by Wiramihardja et~al. (1989).
It is likely that the actual emission-line star or brown dwarf (and the only 
truly young objects) is the visual companion (J2000 coordinates: 05~35~32.38 
--01~12~08.2).

\item[*] StHA~47 is a mid-K-type T~Tauri star according to Downes \& Keyes
(1988). 
It is the fourth strongest X-ray emitter at less than 20\,arcmin to Alnilam,
from {\em XMM-Newton} observations.

\item[*] CVSO~162 is an M1-type variable ($\Delta V$ = 0.23\,mag) star 
with pEW(H$\alpha$) = --3.9\,\AA~and pEW(Li~{\sc i}) = +0.5\,\AA~(Brice\~no 
et~al.~2005).

\item[*] Annizam~363062 is a visual binary with $\rho \approx$ 6.3\,arcsec, 
$\theta \approx$ 243\,deg ($\Delta H$ = 2.41$\pm$0.06\,mag).
Five X-ray events in the surrounding area were tabulated in the catalogue of 
{\em ROSAT} HRI Pointed Observations (ROSAT Team 2000).
It might also be the {\em Einstein} source 2E~B0534--0111.
It is not known which component is the actual X-ray emitter.

\item[*] Haro~5--67 is a G:-type, photometrically variable, T~Tauri star with 
strong Balmer emission detected by several H$\alpha$ objective-prism surveys 
(Haro \& Moreno 1953; Sanduleak 1971; Stephenson 1986) and with {\em IRAS} flux 
excess (Weaver \& Jones 1992).
It has been spectroscopically followed-up by Herbig \& Kaneswara Rao (1972) and
Downes \& Keyes~(1988).

\item[*] Kiso~A--0904~50 has rather blue $I-J$ and $J-K_{\rm s}$ colours.
It might be a variable young star or an active object in the fore-/background.

\item[*] Kiso A--0904~61 could also be the H$\alpha$ emitter {Haro~5--77} 
(suspected variable NSV~2465; Kukarkin et~al.~1981).

\item[*] V583~Ori ({Haro~5--74}) is a variable, emission-line star 
(Haro \& Moreno 1953; Fedorovich 1960; Wiramihardja et~al. 1989).
It is possibly the X-ray source {2E~1423} and might form a Lindroos system
together with the B9V star HD~37187 (Table~\ref{TYC2Malnilam}).

\item[*] HD~36779~B is a K5IV with Li~{\sc i} in absorption in a likely 
Lindroos system with the B-type star HD~36779 (Table~\ref{TYC2Malnilam} -- 
Lindroos 1985; Pallavicini, Pasquini \& Randich 1992; 
Mart\'{\i}n, Magazz\`u \& Rebolo 1992).
See, however, a brief discussion on the position in the H-R diagram and the 
radial velocity of HD~36779~B in Gerbaldi, Faraggiana \& Balin~(2001).

\item[*] 2E~1357 is a K3-type star with pEW(H$\alpha$) = --2.75\,\AA~and 
pEW(Li~{\sc i}) = +0.470$\pm$0.008\,\AA~(Alcal\'a et~al. 1996,~2000).

\item[*] Kiso~A--0904~28 is also the X-ray source        {2E~1340} (McDowell 
1994).
The star was detected by Kraemer et~al. (2003) at 8.3\,$\mu$m, which suggests a 
possible flux excess in the mid-infrared.

\item[*] Kiso~A--0904~30 was previously identified with a fainter, much bluer 
source 24\,arcsec to the~east.

\item[*] 2E~1449 is a K4-type star with pEW(H$\alpha$) = +0.35\,\AA~and 
pEW(Li~{\sc i}) = +0.410$\pm$0.010\,\AA~(Alcal\'a et~al. 1996,~2000).

\item[*] Kiso~A--0904~60 is a K6-type variable ($\Delta V$ = 0.51\,mag) star 
with pEW(H$\alpha$) = --61.9\,\AA~and pEW(Li~{\sc i}) = +0.3\,\AA~(Brice\~no 
et~al.~2005).
It could be a tight binary ($\rho \lesssim$ 1.0\,arcsec)
based on the 2MASS photometry quality flags.

\item[*] Kiso A--0904~65 is a variable ($\Delta V$ = 0.87\,mag) star with 
a very strong Balmer emission, pEW(H$\alpha$) = --400\,\AA, and lithium in 
absorption, pEW(Li~{\sc i}) = +0.3\,\AA~(Brice\~no et~al.~2005).  

\item[*] CVSO~124 is an M3-type variable ($\Delta V$ = 0.19\,mag) star 
with pEW(H$\alpha$) = --17.3\,\AA~and pEW(Li~{\sc i}) = +0.4\,\AA~(Brice\~no 
et~al.~2005).
It also falls in the Mintaka field.

\item[*] Kiso A--0904~34 is also        {[SE2005]~104}. 
It does not show significant periodic variability.

\item[*] Kiso~A--0904~33 has a faint, red, visual companion at $\rho \approx$ 
4.9\,arcsec, $\theta \approx$ 14\,deg, with near-infrared magnitudes $J$ = 
12.55$\pm$0.03\,mag and $K_{\rm s}$ = 11.55$\pm$0.03\,mag.

\item[*] PU~Ori is a pre-main sequence star with H$\alpha$ in strong two-lobe 
emission (Haro \& Moreno 1953; Herbig \& Kameswara Rao 1972; Cohen \& Kuhi 1979;
Wiramihardja et~al. 1989), photometric variability (Fedorovich 1960; Brice\~no 
et~al. 2005), mid-infrared flux excess at the {\em IRAS} passbands (Weintraub 
1990; Weaver \& Jones 1992) and forbidden emission lines ([O~{\sc i}] 
$\lambda$~6300.3\,\AA; Hirth, Mundt \& Solf 1997).
It has an extraordinary red colour of $J-K_{\rm s}$ = 1.77$\pm$0.03\,mag.  

\item[*] StHA~48 is a K4-type T~Tauri star according to Maheswar et~al. (2003).

\end{list}

\paragraph{Notes to Table~\ref{DE2Mmintaka}:}
\begin{list}{}{}

\item[*] CVSO~124 is an M3-type variable ($\Delta V$ = 0.19\,mag) star 
with pEW(H$\alpha$) = --17.3\,\AA~and pEW(Li~{\sc i}) = +0.4\,\AA~(Brice\~no 
et~al.~2005).
It also falls in the Alnilam field.

\item[*] Kiso A--0903~221 has a faint red visual companion at $\rho \approx$ 
6.4\,arcsec, $\theta \approx$ 49\,deg ($\Delta H$ = 5.1$\pm$0.2\,mag).

\item[*] Kiso~A--0904~4 is a K7-type variable ($\Delta V$ = 0.66\,mag) star 
with pEW(H$\alpha$) = --34.2\,\AA~and pEW(Li~{\sc i}) = +0.3\,\AA~(Brice\~no 
et~al.~2005).
It is also        {Kiso~A--0903~228}.

\item[*] IRAS 05307--0038 (also known as        {YSO~CB031YC1} and
       {YSO--C~CB031YC1--I}) is a bright ($K_{\rm s}$ = 8.63$\pm$0.02\,mag)
T~Tauri star embedded in the        {IC~434} Bok globule/reflection nebula
(Dreyer 1895; Hubble 1922; Cederblad 1946; Dorschner \& G\"urtler 1963; Magakian
2003).  
The star has an extended and fuzzy nebulosity in the $J$ band and was classified
as a Class~II object because of a flux excess at 25\,$\mu$m (Yun \& Clemens
1994, 1995).  
Yun et~al. (1997) measured H$\alpha$ and H$\beta$ in emission and Li~{\sc i} in
absorption, and characterised its SED from the $B$ band to the mid-infrared. 
Yun et~al. (1996) discovered two nearby radio sources that could be
associated to the star (there are other two additional radio sources in the
surrounding area, found by Condon et~al. 1998). 
G\'omez et~al. (2006), in a very sensitive survey using NASA 70\,m antenna at
Robledo de Chavela (Spain), failed to detect water maser emission from these 
sources.
See Fig.~\ref{therollingstones}.

\item[*] Kiso A--0904~22 has a nearby visual companion identified as a 
photometric young star candidate, Mantaqah~2385113 (Table~\ref{DE2Omintaka}).

\item[*] Kiso~A--0904~13 is also {Kiso~A--0903~245}.

\item[*] Mantaqah~487126 displayed 14 X-ray events during {\em ROSAT} 
observations (White et~al. 2000 --8 events--; ROSAT 2000 --6 events--).

\item[*] 1AXG~J053127--0021 appears in numerous X-ray catalogs from 
{\em ROSAT} and {\em ASCA} data.
The $i$-band magnitude is from SuperCOSMOS.

\item[*] Mantaqah~148186 is an X-ray source in several catalogs (e.g. Voges 
et~al.~1999).

\item[*] Mantaqah~400105 is an X-ray source in several catalogs (e.g. White 
et~al. 2000).

\item[*] Mantaqah~320042 displayed 12 X-ray events during {\em ROSAT} 
observations (White et~al. 2000 --5 events--; ROSAT 2000 --7 events--).

\item[*] Kiso~A--0904~18 has a $J-K_{\rm s}$ colour redder than 2.0\,mag.

\end{list}

\paragraph{Notes to Table~\ref{DE2O}:}
\begin{list}{}{}
\item[*] E\,Ori~2--1982 has no DENIS or SuperCOSMOS counterpart.
The $i$-band magnitude (actually $I$) is from B\'ejar et~al. (2003b).

\item[*] [OSP2002]~OriI--2N~4 has a tiny (chromospheric?) Balmer emission
(pEW(H$\alpha$) $\approx$ --2.6\,\AA; Ogura, Sugitani \& Pickles~2002).

\item[*] BD--01~969B (``Alnilam~B'', $\epsilon$~Ori~B; $V$ = 10.5\,mag) is 
located at $\rho \approx$ 179.0\,arcsec, $\theta \approx$ 58\,deg, to Alnilam. 
(this value coincides with the original measurements by Burnham in 1879 -- 
Burnham 1906).
No spectral type has been tabulated or measured.

\item[*] AG--00~669 forms a visual double with the foreground solar-like star
StHA~46 (Jeffers, van den Bos \& Greeby 1963).
See Table~\ref{TYC2Nalnilam}.

\item[*] BD--00~983B (``Mintaka~B'', $\delta$~Ori~B; $V$ = 14.0\,mag) is located
at $\rho \approx$ 33.0\,arcsec, $\theta \approx$ 229\,deg, to Mintaka~AE--D 
(this value coincides with the original measurements by Burnham in 1878 -- 
Burnham 1906).
No spectral type has been tabulated or measured.
The star shows no evidence of any significan X-ray emission in deep observations with 
the {\em Chandra} Space Telescope (Miller et~al.~2002).

\item[*] BD--00~993B (``HD~36726~BC''; $V$ = 13.7\,mag) is located at $\rho 
\sim$ 19.8\,arcsec, $\theta \sim$ 213\,deg ($\Delta H$ = 1.30$\pm$0.05\,mag) to 
the A0Vm-type star HD~36726 (Table~\ref{TYC2Mmintaka}). 
BD--00~993B is, in its turn, a close binary with $\rho \sim$ 0.8\,arcsec, 
$\theta \sim$ 64\,deg (Dommanget \& Nys~1994).
\end{list}

\paragraph{Notes to Table~\ref{DE2N}:}
\begin{list}{}{}
\item[*] X~Ori is the reddest object in the studied area.
It is an M8--9-type Mira~Cet variable found by Wolf (1904) with $P$ =
424.15$\pm$1.77\,d (Templeton, Mattei \& Willson 2005) and silicate dust 
emission (Sloan \& Price 1998; Speck et~al. 2000).
Although X~Ori is even brighter in the near- and mid-infrared than Alnilam and
Mintaka ($K_{\rm s} \sim$ 0.9\,mag), its extremely red $V-K_{\rm s}$ colour, of
more than 10\,mag, prevented its detection in the Tycho-2~catalogue. 

\item[*] Ruber~1 has a proper motion of --12$\pm$5, --49$\pm$2\,mas\,s$^{-1}$,
measured by us using POSSI, UKST blue, red and infrared, DENIS and 2MASS
(the six epochs cover 47 years; see method details in Caballero 2007b).
Ruber~1 probably is a late M dwarf in the foreground.

\item[*] G~99--18 is a high proper-motion star, with $\mu$ = 280\,mas\,a$^{-1}$.
It falls in the tiny overlapping region between the Alnilam and Mintaka fields.

\item[*] Ruber~2 has very peculiar colours: the near-infrared colours are very 
red (e.g. $J-K_{\rm s}$ = 1.33$\pm$0.04\,mag), typical of early and 
intermediate L dwarfs, very late M giants or carbon stars.
Its $I-J$ colour is red enough, as well, to be selected as an association
member candidate.
The optical colours are, however, contradictory and variable.
The SuperCOSMOS Science Archive tabulates photographic magnitudes
$B$(1988.0) = 17.643\,mag, $R_1$(1951.9) = 19.148\,mag and 
$R_2$(1989.0) = 15.713\,mag. 
Photographic $R$ bands are separated by 37.1 years and probably reflect
intrinsic (high-amplitude, long-time-scale) photometric variability of the 
object.
SuperCOSMOS and USNO-B1 tabulate appreciable proper motions for Ruber~2.
However, after a careful astrometric study using the original plate 
digitisations, DENIS and 2MASS, we conclude that the proper motion of the object
is null within uncertainties of 10\,mas\,s$^{-1}$.
Ruber~2 is at a separation of only $\sim$9\,arcmin to the core of the unusual 
Berkeley~20 cluster (see Section~\ref{introduction}). 
Therefore, Ruber~2 probably is a pulsating late M giant of that cluster, at an
heliocentric distance of~8.4\,kpc.
\end{list}

\paragraph{Notes to Table~\ref{gx2Malnilam}:}
\begin{list}{}{}
\item[*] 2MASS J05345451--0143256 has a 2MASS double detection, with quality 
flags AUU and UEA.
It is not in the NASA/IPAC Extragalactic Database (NED).
It could be an unresolved stellar binary instead of a~galaxy.

\item[*] 2MASX J05332498--0106242 is also the NED object {LCSB~S0895N} 
(Monnier Ragaigne et~al. 2003).

\item[*] 2MASX J05365804--0042413 is the infrared source IRAS~05344--0044, very 
close to the young early-type star HD~37235.

\item[*] 2MASX J05364723--0039144 is also NED object {LCSB~S0899N} 
(Monnier Ragaigne et~al. 2003).

\item[*] 2MASS J05364746--0039110 is located at 4.3\,arcsec to the centre of 
2MASX~J05364723--0039144 (see just above), in the plane of the galaxy.
This source is probably an~artifact.
\end{list}

\paragraph{Notes to Table~\ref{gx2Mmintaka}:}
\begin{list}{}{}
\item[*] 2MASX J05322266--0000555 is also the NED object        {LEDA~147610} 
(Klemola, Jones \& Hanson 1987; Paturel et~al.~1989). 

\item[*] 2MASS~J05341337--0044087 is        {PMN~J0534--0044}, a powerful radio 
source discovered in many surveys (e.g. Becker, White \& Edwards 1991; Griffith
et~al. 1995; Douglas et~al. 1996; Condon et~al. 1998). 
Its optical/near infrared counterpart is faint ($i$ = 16.82$\pm$0.11\,mag) and
relatively blue ($i-K_{\rm s}$ = 1.2$\pm$0.2\,mag).
\end{list}

\begin{figure}
\centering
\caption{False-colour composite image, 7.5\,$\times$\,7.5\,arcmin$^2$ wide, of
the IC~434 Bok globule, centred on IRAS~05307--0038 (labelled as YSO~CB031YC1). 
Blue, green and red are for photographic $B_J$, $R_F$ and $I_N$, respectively.
North is up, east is left.
The six reddened sources in the IC~434 Bok globule (Table~\ref{DE2Mred}) are
indicated with (red) circles.
{\em [NOTE to the reader: this image is only available at A\&A]}.} 
\label{therollingstones}
\end{figure}

\paragraph{Notes to Table~\ref{DE2Oalnilam}:}
\begin{list}{}{}
\item[*] Annizam~2473146 is at $\sim$4\,arcsec to the east of an extended 
source with a galactic appearance.

\item[*] Annizam~2464138 could be reddened by the nearby Ori~I--2 Bok globule
and be a background~star. 

\item[*] Annizam~1840146 was subject of a dedicated astrometric study using 
public data (plate digitisations, DENIS and 2MASS).
There seems to be an artifact in the POSSI Schmidt plate of 1951 that causes a 
false proper motion of more than 100\,mas\,s$^{-1}$ in the astrometric catalogs 
USNO-B1 and SuperCOSMOS Science Archive.
The very red object has, however, null proper motion within the uncertainty of 
10\,mas\,s$^{-1}$ using seven astrometric epochs between 1987 and 2000 
(see Caballero 2007b for details of the astrometric analysis).

\item[*] Annizam~1106127 has 2MASS photometry quality flags ``EEE''; 
it might indicate that it is an unresolved binary with $\rho \sim$ 1--2\,arcsec.

\item[*] Annizam~1415101 is surrounded by a galaxy arm-like structure.
It could be the point-like core of a background galaxy.

\item[*] Annizam~2042095 has a bluer, fainter, visual companion at about 
3\,arcsec to the~northeast.

\item[*] Annizam~2611268 is at $\sim$8\,arcsec to the south of a probable 
foreground star, about $\sim$3.3\,mag brighter in the $i$~band.

\item[*] Annizam~123132 is in the glare of Alnilam ($\rho \sim$ 
2.0\,arcmin, $\theta \sim$ 130\,deg).

\item[*] Annizam~1751268 is probably the {\em Einstein} X-ray source
{2E~1458}. 

\item[*] Annizam~1748267 is at only $\rho \sim$ 8.4\,arcsec, $\theta
\sim$ 340\,deg to Annizam~1751268 (see just~above).

\item[*] V993~Ori is a long-time-known variable star discovered by 
Luyten~(1932).
It has a very red colour $J-K_{\rm s}$ = 1.26$\pm$0.04\,mag for its brightness
($H$ = 9.82$\pm$0.02\,mag).

\item[*] Annizam~798196 could be associated to the X-ray source 
{1WGA~J0536.4--0059}, found in several {\em ROSAT} HRI and PSPC catalogues
(Moran et~al. 1996; {\em ROSAT} Consortium 2000; Flesch \& Hardcastle 2004). 
It might be a background source associated to the (extragalactic?) radio 
source {NVSS~053627--005937} in the 1.4\,GHz NRAO VLA Sky Survey (Condon et~al.
1998), located at $\rho \sim$ 18\,arcsec, $\theta \sim$~200\,deg. 

\item[*] {[SE2005]~120} (Mantaqah~1357158) is a photometric candidate member of
the ``$\epsilon$~Orionis cluster'' firstly identified by Scholz \& Eisl\"offel
(2005). 
It is the most variable star of the five objects with high-amplitude ($A_I$ =
0.952\.mag), irregular variations.
It also have a significant periodic variability of $P$ = 82$\pm$3\,h, with
superimposed short-term fluctuations.
Scholz \& Eisl\"offel (2005) classified it as very low-mass analogue of
classical T~Tau stars affected by intrinsic reddening.
For explaining the large variations, they proposed two possible scenarios
involving a eclipsing ``hot Jupiter'' in close orbit and ``hot spots formed
by matter flow from an accretion disc onto the central object''.
It might be, however, a typical eclipsing binary in the fore-/background.

\item[*] {[SE2005]~126} (Mantaqah~1582164) is other photometric candidate member
of the ``$\epsilon$~Orionis cluster'' with significant periodic variability in
the work by Scholz \& Eisl\"offel (2005).
In this case, the object has a low-amplitude variability ($A_I$ = 0.016\,mag)
with a very short period ($P$ = 4.06$\pm$0.05\,mag).
The values are consistent with pulsations induced by deuterium-burning in young
brown dwarfs (Caballero et~al. 2004; Palla \& Baraffe~2005).

\item[*] [SE2005]~71 (Annizam~2446149) is a non-variable photometric candidate
member of the ``$\epsilon$~Orionis~cluster'' (Scholz \& Eisl\"offel 2005). 
It is located at $\rho \sim$ 14\,arcsec, $\theta \sim$ 270\,deg, to V472~Ori.
\end{list}

\paragraph{Notes to Table~\ref{DE2Omintaka}:}
\begin{list}{}{}
\item[*] Mantaqah~2041159 is surrounded by galaxy arm-like structures.
It could be the point-like core of a background galaxy.

\item[*] {[SE2005]~44} (Mantaqah~2627125) is a photometric candidate 
member of the ``$\epsilon$~Orionis cluster'' with significant periodic
variability ($A_I$ = 0.027\,mag, $P$ = 31.0$\pm$1.8\,h; Scholz \&
Eisl\"offel 2005). 

\item[*] Mantaqah~2216132 has a fainter redder visual companion at $\rho \sim$ 
5.6\,arcsec, $\theta \sim$ 210\,deg.

\item[*] Mantaqah~2385113 is at $\rho \approx$ 6.87\,arcsec, $\theta \approx$
352\,deg to Kiso~A--0904~22, together with it could form a $\sim$2700\,AU-wide
low-mass binary.

\item[*] Mantaqah~941101 has a brighter bluer visual companion at $\rho \sim$ 
4.5\,arcsec, $\theta \sim$ 230\,deg.

\item[*] Mantaqah~1926266 is at $\rho \approx$ 15.1\,arcmin, $\theta \approx$ 
224\,deg, to Kiso~A--0903~183.

\item[*] Mantaqah~161138 is in the glare of Mintaka ($\rho \sim$ 
2.7\,arcmin, $\theta \sim$ 140\,deg).

\item[*] Mantaqah~104142 is in the glare of Mintaka ($\rho \sim$ 
1.7\,arcmin, $\theta \sim$ 140\,deg).

\item[*] Mantaqah~67144 is in the glare of Mintaka ($\rho \sim$ 
1.1\,arcmin, $\theta \sim$ 145\,deg).
Its $J$-band magnitude is strongly affected.

\item[*] Mantaqah~1982094 has a blue visual companion  at $\rho \sim$ 
6.6\,arcsec, $\theta \sim$ 60\,deg. 
\end{list}

\paragraph{Notes to Table~\ref{remarkable}:}
\begin{list}{}{}
\item[*] Albus~1 has by far the bluest colour among all the investigated objects
($V_T-K_{\rm s}$ = --0.95$\pm$0.14\,mag) and a measurable proper motion $\mu$ =
19\,mas\,a$^{-1}$.
Although Albus~1 was detected from the data presented in this work, it was
followed-up with additional photometric data and discussed in detail in
Caballero \& Solano (2007). 
Very recently, Vennes, Kawka \& Allyn Smith (2007) obtained a series of optical
spectra, showing that it is a peculiar, bright, helium-rich B3 subdwarf ($n_{\rm
He}$ = 0.6$\pm$0.1).
\item[*] TYC~5360~681~1, the stellar counterpart of an {\em IRAS} source, has an 
appreciable proper motion tabulated by Tycho-2: ($\mu_\alpha \cos{\delta}$, 
$\mu_\delta$) = (--4$\pm$2, +16$\pm$3)\,mas\,a$^{-1}$. 
Assuming an heliocentric distance of $d \gtrsim$ 1\,kpc, typical of a giant, it
would have a very large tangential velocity of V$_t \gtrsim$ 80\,km\,s$^{-1}$. 
\item[*] BD--13~1293 is the reddest {\em IRAS} source in this list.
The star, investigated here for the first time, displays very strong wide 
absorption bands, especially at $\sim$1\,$\mu$m  (between the $i$ and $J$ bands).
It is a photometric variable: from the Hipparcos catalogue, its $V_T$ magnitude 
varies between $V_T$ = 9.41 and 10.00\,mag (15 and 85\,\% percentiles, 
respectively). 
\end{list}

\begin{longtable}{llccccc}
\caption{\label{allTYC2Malnilam} Photometry of correlated Tycho-2/2MASS sources in the Alnilam field.}\\ %
\hline
            \noalign{\smallskip}
Name		& TYC		& $B_T$		& $V_T$		& $J \pm \delta J$	& $H \pm \delta H$	& $K_{\rm s} \pm \delta K_{\rm s}$ \\  
		& 		& (mag)		& (mag)		& (mag)			& (mag)			& (mag) \\  
            \noalign{\smallskip}
\hline
            \noalign{\smallskip}
\endfirsthead
\caption{Photometry of correlated Tycho-2/2MASS sources in the Alnilam field (cont.).}\\
\hline
\hline
            \noalign{\smallskip}
Name			& TYC		& $B_T$		& $V_T$		& $J \pm \delta J$	& $H \pm \delta H$	& $K_{\rm s} \pm \delta K_{\rm s}$ \\  
			& 		& (mag)		& (mag)		& (mag)			& (mag)			& (mag) \\  
            \noalign{\smallskip}
\hline
            \noalign{\smallskip}
\endhead
\hline
\endfoot
HD 290690		& 4770 1664 1	& 11.423 	& 10.847        & 9.771   0.023 & 9.513   0.033 & 9.445   0.021 \\ %
...			& 4770 1072 1	& 11.747 	& 10.271        & 8.056   0.018 & 7.479   0.034 & 7.339   0.018 \\ %
HD 36980 AB		& 4770 1635 1	& 9.044  	& 9.007         & 8.797   0.024 & 8.898   0.042 & 8.812   0.019 \\ %
RX J0535.6--0152 AB	& 4770 1685 1	& 11.660 	& 12.113        & 10.541  0.022 & 10.223  0.028 & 10.167  0.021 \\ %
...			& 4766 2029 1	& 12.161 	& 10.987        & 9.678   0.023 & 9.240   0.027 & 9.127   0.019 \\ %
HD 290691		& 4766 1575 1	& 10.505 	& 10.244        & 9.926   0.024 & 9.817   0.030 & 9.821   0.021 \\ %
...			& 4767 3   1	& 12.214 	& 10.230        & 7.132   0.023 & 6.381   0.031 & 6.111   0.020 \\ %
HD 290686 		& 4766 1897 1	& 11.290 	& 11.048        & 10.372  0.019 & 10.343  0.031 & 10.221  0.023 \\
\end{longtable}

\begin{longtable}{llccccc}
\caption{\label{allTYC2Mmintaka} Photometry of correlated Tycho-2/2MASS sources in the Mintaka field.}\\ %
\hline
            \noalign{\smallskip}
Name		& TYC		& $B_T$		& $V_T$		& $J \pm \delta J$	& $H \pm \delta H$	& $K_{\rm s} \pm \delta K_{\rm s}$ \\  
		& 		& (mag)		& (mag)		& (mag)			& (mag)			& (mag) \\  
            \noalign{\smallskip}
\hline
            \noalign{\smallskip}
\endfirsthead
\caption{Photometry of correlated Tycho-2/2MASS sources in the Mintaka field (cont.).}\\
\hline
\hline
            \noalign{\smallskip}
Name			& TYC		& $B_T$		& $V_T$		& $J \pm \delta J$	& $H \pm \delta H$	& $K_{\rm s} \pm \delta K_{\rm s}$ \\  
			& 		& (mag)		& (mag)		& (mag)			& (mag)			& (mag) \\  
            \noalign{\smallskip}
\hline
            \noalign{\smallskip}
\endhead
\hline
\endfoot
...     	     	& 4766 590  1   & 11.055	& 10.325	& 9.108   0.020 & 8.841   0.059 & 8.722   0.019 \\ %
HD 290585     	     	& 4766 602  1   & 11.329	& 11.012	& 10.355  0.029 & 10.241  0.035 & 10.196  0.029 \\ %
...	     		& 4766 644  1   & 12.670	& 11.104	& 8.727   0.037 & 8.035   0.034 & 7.870   0.024 \\ %
...     	     	& 4766 687  1   & 13.049	& 11.051	& 8.192   0.020 & 7.439   0.059 & 7.183   0.027 \\ %
...     	     	& 4766 733  1   & 12.498	& 12.726	& 10.954  0.023 & 10.651  0.026 & 10.575  0.023 \\ %
...     	     	& 4766 744  1   & 10.966	& 10.382	& 9.457   0.023 & 9.247   0.023 & 9.133   0.021 \\ %
...     	     	& 4766 790  1   & 13.208	& 12.133	& 10.803  0.030 & 10.403  0.026 & 10.284  0.026 \\ %
HD 290513     	     	& 4766 828  1   & 10.852	& 10.245	& 8.569   0.019 & 8.270   0.044 & 8.239   0.023 \\
\end{longtable}

   \begin{table*}
      \caption[]{Known young stars in the Alnilam field and in Tycho-2/2MASS.} 
         \label{TYC2Malnilam}
     $$ 
         \begin{tabular}{lllccll}
            \hline
            \hline
            \noalign{\smallskip}
Name			&	& Sp.       	& $\alpha$	& $\delta$	& Features 		& Reference(s)\\  
			&	& type	        & (J2000)	& (J2000)	& of youth		& \\
            \noalign{\smallskip}
            \hline
            \noalign{\smallskip}
       {HD 36980}~AB	& *     & B9.5V	      	& 05 35 15.17	& --01 54 07.6	& OBA		     	& SCo71, Gu76, Gi83 \\ 
       {RX J0535.6--0152}~AB& * & G6V   	& 05 35 33.39	& --01 52 35.7	& T~Tau, Li~{\sc i}, X-rays & Al96, Al00, Ma07 \\ %
       {HD 290691}	&       & B9V	      	& 05 35 58.54	& --01 51 46.6	& OBA		      	& Ne95 \\ %
       {HD 290686}	&	& A5V		& 05 36 18.37	& --01 50 49.7	& OBA			& Ne95 \\ 
       {HD 290684}	&	& A2V		& 05 35 35.83	& --01 46 51.7	& OBA			& AK59, Gu76, Gi83 \\ 
       {HD 37285}~AB	& *	& A0V		& 05 37 20.41	& --01 46 16.3	& OBA			& Gu76, Gi83 \\
\hline
         \end{tabular}
     $$ 
   \end{table*}

   \begin{table*}
      \caption[]{Known young stars in the Mintaka field and in Tycho-2/2MASS.} 
         \label{TYC2Mmintaka}
     $$ 
         \begin{tabular}{lllccll}
            \hline
            \hline
            \noalign{\smallskip}
Name			&	& Sp.       	& $\alpha$	& $\delta$	& Features 		& Reference(s)\\  
			&	& type	        & (J2000)	& (J2000)	& of youth		& \\
            \noalign{\smallskip}
            \hline
            \noalign{\smallskip}
       {HD 290516}    	&       & B9.5V	      	& 05 30 16.50	& --00 53 29.4	& OBA		      	& WH78, Gu81 \\  
       {HD 290515}    	& *     & A0V	      	& 05 30 29.15	& --00 50 50.7	& OBA		      	& AK59, Gu81 \\  
       {HD 36669}     	&       & B9V	      	& 05 33 25.00	& --00 47 01.5	& OBA		      	& Gu76, AL77 \\ 
       {HD 36605}     	&       & B8V	      	& 05 32 49.65	& --00 42 47.2	& OBA		      	& CB66, Gu76 \\ 
       {HD 290578}    	&       & B9V	      	& 05 32 41.24	& --00 38 05.6	& OBA		      	& Gu79, Ne95 \\ 
       {HD 290492}~AB 	& *     & A0.5Vb+    	& 05 31 18.75	& --00 29 24.2	& OBA			& WH78, PG97 \\
\hline
         \end{tabular}
     $$ 
   \end{table*}

   \begin{table*}
      \caption[]{Stars of unknown association membership status in the Alnilam 
      field and in Tycho-2/2MASS.} 
         \label{TYC2Oalnilam}
     $$ 
         \begin{tabular}{lllccccll}
            \hline
            \hline
            \noalign{\smallskip}
Name				&	& Sp.   & $\alpha$	& $\delta$	& $\mu_\alpha \cos{\delta}$	& $\mu_\delta$  	& Remarks	& Reference(s) \\  
				&	& type  & (J2000)	& (J2000)	& (mas\,a$^{-1}$)		& (mas\,a$^{-1}$)	&		& \\  
            \noalign{\smallskip}
            \hline
            \noalign{\smallskip}
       {HD 37038}~AB		& *	& F0V	& 05 35 32.87	& --01 50 46.1	&  +4.3$\pm$1.0			&  +5.5$\pm$1.0		&		& SCo71, No97, No04 \\ 
       {TYC 4766 1605 1}	&	& --	& 05 35 08.00	& --01 50 19.2	& --1.6$\pm$1.5 		&  +3.4$\pm$1.5 	&		& {\em cs08} \\ %
       {TYC 4767  935 1}	&	& --	& 05 37 32.72	& --01 48 23.9	& --6.1$\pm$2.0 		&  +9.1$\pm$2.1 	&		& {\em cs08} \\ %
       {HD 36863}		& *	& A3V	& 05 34 34.34	& --01 44 37.4	& --7.5$\pm$1.2			& --6.8$\pm$1.1		&		& Lee68, WH78, Gi83 	\\ 
       {TYC 4766 1699 1}	&	& --	& 05 36 02.32	& --01 44 05.2	&  +1.9$\pm$1.4 		& --3.9$\pm$1.4 	&		& {\em cs08} \\ %
       {TYC 4766  121 1}	&	& --	& 05 34 42.31	& --01 40 48.8	&  +1.6$\pm$2.3 		& --0.6$\pm$2.2 	&		& {\em cs08} \\
\hline
         \end{tabular}
     $$ 
   \end{table*}

   \begin{table*}
      \caption[]{Stars of unknown association membership status in the Mintaka 
      field and in Tycho-2/2MASS.} 
         \label{TYC2Omintaka}
     $$ 
         \begin{tabular}{lllccccll}
            \hline
            \hline
            \noalign{\smallskip}
Name				&	& Sp.   & $\alpha$	& $\delta$	& $\mu_\alpha \cos{\delta}$	& $\mu_\delta$  	& Remarks	& Reference(s) \\  
				&	& type  & (J2000)	& (J2000)	& (mas\,a$^{-1}$)		& (mas\,a$^{-1}$)	&		& \\  
            \noalign{\smallskip}
            \hline
            \noalign{\smallskip}
       {HD 290585}	    	& *	& F1V	& 05 32 21.79	& --01 01 19.2	&  +0.6$\pm$1.4			&  +1.7$\pm$1.3	   	&		& Ne95, He06 \\ 
       {TYC 4766  733 1}	&	& --	& 05 31 52.22	& --00 58 37.2	&  +2.2$\pm$2.3		   	& --0.9$\pm$2.5		&		& {\em cs08} \\ 
       {HD 290513}		& *	& F0V? 	& 05 30 43.18	& --00 56 58.9	& --1.8$\pm$1.3 		&--10.7$\pm$1.3 	&		& Ne95 \\ 
       {TYC 4766  790 1}	& *	& --	& 05 33 10.86	& --00 57 30.7	& 0		   		& 0			&		& {\em cs08} \\ 
       {HD 290582}		&	& F0V	& 05 33 15.51	& --00 55 15.7	& --1.0$\pm$0.9		   	&  +2.2$\pm$0.9		&		& Ne95 \\ 
       {HD 290508}		&	& F5V	& 05 31 03.79	& --00 53 55.1	& --1.5$\pm$1.1		   	&  +2.0$\pm$1.1		&		& Ne95 \\
\hline
         \end{tabular}
     $$ 
   \end{table*}

\begin{longtable}{lllccccl}
\caption{\label{TYC2Nalnilam} Fore- and background stars in the Alnilam field 
and in Tycho-2/2MASS.}\\ 
\hline
\hline
            \noalign{\smallskip}
Name				&	& Sp.   & $\alpha$	& $\delta$	& $\mu_\alpha \cos{\delta}$	& $\mu_\delta$  	& Reference(s) \\  
				&	& type  & (J2000)	& (J2000)	& (mas\,a$^{-1}$)		& (mas\,a$^{-1}$)	& \\  
            \noalign{\smallskip}
\hline
            \noalign{\smallskip}
\endfirsthead
\caption{Fore- and background stars in the Alnilam field and in 
Tycho-2/2MASS (cont.).}\\
\hline
\hline
            \noalign{\smallskip}
Name				&	& Sp.   & $\alpha$	& $\delta$	& $\mu_\alpha \cos{\delta}$	& $\mu_\delta$  	& Reference(s) \\  
				&	& type  & (J2000)	& (J2000)	& (mas\,a$^{-1}$)		& (mas\,a$^{-1}$)	& \\  
            \noalign{\smallskip}
\hline
            \noalign{\smallskip}
\endhead
\hline
\endfoot
       {HD 290690}		&	& G0V	& 05 35 44.98	& --01 56 03.8	& --4.9$\pm$1.3			&  +3.9$\pm$1.3		& Ne95 			\\ 
       {TYC 4770 1072 1}	&	& K--M:	& 05 36 26.83	& --01 56 03.1	& --1.5$\pm$1.3 		&  +9.6$\pm$1.3 	& {\em cs08} 		\\ 
       {TYC 4766 2029 1}	&	& F--G:	& 05 34 48.41	& --01 51 46.8	&--11.1$\pm$1.8			&  +8.3$\pm$1.8		& {\em cs08} 		\\ 
       {TYC 4767    3 1}	&	& K--M:	& 05 37 38.35	& --01 51 22.9	&  +0.9$\pm$1.3			& --9.4$\pm$1.3		& Kr03, {\em cs08} 	\\ 
       {TYC 4766 1891 1}	&	& F--G:	& 05 36 41.21	& --01 46 34.2	& --6.0$\pm$2.0			&--33.5$\pm$2.1		& {\em cs08} 		\\ 
       {HD 290685}		&	& K0V	& 05 36 28.07	& --01 45 15.5	& --9.9$\pm$1.0			&  +3.3$\pm$1.1		& Ne95 			\\ 
       {HD 290683}		&	& F8V	& 05 35 51.32	& --01 42 20.7	&--10.6$\pm$1.5			& +26.7$\pm$1.5		& Ne95 			\\ 
       {IRAS 05354--0142}	& *	& C/S?  & 05 37 56.58	& --01 40 50.2	&  +1.2$\pm$2.0			&  +5.8$\pm$2.0		& Kr03, {\em cs08} 	\\
\end{longtable}

\begin{longtable}{lllccccl}
\caption{\label{TYC2Nmintaka} Fore- and background stars in the Mintaka field 
and in Tycho-2/2MASS.}\\ 
\hline
\hline
            \noalign{\smallskip}
Name				&	& Sp.   & $\alpha$	& $\delta$	& $\mu_\alpha \cos{\delta}$	& $\mu_\delta$  	& Reference(s) \\  
				&	& type  & (J2000)	& (J2000)	& (mas\,a$^{-1}$)		& (mas\,a$^{-1}$)	& \\  
            \noalign{\smallskip}
\hline
            \noalign{\smallskip}
\endfirsthead
\caption{Fore- and background stars in the Mintaka field and in 
Tycho-2/2MASS (cont.).}\\
\hline
\hline
            \noalign{\smallskip}
Name				&	& Sp.   & $\alpha$	& $\delta$	& $\mu_\alpha \cos{\delta}$	& $\mu_\delta$  	& Reference(s) \\  
				&	& type  & (J2000)	& (J2000)	& (mas\,a$^{-1}$)		& (mas\,a$^{-1}$)	& \\  
            \noalign{\smallskip}
\hline
            \noalign{\smallskip}
\endhead
\hline
\endfoot
       {HD 290586}		&	& G5V	& 05 32 09.22	& --01 01 31.5	&  +1.9$\pm$1.3 		&--45.9$\pm$1.3 	& Ne95 		\\ 
       {TYC 4766  644 1}	&	& K--M: & 05 32 29.78	& --01 00 38.8	& --2.7$\pm$1.5 		&  +1.2$\pm$1.5 	& {\em cs08} 	\\ 
       {TYC 4766  687 1}	&	& K--M: & 05 31 35.28	& --00 59 42.9	&  +2.7$\pm$1.5 		&  +1.5$\pm$1.5 	& {\em cs08} 	\\ 
       {HD 290509}		&	& F0V	& 05 31 26.41	& --00 58 33.1	& +29.4$\pm$1.1 		& --9.5$\pm$1.1 	& Ne95 		\\ 
       {TYC 4766  716 1}	&	& F--G:	& 05 33 09.06	& --00 52 30.8	&  +8.5$\pm$1.3 		&--37.3$\pm$1.4 	& {\em cs08} 	\\ 
       {HD 290514}		&	& F8V	& 05 30 38.34	& --00 52 09.0	& +13.0$\pm$1.2 		&--13.6$\pm$1.2 	& Ne95 		\\ 
       {HD 290581}		&	& G0V 	& 05 33 18.73	& --00 50 12.4	& +57.0$\pm$2.1 		&--57.8$\pm$1.9 	& Ne95, H{\o}g98 \\ 
       {HD 290584}		&	& G5V   & 05 31 54.41	& --00 49 56.0	& +28.8$\pm$1.3 	       &--111.3$\pm$1.3 	& Ne95 		\\
\end{longtable}

\newpage 
   \begin{table*}
      \caption[]{Known and new young stars in the Alnilam field and in DENIS/2MASS.}  
         \label{DE2Malnilam}
     $$ 
         \begin{tabular}{lcccccll}
            \hline
            \hline
            \noalign{\smallskip}
Name			&	& $\alpha$  	& $\delta$     & $i \pm \delta i$      & $K_{\rm s} \pm \delta K_{\rm s}$	& Features       		& Reference(s) \\ %
			&	& (J2000)  	& (J2000)      & (mag)  	       & (mag)  			 	& of youth			& \\  
            \noalign{\smallskip}
            \hline
            \noalign{\smallskip}
       {E\,Ori 2--1328}		& *	& 05 36 26.09	& --01 39 46.8  &  16.380 1.00  &  13.792 0.040 			& Li~{\sc i}, low $g$		& B\'e03a	\\ 
       {E\,Ori 2--1868}		& *	& 05 36 21.56	& --01 37 49.4  &  18.0   0.3  	&  15.266 0.115 			& low $g$			& B\'e03b	\\ 
       {E\,Ori 1--388}		& *	& 05 37 54.30	& --01 37 20.6  &  17.397 1.00  &  13.908 0.049  			& low $g$			& B\'e03a	\\ 
       {E\,Ori 1--1644}		& *	& 05 36 57.12	& --01 36 29.4  &  16.752 1.00  &  13.884 0.044 			& Li~{\sc i}			& B\'e03a	\\ 
       {E\,Ori 2--878}		& * 	& 05 36 18.28	& --01 27 08.3  &  17.353 1.00  &  14.567 0.066 			& low $g$			& B\'e03b	\\ 
       {E\,Ori 2--705}		& * 	& 05 35 59.47	& --01 25 38.9  &  16.8   0.3  	&  14.336 0.059 			& low $g$			& B\'e03b	\\
\hline
         \end{tabular}
     $$ 
   \end{table*}

   \begin{table*}
      \caption[]{Known and new young stars in the Mintaka field and in DENIS/2MASS.}  
         \label{DE2Mmintaka}
     $$ 
         \begin{tabular}{lcccccll}
            \hline
            \hline
            \noalign{\smallskip}
Name			&	& $\alpha$  	& $\delta$     & $i \pm \delta i$      & $K_{\rm s} \pm \delta K_{\rm s}$	& Features       		& Reference(s) \\ %
			&	& (J2000)  	& (J2000)      & (mag)  	       & (mag)  			 	& of youth			& \\  
            \noalign{\smallskip}
            \hline
            \noalign{\smallskip}
       {CVSO 124}		& *	& 05 33 49.04  	& --00 46 26.0  &  13.479 0.03  &  11.119 0.021  			& Li~{\sc i}, H$\alpha$		& Br05, McG06  	\\ 
       {Kiso A--0903 231}	& 	& 05 32 29.42  	& --00 41 39.7  &  13.209 0.03  &  11.329 0.021  			& H$\alpha$		     	& Ko89		\\ 
       {Kiso A--0903 221}	& *	& 05 32 10.16 	& --00 37 12.4  &  12.488 0.03  &  10.662 0.021  			& H$\alpha$		     	& Ko89		\\ 
       {Kiso A--0904 4}		& *	& 05 32 25.78   & --00 36 53.3  &  12.895 0.03  &  10.287 0.021  			& Li~{\sc i}, H$\alpha$		& Wi89, Ko89, Br05 \\ 
       {Kiso A--0904 10}	& 	& 05 33 00.78   & --00 36 20.8  &  11.271 0.03  &   9.170 0.021 			& H$\alpha$		     	& Ko89		\\ 
       {IRAS 05307--0038}	& *	& 05 33 18.68	& --00 36 14.1	&  11.859 0.03  &   8.627 0.021       			& Li~{\sc i}, H$\alpha$, H$\beta$, mIR 	& YC94, Yun97	\\
\hline
         \end{tabular}
     $$ 
   \end{table*}

\newpage 
   \begin{table*}
      \caption[]{Stars of unknown status in the Alnilam and Mintaka fields 
      and in DENIS/2MASS.}  
         \label{DE2O}
     $$ 
         \begin{tabular}{lccccclll}
            \hline
            \hline
            \noalign{\smallskip}
Name 				&	& $\alpha$	& $\delta$	& $i \pm \delta i$	& $K_{\rm s} \pm \delta K_{\rm s}$	& Remarks				& Reference(s)		& Field 	\\  
 				&	& (J2000)	& (J2000)	& (mag)			& (mag)					& 					& 			&		\\  
            \noalign{\smallskip}
            \hline
            \noalign{\smallskip}
       {E\,Ori 1--726}		& 	& 05 37 39.00	& --01 39 46.8  & 17.966 1.00  		& 15.035 0.150 				& M6.0					& B\'e03a		& Alnilam 	\\ 
       {E\,Ori 2--1982}		& *	& 05 36 05.54	& --01 39 00.5  & 19.0   0.3  		& 15.152 0.110 				& M8.0:					& B\'e03b		& Alnilam 	\\ 
       {[NYS99] B--16}		& 	& 05 38 35.98   & --01 38 31.6  & 15.426 0.05   	& 13.532 0.034   			& X-rays? 				& Na99  		& Alnilam 	\\ %
       {[OSP2002] OriI--2N 4}	& *	& 05 38 02.60   & --01 34 39.3  & 16.560 1.00   	& 14.096 0.059   			& H$\alpha$? 				& OGP02  		& Alnilam 	\\ %
       {2E 1343}		& 	& 05 34 58.20   & --01 12 37.9  & 11.635 0.02   	& 10.296 0.021   			& X-rays? 				& McD94  		& Alnilam 	\\ %
       {BD--01~969B}		& *	& 05 36 22.96   & --01 10 30.9  &  9.968 1.00   	&  7.643 0.027   			& ``Alnilam B'' 			& Sc1886  		& Alnilam 	\\
\hline
         \end{tabular}
     $$ 
   \end{table*}
%

   \begin{table*}
      \caption[]{Fore- and background objects in the Alnilam and Mintaka 
      fields and in DENIS/2MASS.}  
         \label{DE2N}
     $$ 
         \begin{tabular}{lccccclll}
            \hline
            \hline
            \noalign{\smallskip}
Name 			&	& $\alpha$	& $\delta$	& $i \pm \delta i$	& $K_{\rm s} \pm \delta K_{\rm s}$	& Sp.		& References	& Field 	\\  
 			&	& (J2000)	& (J2000)	& (mag)			& (mag)					& type		& 		&		\\  
            \noalign{\smallskip}
            \hline
            \noalign{\smallskip}
       {X~Ori}  	& *	& 05 37 38.65	& --01 46 16.7	& 11.4--15.5 ($V_T$)	&  1.55  0.05	   			& M8  		& Wo1904, TMW05	& Alnilam	\\ 
       {HD~290769}  	& 	& 05 37 52.38	& --01 27 37.2	&  10.25  1.0		&  9.735 0.023	   			& F8 		& Ne95		& Alnilam	\\ 
       {Ruber 1}	& * 	& 05 31 51.40	& --00 45 51.2  &  15.717 0.05		& 12.927 0.027	   			& late MV: 	& {\em cs08} 	& Mintaka 	\\ 
       {G~99--18}  	& *	& 05 34 00.01	& --00 42 03.5	&  11.941 0.03 		& 10.056 0.019	   			& K--M:   	& GBT61, SG03	& Aln.+Min.     \\ %
       {HD~290571}  	& 	& 05 34 00.31	& --00 03 49.6	&  10.634 0.03 		&  9.989 0.023	   			& F9 		& Ne95		& Mintaka	\\ 
       {HD~290562}  	& 	& 05 32 39.45	&  +00 11 30.3	&  11.068 0.04		& 10.433 0.022	   			& F5 		& Ne95		& Mintaka	\\ 
       {Ruber 2}	& *	& 05 33 11.20	&  +00 13 58.4 	&  15.346 0.06		& 12.144 0.028	   			& late MIII: 	& {\em cs08} 	& Mintaka 	\\ 
       \noalign{\smallskip}
            \hline
         \end{tabular}
     $$ 
   \end{table*}
   \begin{table*}
      \caption[]{Probable reddened stars in DENIS/2MASS associated to remnant
      molecular clouds in the Alnilam and Mintaka fields.}  
         \label{DE2Mred}
     $$ 
         \begin{tabular}{lccccll}
            \hline
            \hline
            \noalign{\smallskip}
2MASS 					& $i \pm \delta i$	& $J \pm \delta J$	& $H \pm \delta H$	& $K_{\rm s} \pm \delta K_{\rm s}$ 	& Associated 	& Field	\\  
designation				& (mag)			& (mag)			& (mag)			& (mag)					& to cloud 	&	\\  
            \noalign{\smallskip}
            \hline
            \noalign{\smallskip}
       {2MASS J05380477--0145322}	&  16.743 1.00  	& 13.557 0.042  	&  12.735 0.035  	&  12.164 0.030 			& ORI I--2 	& Alnilam \\ 
       {2MASS J05332555--0038348}       &  16.691 0.10  	& 14.120 0.030  	&  12.799 0.032  	&  12.285 0.026       			& IC 423 	& Mintaka \\ 
       {2MASS J05332310--0038196}       &  18.251 0.21 	 	& 15.961 0.076  	&  14.853 0.055  	&  14.291 0.077       			& IC 423 	& Mintaka \\ 
       {2MASS J05332631--0037462}       &  16.335 0.08  	& 14.272 0.033  	&  13.057 0.029  	&  12.590 0.027       			& IC 423 	& Mintaka \\ 
       {2MASS J05332263--0036565}       &  18.258 0.21  	& 15.425 0.049  	&  14.072 0.032  	&  13.397 0.029       			& IC 423 	& Mintaka \\ 
       {2MASS J05332198--0036306}       &  16.912 0.11  	& 14.896 0.033  	&  13.757 0.027  	&  13.174 0.021       			& IC 423 	& Mintaka \\ 
       {2MASS J05332038--0036225}       &  16.251 0.08  	& 14.415 0.026  	&  13.489 0.022  	&  13.065 0.036       			& IC 423 	& Mintaka \\ 
        \noalign{\smallskip}
            \hline
         \end{tabular}
     $$ 
   \end{table*}
%

   \begin{table*}
      \caption[]{Galaxies in the Alnilam field and in 2MASS.}  
         \label{gx2Malnilam}
     $$ 
         \begin{tabular}{llllll}
            \hline
            \hline
            \noalign{\smallskip}
Name				&   &  
Name				&   &  
Name				&   \\ %
            \noalign{\smallskip}
            \hline
            \noalign{\smallskip}
       {2MASX J05355304--0156495}&   &   
       {2MASX J05363979--0156145}&   &  	
       {2MASX J05362575--0154324}&   \\  	
       {2MASX J05364032--0152595}&   &         
       {2MASX J05350166--0152546}&   &         
       {2MASX J05360480--0151535}&   \\  	
       {2MASX J05360091--0151165}&   &         
       {2MASX J05365752--0150543}&   &         
       {2MASX J05374550--0150164}&   \\  	
       {2MASX J05375584--0146454}&   &  
       {2MASX J05344538--0145356}&   &         
       {2MASX J05381044--0145266}&   \\  	
       {2MASX J05371064--0145203}&   &  
       {2MASX J05352182--0144308}&   &         
       {2MASX J05371671--0144223}&   \\  	
       {2MASX J05365111--0143514}&   &  	
       {2MASS J05345451--0143256}& * &  
       {2MASX J05372418--0141053}&   \\  	
       {2MASX J05381458--0139166}&   &         
       {2MASX J05365376--0138594}&   &  
       {2MASX J05350345--0138506}&   \\
\hline
         \end{tabular}
     $$ 
   \end{table*}
%

   \begin{table*}
      \caption[]{Galaxies in the Mintaka field and in 2MASS.}  
         \label{gx2Mmintaka}
     $$ 
         \begin{tabular}{llllllll}
            \hline
            \hline
            \noalign{\smallskip}
Name				&   &  
Name				&   &  
Name				&   \\ %
            \noalign{\smallskip}
            \hline
            \noalign{\smallskip}
       {2MASX J05325501--0053516}&   &       
       {2MASX J05314416--0053202}&   &       
       {2MASX J05305116--0052384}&   \\       
       {2MASX J05335414--0052272}&   &       
       {2MASX J05314556--0050562}&   &       
       {2MASS J05315182--0050178}&   \\ 
       {2MASS J05314896--0049143}&   & 
       {2MASS J05314022--0049037}&   & 
       {2MASX J05323724--0048535}&   \\       
       {2MASX J05325934--0048476}&   &       
       {2MASX J05323377--0048285}&   &       
       {2MASX J05325308--0048086}&   \\       
       {2MASX J05313329--0047492}&   &       
       {2MASX J05320003--0045342}&   & 
       {2MASX J05321351--0045065}&   \\       
       {2MASX J05335608--0042102}&   &       
       {2MASX J05343163--0041185}&   &       
       {2MASS J05330941--0040445}&   \\ 
       {2MASX J05330015--0039056}&   &       
       {2MASX J05301419--0037282}&   &       
       {2MASX J05314250--0028332}&   \\
\hline
         \end{tabular}
     $$ 
   \end{table*}

\begin{longtable}{lccccccc}
\caption{\label{DE2Oalnilam} Young star and brown dwarf candidates in the
Alnilam field and in DENIS/2MASS.}\\ 
\hline
\hline
            \noalign{\smallskip}
Identification		&	& $\alpha$  	& $\delta$ 	& $i \pm \delta i$	& $J \pm \delta J$	& $H \pm \delta H$	& $K_{\rm s} \pm \delta K_{\rm s}$ 	\\ %
			&	& (J2000)  	& (J2000)  	& (mag)  		& (mag)  		& (mag)  		& (mag)  				\\  
            \noalign{\smallskip}
\hline
            \noalign{\smallskip}
\endfirsthead
\caption{Young star and brown dwarf candidates in the
Alnilam field and in DENIS/2MASS (cont.).}\\
\hline
\hline
            \noalign{\smallskip}
Identification		&	& $\alpha$  	& $\delta$ 	& $i \pm \delta i$	& $J \pm \delta J$	& $H \pm \delta H$	& $K_{\rm s} \pm \delta K_{\rm s}$ 	\\ %
			&	& (J2000)  	& (J2000)  	& (mag)  		& (mag)  		& (mag)  		& (mag)  				\\  
            \noalign{\smallskip}
\hline
            \noalign{\smallskip}
\endhead
\hline
\endfoot
       {Annizam  2511174}	&	& 05 36 30.72	& --01 53 43.9  &  14.669 1.00  &  13.059 0.026  &  12.321 0.032  &  12.103 0.026 \\ 
       {Annizam  2445191}	&	& 05 35 42.05	& --01 52 08.2  &  16.014 0.06  &  14.264 0.031  &  13.629 0.040  &  13.279 0.032 \\ 
       {Annizam  2557157}	&	& 05 37 19.82	& --01 51 18.7  &  14.686 1.00  &  13.116 0.023  &  12.483 0.028  &  12.176 0.021 \\ 
       {Annizam  2609154}	&	& 05 37 28.93	& --01 51 13.2  &  17.777 1.00  &  15.178 0.039  &  14.567 0.035  &  14.163 0.064 \\ 
       {Annizam  2511204}	&	& 05 35 05.80	& --01 50 28.3  &  16.424 0.07  &  14.597 0.035  &  13.960 0.043  &  13.726 0.039 \\ 
       {Annizam  2306164}	&	& 05 36 55.48	& --01 49 02.0  &  16.444 1.00  &  14.739 0.030  &  14.133 0.038  &  13.746 0.035 \\ 
       {Annizam  2292164}	&	& 05 36 54.01	& --01 48 54.2  &  16.538 1.00  &  14.459 0.028  &  13.773 0.038  &  13.505 0.035 \\ 
       {Annizam  2684144}	&	& 05 37 57.03	& --01 48 29.8  &  16.584 1.00  &  14.603 0.035  &  13.962 0.032  &  13.723 0.054 \\
\end{longtable}

\begin{longtable}{llcccccc}
\caption{\label{DE2Omintaka} Young star and brown dwarf candidates in the
Mintaka field and in DENIS/2MASS.}\\ 
\hline
\hline
            \noalign{\smallskip}
Identification			&	& $\alpha$  	& $\delta$ 	& $i \pm \delta i$	& $J \pm \delta J$	& $H \pm \delta H$	& $K_{\rm s} \pm \delta K_{\rm s}$ 	\\ %
				&	& (J2000)  	& (J2000)  	& (mag)  		& (mag)  		& (mag)  		& (mag)  				\\  
            \noalign{\smallskip}
\hline
            \noalign{\smallskip}
\endfirsthead
\caption{Young star and brown dwarf candidates in the
Mintaka field and in DENIS/2MASS (cont.).}\\
\hline
\hline
            \noalign{\smallskip}
Identification			&	& $\alpha$  	& $\delta$ 	& $i \pm \delta i$	& $J \pm \delta J$	& $H \pm \delta H$	& $K_{\rm s} \pm \delta K_{\rm s}$ 	\\ %
				&	& (J2000)  	& (J2000)  	& (mag)  		& (mag)  		& (mag)  		& (mag)  				\\  
            \noalign{\smallskip}
\hline
            \noalign{\smallskip}
\endhead
\hline
\endfoot
       {Mantaqah 2389165}     &       & 05 32 42.57   & --00 56 20.5  &  17.036 0.11  &  14.988 0.035  &  14.474 0.052  &  14.011 0.058       \\ %
       {Mantaqah 2668148}     &       & 05 33 35.14   & --00 55 35.1  &  15.176 0.05  &  13.693 0.026  &  12.718 0.024  &  12.159 0.024       \\ 
       {Mantaqah 2473149}     &       & 05 33 24.52   & --00 53 23.9  &  14.933 0.05  &  13.358 0.026  &  12.737 0.027  &  12.430 0.024       \\ %
       {Mantaqah 1995178}     &       & 05 32 05.67   & --00 51 09.7  &  15.705 0.05  &  14.021 0.022  &  13.357 0.023  &  13.139 0.037       \\ %
       {Mantaqah 2691223}     &       & 05 29 58.45   & --00 50 50.3  &  18.219 1.00  &  15.521 0.056  &  14.783 0.056  &  14.503 0.086       \\ 
       {Mantaqah 1970186}     &       & 05 31 45.96   & --00 50 35.0  &  14.409 0.03  &  12.785 0.024  &  12.103 0.027  &  11.836 0.026       \\ 
       {Mantaqah 2445218}     &       & 05 30 19.69   & --00 49 59.7  &  16.398 1.00  &  14.254 0.032  &  13.532 0.032  &  13.225 0.034       \\ %
       {Mantaqah 2041159}     & *     & 05 32 48.68   & --00 49 44.8  &  14.644 0.04  &  12.969 0.033  &  12.396 0.040  &  12.088 0.033       \\
\end{longtable}

   \begin{table*}
      \caption[]{Remarkable fore- and background objects in the comparison
      fields.}  
         \label{remarkable}
     $$ 
         \begin{tabular}{lcccccll}
            \hline
            \hline
            \noalign{\smallskip}
Name 			&	& $\alpha$	& $\delta$	& $V_T$			& $K_{\rm s}$		& Remarks			& References		\\  
 			&	& (J2000)	& (J2000)	& (mag)			& (mag)			& 				& 			\\  
            \noalign{\smallskip}
            \hline
            \noalign{\smallskip}
       {TYC~5945~228~1}	&	& 06 09 19.74	& --21 04 36.2	& 11.16$\pm$0.07	&  5.66$\pm$0.03	& IRAS				& {\em cs08}		\\  
       {Albus~1}	& *	& 06 06 13.39	& --20 21 07.3	& 11.80$\pm$0.14	& 12.76$\pm$0.03	& bright He-B subdwarf		& CS07, VKAS07 		\\  
       {$19$~Lep}  	& 	& 06 07 41.64	& --19 09 57.1	& 5.480$\pm$0.009	&  1.05$\pm$0.06	& M1III, var., $d$ = 220$\pm$30\,pc & Ad26 		\\  
       {CSS~205}	&	& 06 15 33.16	& --18 42 16.6	& 11.63$\pm$0.10	& 5.850$\pm$0.017	& IRAS, S-type star		& St84			\\  
       {HD~43429}  	& 	& 06 15 17.72	& --18 28 37.8	& 6.102$\pm$0.009	&   3.6$\pm$0.2		& K1III, $d$ = 64$\pm$3\,pc	& Ha76 			\\  
       {IRAS~06101--1659}&	& 06 12 20.36	& --16 59 52.4	& 10.97$\pm$0.05	& 4.885$\pm$0.018	& IRAS				& {\em cs08}		\\
\hline
         \end{tabular}
     $$ 
   \end{table*}

\begin{longtable}{ll ll}
\caption{\label{references} Reference abbreviations used in this Appendix.}\\
\hline
\hline
            \noalign{\smallskip}
Abbreviation	& Reference 			& Abbreviation	& Reference 			\\ %
            \noalign{\smallskip}
\hline
            \noalign{\smallskip}
\endfirsthead
\caption{Reference abbreviations used in this Appendix.}\\
\hline
\hline
            \noalign{\smallskip}
Abbreviation	& Reference 			& Abbreviation	& Reference 			\\ %
            \noalign{\smallskip}
\hline
            \noalign{\smallskip}
\endhead
\hline
\endfoot
Sc1886		& Schoenfeld 1886 		& Re88  	  & Renson 1988 		  \\ %
Ha1904		& Hartmann 1904 		& Ko89  	  & Kogure et~al. 1989  	  \\ %
Wo1904		& Wolf 1904 			& Wi89  	  & Wiramihardja et~al. 1989	  \\ %
Fr09		& Frost 1909 			& DHu91 	  & Dong \& Hu 1991		  \\ %
Da16		& Daniel 1916 			& ML91  	  & Morrell \& Levato 1991	  \\ %
AJ17		& Adams \& Joy 1917 		& Wi91  	  & Wiramihardja et~al. 1991	  \\ %
CP18	       & Cannon \& Pickering 1918--1924 & Gr92  	  & Grillo et~al. 1992  	  \\ %
Ad26		& Adams et~al. 1926 		& MMR92   & Mart\'{\i}n, Magazz\`u \& Rebolo 1992 \\
\end{longtable}

\end{document}